\documentclass[preprint,12pt]{elsarticle}

\usepackage{graphicx}
\usepackage[update,prepend]{epstopdf}

\usepackage{amssymb}
\usepackage{amsmath}

\usepackage{url}

\usepackage[includeheadfoot,twoside,
            inner=2.5cm,outer=2.5cm,
           top=2cm,bottom=2cm]{geometry}

\usepackage{pdfpages}

\usepackage{fancyhdr}
\usepackage{titlesec}

\usepackage{nomencl}
\makenomenclature

\fancypagestyle{noheader}{
  \fancyhf{}
  \fancyfoot[R]{\thepage}
}

\newcommand{\bfu}{\mathbf{u}}

\newcommand{\rr}{\rho_r}
\newcommand{\zzero}{\zeta^{(0)}}
\newcommand{\zone}{\zeta^{(1)}}
\newcommand{\ztwo}{\zeta^{(2)}}
\newcommand{\zhat}{\hat{\zeta}}

\newcommand{\D}{\mathrm{d}}

\newcommand{\Uzero}{U^{(0)}}
\newcommand{\Uone}{U^{(1)}}
\newcommand{\Utwo}{U^{(2)}}

\newcommand{\eps}{\varepsilon}
\DeclareMathOperator{\sech}{sech}

\newcommand{\tmin}{t_\mathrm{min}}
\newcommand{\tmax}{t_\mathrm{max}}

\newcommand{\ximin}{\xi_\mathrm{min}}
\newcommand{\ximax}{\xi_\mathrm{max}}

\usepackage[normalem]{ulem}

\journal{Physica D}

\begin{document}

\begin{frontmatter}



{\it Dedicated to the memory of Vladimir Zakharov}\\

\title{Internal {\color{black}solitary} and cnoidal waves of moderate amplitude in a two-layer fluid: the extended KdV equation approximation}


\author{Nerijus Sidorovas$^a$, Dmitri Tseluiko$^a$, Wooyoung Choi$^b$, Karima Khusnutdinova$^{a, *}$} 

\affiliation{organization={Department of Mathematical Sciences, Loughborough University},
            city={Loughborough},
            postcode={LE11 3TU}, 
            country={UK}}
            
\affiliation{organization={Department of Mathematical Sciences, New Jersey Institute of Technology},
            city={Newark},
            postcode={NJ 07102-1982}, 
            country={USA}}

\begin{abstract}
We consider travelling internal waves in a two-layer fluid with linear shear currents from the viewpoint of the extended Korteweg-de Vries (eKdV) equation derived from a strongly-nonlinear long-wave model. Using an asymptotic Kodama-Fokas-Liu near-identity transformation, we map the eKdV equation to the Gardner equation. This improved Gardner equation has a different cubic nonlinearity coefficient and an additional transport term compared to the frequently used truncated Gardner equation. We then construct approximate solitary and cnoidal wave solutions of the eKdV equation using this mapping and test validity and performance of these approximations, as well as performance of the truncated and improved Gardner and eKdV equations, by comparison with direct numerical simulations of the strongly-nonlinear two-layer long-wave parent system in the absence of currents.
\end{abstract}

\begin{graphicalabstract}
\includegraphics[scale=0.2]{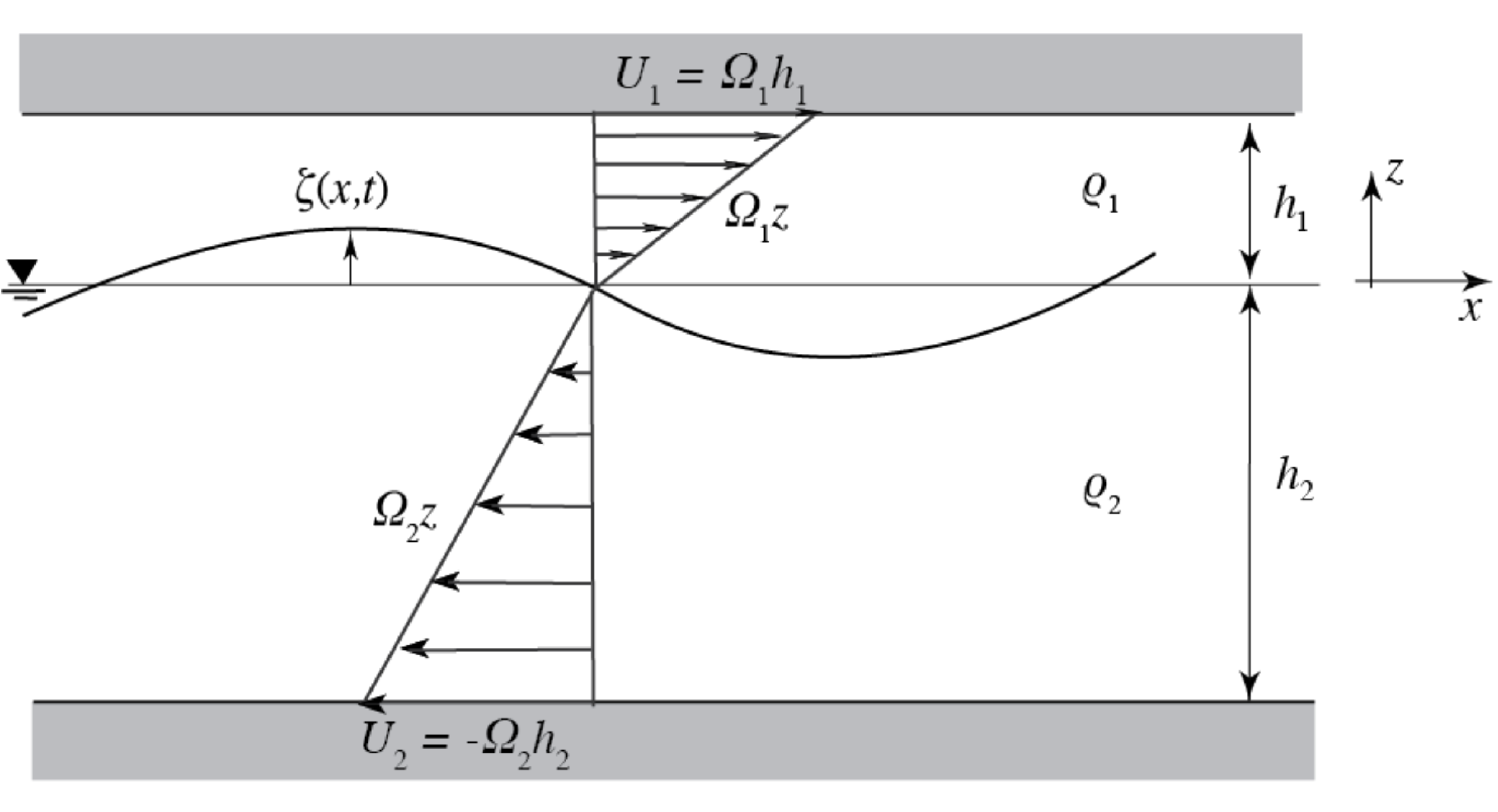}
\end{graphicalabstract}


\begin{keyword} Internal wave, extended KdV equation, solitary wave, cnoidal wave\\[2ex]


$*$ {\it Corresponding author}, email: K.Khusnutdinova@lboro.ac.uk
\end{keyword}

\end{frontmatter}



\section{Introduction}
\label{chap:chapter_4}

A two-layer fluid serves  as a convenient testing ground both for the reduced mathematical models and experimental studies of internal waves. In the absence of an underlying current, the parameter space of the model includes the densities of the top and bottom layers, $\rho_1$ and $\rho_2$, and their unperturbed depths, $h_1$ and $h_2$, respectively. In the shallow water configuration considered below, the ratio of the two depths, $h_r = h_1/h_2$ is $O(1)$. For typical oceanic conditions, when the density difference is small, the `rigid lid' approximation filters out the surface wave (barotropic mode), but has little effect on the internal wave (baroclinic mode). In what follows we consider the rigid-lid approximation.\\

Most of the initial theoretical progress has been made within the scope of the weakly-nonlinear long-wave models developed {\color{black} from} the Euler equations and applicable {\color{black} when both the amplitude parameter, $\varepsilon = a/h_2$,  and the wavelength parameter, $\delta = h_2/L$, are small, with $a$ and $L$ being a typical wave amplitude and a typical wavelength, respectively.} The `maximal balance' condition $\varepsilon = O(\delta^2)$ {\color{black} (balancing the leading order nonlinear and dispersive corrections at the same order)} allows one to derive the Korteweg-de Vries (KdV) equation 
\begin{equation}
A_T + \alpha A A_\xi + \beta A_{\xi \xi \xi} = 0, \quad \mbox{with} \quad \xi = x - ct, T = \varepsilon t
\end{equation}
as a uni-directional model  for   the leading-order term   $A(T, \xi)$ of the interfacial displacement as a function of the slow time variable $T$ and the characteristic variable $\xi$  \cite{B1966, Ben1966}.  Here, 
\begin{equation}
c = \left( \frac{g (\rho_2 - \rho_1) h_1 h_2}{\rho_2 h_1 + \rho_1 h_2} \right)^{1/2}
\end{equation}
 is the linear long-wave speed, and the nonlinearity and dispersion coefficients are given by
\begin{equation}
\alpha = \frac{3c(\rho_2 h_1^2 - \rho_1 h_2^2)}{2 h_1 h_2 (\rho_2 h_1 + \rho_1 h_2)}, \quad \beta = \frac{ch_1 h_2 (\rho_2 h_2 + \rho_1 h_1)}{6 (\rho_2 h_1 + \rho_1 h_2)}.
\end{equation}
The KdV equation was also derived for an arbitrary stratification and arbitrary depth-dependent parallel shear flow and generalised to account for a variable environment (see \cite{LB1974, MR1980, G1981, G2001} and references therein). \\

The extended KdV (eKdV) equation 
\begin{equation}
A_T + \alpha A A_\xi + \beta A_{\xi \xi \xi} + \varepsilon (\alpha_1 A^2 A_{\xi} + \gamma_1 A A_{\xi \xi \xi} + \gamma_2 A_{\xi} A_{\xi \xi} + \beta_1 A_{\xi \xi \xi \xi \xi})  = 0
\end{equation}
was introduced for a two-layer fluid to include higher-order nonlinear and dispersive effects  \cite{KB1981} (see also \cite{KRI2014, KST2018, HFS2022} and references therein for other settings). Later, it was also derived from the Euler equations for an arbitrary stratification and arbitrary depth-dependent parallel shear flow (see \cite{GPP2002} and references therein). However, {\color{black} computing}  the coefficients for a given density and current is not an easy task. The additional coefficients for the two-layer configuration can be found in \cite{GKKS2014} (see also \cite{KST2018}). Most studies of internal waves were devoted to the truncated version of this model, the Gardner equation {\color{black} (in what follows we will refer to this equation as the truncated Gardner equation)}:
\begin{equation}
A_T + \alpha A A_\xi + \beta A_{\xi \xi \xi} + \varepsilon \alpha_1 A^2 A_{\xi}  = 0,
\end{equation}
which is particularly useful when the nonlinearity coefficient $\alpha$ is small or vanishing (e.g.\ \cite{FO1986, GPT1999, TPLGH1999, HM2006, RKK2014,  OPSS2015, OPSS2024}). In particular,  in the ``Gardner minus" equation {\color{black} (with $\alpha_1<0$, assuming that  $\beta > 0$)}, solitons range from the small-amplitude KdV-type {\color{black} solitary waves} to the moderate-amplitude table-top {\color{black} solitary waves}, in qualitative agreement with the waves observed in the oceanic and laboratory settings.  We note that a different version of the extended KdV equation,  balancing the dispersive term of the KdV equation and the cubic term of the Gardner equation at the same order,  was derived and studied for the cases when the quadratic nonlinearity coefficient is small or vanishing in \cite{PS2006}. {\color{black} It was shown that this equation was asymptotically equivalent to an extended Gardner-type equation containing the additional quartic and quintic nonlinear terms and coinciding with the truncated Gardner equation to leading order, which indicated the significance of the truncated Gardner equation when the quadratic nonlinearity coefficient was small or vanishing.}   \\

The strongly-nonlinear two-layer model was introduced in order to improve the quantitative agreement with the available observations and numerical modelling of internal waves of large amplitude independently by Miyata \cite{M1985}, Maltseva \cite{Ma1989} and Choi and Camassa \cite{CC1999}. In what follows we will refer to this model as Miyata-Maltseva-Choi-Camassa (MMCC) model instead of the commonly used name `MCC model', acknowledging the previously unknown contribution by Maltseva \cite{Ma1989}. 
The MMCC model was  extended to the case of linear shear current in \cite{C2006}.  In the absence of background currents, the eKdV equation was derived from the MMCC model in \cite{CC1999}. 
\\

Alongside these developments, in different applied contexts there {\color{black} have been} studies providing some evidence that the eKdV equation could also be useful in generic situations when all coefficients of the leading-order KdV equation are non-vanishing, as a suitable model for the description of waves of moderate amplitude (see \cite{MS1996}-\cite{BHF2025} and references therein). In particular, in \cite{STCK2024}, we performed extensive long-time numerical experiments for concentric water waves within the scope of the 2D Boussinesq system and showed that the extended concentric KdV (cKdV) model indeed extended the range of validity of the weakly-nonlinear modelling to the waves of moderate amplitude, with the amplitude parameter $\eps \sim 0.5$.\\

The aim of the present study is twofold. Firstly, we derive the eKdV equation from the strongly-nonlinear two-layer model with linear shear currents  \cite{C2006} and derive its approximate solitary and cnoidal wave solutions with the help of a Kodama-Fokas-Liu near-identity transformation \cite{K1985a, K1985b, FL1996} to the Gardner equation, extending the results in \cite{GBK2020}. {\color{black} Secondly, we compare the performance of the reduced models} with the results of direct numerical simulations for the MMCC model in the absence of currents, illustrating their range of validity. The constructed approximate solutions can be used to study the effects of linear shear currents {\color{black} on the waves of small and moderate amplitude}. \\

The paper is organised as follows. In Section~2 we discuss the derivation of the eKdV equation from the strongly-nonlinear model developed in \cite{C2006}. In Section~3 we discuss the near-identity transformation to the Gardner equation and streamline the approach developed in \cite{GBK2020} to obtain the approximate solitary wave solution of the eKdV equation. In Section~4 we test the constructed approximation by performing numerical simulations for the parent strongly-nonlinear system in the absence of currents. In Section~5 we use the same near-identity transformation to obtain the approximate cnoidal wave solution of the eKdV equation. In Section~6 we test this approximation in numerical runs for the parent system in the absence of currents. 
 We conclude in Section~7.\\

\section{Derivation of eKdV equation from strongly-nonlinear model}
\label{sec:Derivation}

\begin{figure}
\centerline{\includegraphics[scale=0.2]{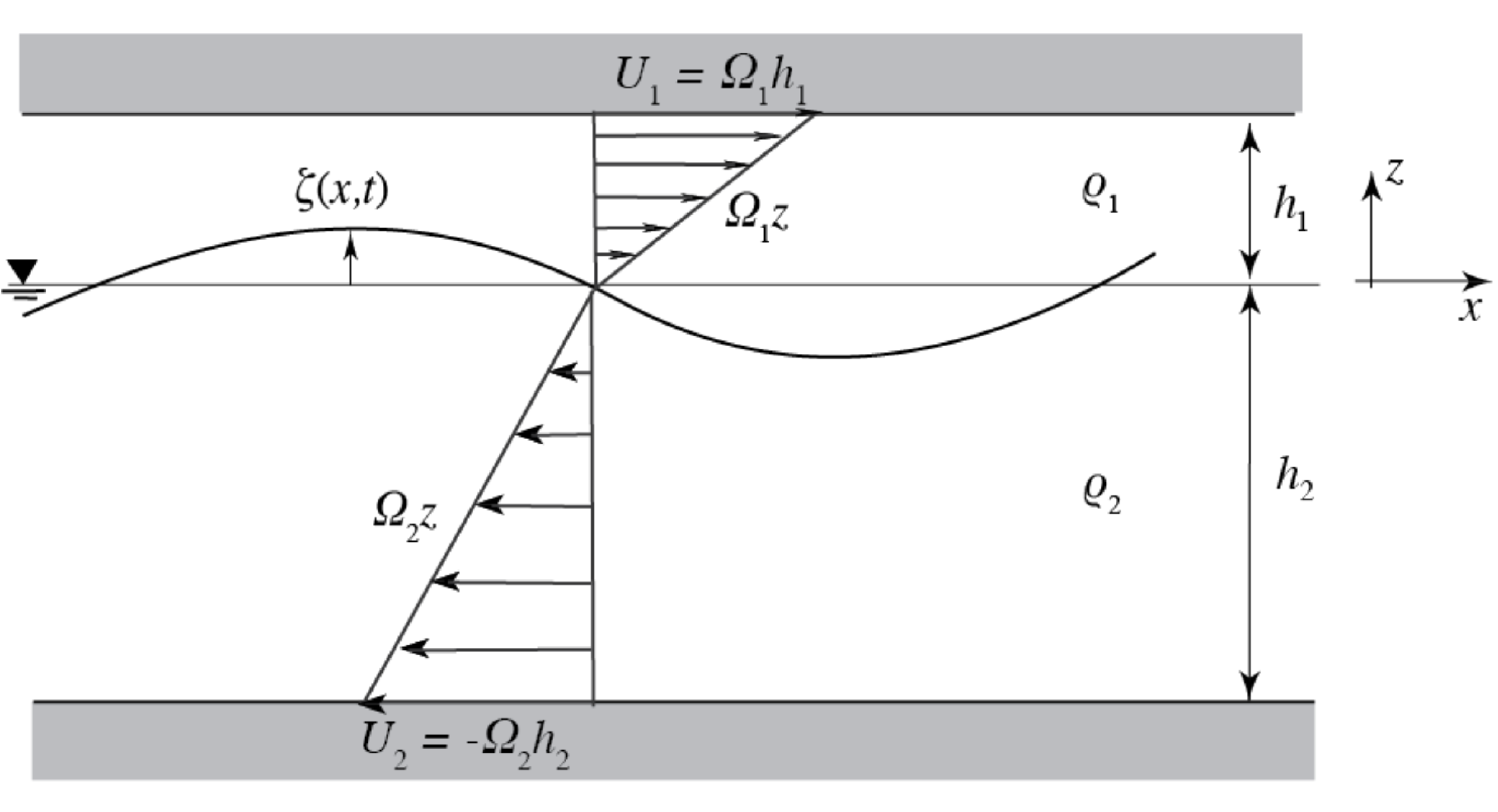}}
  \caption{Schematic of a two-layer fluid with linear shear currents.  }
  \label{fig:Fig2}
\end{figure}

In this section, we derive the eKdV equation describing the time evolution of the interfacial displacement $\zeta(x,t)$ for internal waves in a two-layer fluid with linear currents, see Figure~\ref{fig:Fig2}. Here, the top and bottom layers have constant undisturbed depths $h_{1,2}$ and constant densities $\rho_{1,2}$, respectively. Each layer has a linear shear current of strength $\Omega_{1,2} > 0$ which represents constant vorticity of the background uniform shear. The velocities at top and bottom boundaries can then be written as

\begin{equation}
U_1 = \Omega_1 h_1 > 0, \qquad U_2 = - \Omega_2 h_2 < 0.
\end{equation}
The thicknesses of the top and bottom layers are defined as
\begin{equation}
\eta_1 = h_1 - \zeta, \qquad \eta_2 = h_2 + \zeta.
\end{equation}
We also introduce $\bar{u}_{1,2}(x,t)$ to denote the depth-averaged perturbation velocities to background linear shear currents in the top and bottom layers. \\

The full strongly-nonlinear model with background linear shear currents involved can be found in \cite{C2006}, and we write it here in its dimensional form as
\begin{eqnarray}
\label{mcc_shear_1}
&\eta_{1t} + (\eta_1 \bar{u}_1)_x + U_1 \eta_{1x} \left( 1 - \dfrac{\eta_1}{h_1} \right) = 0, \\
\label{mcc_shear_2}
&\bar{u}_{1t} + \bar{u}_1 \bar{u}_{1x} + g \zeta_x + U_1 \bar{u}_{1x} = -\dfrac{P_x}{\rho_1} + \dfrac{1}{3 \eta_1} \left( \eta_1^3 \left[ G_1 + U_1 \bar{u}_{1xx} \left(1 - \dfrac{\eta_1}{h_1} \right) \right] \right)_x + O(\delta^4), \\
\label{mcc_shear_3}
&\eta_{2t} + (\eta_2 \bar{u}_2)_x + U_2 \eta_{2x} \left( 1 - \dfrac{\eta_2}{h_2} \right) = 0, \\
\label{mcc_shear_4}
&\bar{u}_{2t} + \bar{u}_2 \bar{u}_{2x} + g \zeta_x + U_2 \bar{u}_{2x} = -\dfrac{P_x}{\rho_2} + \dfrac{1}{3 \eta_2} \left( \eta_2^3 \left[ G_2 + U_2 \bar{u}_{2xx} \left(1 - \dfrac{\eta_2}{h_2} \right) \right] \right)_x + O(\delta^4),
\end{eqnarray}
where the auxiliary function $G_i$ is then defined as
\begin{equation}
G_i(x,t) = \bar{u}_{itx} + \bar{u}_i \bar{u}_{ixx} - \bar{u}_{ix}^2.
\end{equation}

Equations \eqref{mcc_shear_1} and \eqref{mcc_shear_3} represent conservation of mass in the two layers, whereas \eqref{mcc_shear_2}  and \eqref{mcc_shear_4} represent conservation of linear momentum. These momentum equations have small error terms of order $O(\delta^4)$ that we will omit. Additionally, the momentum equations also have $P(x,t)$ which describes the pressure at the interface, and this is how the two layers are linked together. In order to derive the eKdV equation, we will begin by eliminating $P$ and $\bar{u}_1$ to obtain two evolution equations for $\zeta$ and $\bar{u}_2$. Afterwards, we shall nondimensionalise these two equations before using multiple-scale expansions to derive the dimensionless eKdV equation. \\

Firstly, we can eliminate $\bar{u}_1$ by adding equations \eqref{mcc_shear_1} and \eqref{mcc_shear_3}, and rewriting 
\begin{equation}
\eta_i{_x} \left( 1 - \dfrac{\eta_i}{h_i} \right) \equiv \bigg[\eta_i \left( 1 - \dfrac{\eta_i}{2 h_i} \right) \bigg]_x
\end{equation}
to obtain
\begin{equation}
\left[ \eta_1 \bar{u}_1 + \eta_2 \bar{u}_2 + U_1 \eta_1 \left( 1 - \dfrac{\eta_1}{2 h_1} \right) + U_2 \eta_2 \left( 1 - \dfrac{\eta_2}{2 h_2} \right) \right]_x = 0 ,
\end{equation}
where we have used the fact that $\eta_1 + \eta_2 = h_1 + h_2 = \text{const.}$ By integrating both sides, and assuming undisturbed conditions in the far-field (so that $\bar{u}_{i} \to 0$, $\eta_i \to h_i$ as $x\to \infty$), we obtain
\begin{equation}
\eta_1 \bar{u}_1 + \eta_2 \bar{u}_2 + U_1 \eta_1 \left( 1 - \dfrac{\eta_1}{2 h_1} \right) + U_2 \eta_2 \left( 1 - \dfrac{\eta_2}{2 h_2} \right) = \dfrac{U_1 h_1 + U_2 h_2}{2}.
\end{equation}
From this, we deduce
\begin{equation}
\bar{u}_1 = - \dfrac{\eta_2}{\eta_1} \bar{u}_2 + \dfrac{U_1 h_1 + U_2 h_2}{2 \eta_1} - U_1 \left( 1 - \dfrac{\eta_1}{2 h_1} \right) - U_2 \dfrac{\eta_2}{\eta_1} \left( 1 - \dfrac{\eta_2}{2 h_2} \right).
\label{eq:u1}
\end{equation}
Pressure $P$ can be eliminated by introducing the dimensionless density ratio $\rr = \rho_1/\rho_2$ and subtracting \eqref{mcc_shear_2} multiplied by $\rr$ from \eqref{mcc_shear_4} to get
\begin{equation}
\label{mcc_shear_24}
\begin{aligned}
\bar{u}_{2t} & - \rr \bar{u}_{1t} + \bar{u}_2 \bar{u}_{2x} - \rr \bar{u}_1 \bar{u}_{1x} + g (1-\rr) \zeta_x + U_2 \bar{u}_{2x} - \rr U_1 \bar{u}_{1x} \\ & = \dfrac{1}{3 \eta_2} \left( \eta_2^3 \left[ G_2 + U_2 \bar{u}_{2xx} \left(1 - \dfrac{\eta_2}{h_2} \right) \right] \right)_x - \dfrac{\rr}{3 \eta_1} \left( \eta_1^3 \left[ G_1 + U_1 \bar{u}_{1xx} \left(1 - \dfrac{\eta_1}{h_1} \right) \right] \right)_x.
\end{aligned}
\end{equation}
Equations \eqref{mcc_shear_3} and \eqref{mcc_shear_24} (with $u_1$ eliminated using \eqref{eq:u1}) are now our dimensional system for two unknowns $\zeta$ and $\bar{u}_2$. In addition to $h_{1,2}$, we also introduce $a,\,\lambda$, and $c_0$ as the characteristic amplitude, wavelength, and speed of the internal wave, respectively.  Here, we use $c_0 = \sqrt{g h_2 (1-\rr)}$.  We introduce the amplitude and wavelength parameters,  
\begin{equation}
\eps = \dfrac{a}{h_2}, \quad \delta = \dfrac{h_2}{\lambda},
\end{equation}
respectively, and nondimensionalise the variables as follows:
\begin{equation}
\label{scalings}
 x = \lambda x^*, \quad t = \dfrac{\lambda}{c_0} t^*, \quad \bar{u}_i = \eps {\color{black} c_0} \bar{u}_i^*, \quad \zeta = a \zeta^*, \quad U_i = c_0 U^*_i.
\end{equation}
In what follows, we omit the asterisks. Applying these scalings to \eqref{mcc_shear_3} and \eqref{mcc_shear_24} gives us 
the system
\begin{eqnarray}
\label{dimless_shear_mcc1}
&\zeta_t + ( \eta_2 \bar{u}_2 )_x - \eps U_2 \zeta \zeta_x = 0, \\
\label{dimless_shear_mcc2}
&\begin{aligned}
(\bar{u}_{2} & - \rr \bar{u}_{1})_t + \eps ( \bar{u}_2 \bar{u}_{2x} - \rr \bar{u}_1 \bar{u}_{1x} ) + \zeta_x  + (U_2 \bar{u}_2 - \rr U_1 \bar{u}_1)_x \\ 
 =  & \dfrac{\delta^2}{3} \left[ \dfrac{\left( \eta_2^3 \left[ \mathcal{G}_2 - \eps U_2 \zeta \bar{u}_{2xx} \right] \right)_x}{\eta_2} - \rr \dfrac{\left( \eta_1^3 \left[ h_r \mathcal{G}_1 + \eps U_1 \zeta \bar{u}_{1xx} \right] \right)_x}{h_r \eta_1} \right],
\end{aligned}
\end{eqnarray}
where the dimensionless thicknesses of the layers $\eta_i$ are given by
\begin{equation}
\label{dimless_etas}
\eta_1 = h_r - \eps \zeta, \qquad \eta_2 = 1 + \eps \zeta,
\end{equation}
with $h_r = h_1/h_2$ being the depth ratio, and the dimensionless auxiliary functions $\mathcal{G}_i$ here are
\begin{equation}
\label{dimless_Gs}
\mathcal{G}_i(x,t) = \bar{u}_{itx} + \eps \bar{u}_i \bar{u}_{ixx} - \eps \bar{u}_{ix}^2.
\end{equation}

In this context, we must have $\rr < 1$ for stable stratification, and the sign of $h_r-1$ determines whether we have waves of elevation or waves of depression. We consider the KdV regime $\eps = \delta^2$, but aim to consider waves of greater amplitude. \\

Next, we seek solutions of system \eqref{dimless_shear_mcc1} - \eqref{dimless_shear_mcc2} in the form of asymptotic multiple-scale expansions
\begin{align}
\label{zetaexpansion} \zeta & = \zzero + \eps \zone + \eps^2 \ztwo + {\color{black} O(\eps^3)} \\[2mm]
\label{u2expansion} \bar{u}_2 & = \Uzero + \eps \Uone + \eps^2 \Utwo +  {\color{black} O(\eps^3)}
\end{align}
{\color{black} as $\eps \to 0$}, where all $\zeta^{(i)}$ and $u^{(i)}$ depend on the slow time variable $T = \eps t,$ and on the characteristic variable $\xi = x - v t.$ Here, $v>0$ denotes the characteristic speed of the right-propagating interfacial wave. Upon substituting these expansions into system \eqref{dimless_shear_mcc1}-\eqref{dimless_shear_mcc2}, we can now consider equations at each order of $\eps$ and derive the evolution equations for $\zeta^{(i)}$ and $u^{(i)}$. At leading order, i.e.  $O(1)$, we obtain:
\begin{equation}
\label{eKdVorderO_shear}
\begin{cases}
\Uzero_\xi - v \zzero_\xi = 0, \\[2mm]
\dfrac{h_r U_2 + \rr U_1 - v (h_r + \rho_r)}{h_r} \Uzero_\xi + \zzero_\xi = 0,
\end{cases}
\end{equation}
which is a linear system for $\zzero_\xi$ and $\Uzero_\xi$. Requiring that the determinant of this linear system is zero gives the {\color{black} linear long-wave speed}  
\begin{equation}
\label{long_wave_speed}
v = \dfrac{ h_r U_2 + \rr U_1 + \sqrt{ (h_r U_2 + \rr U_1)^2 + 4 h_r (h_r + \rr) } }{ 2(h_r + \rr) },
\end{equation}
which is positive for right-propagating waves. For the left-propagating waves, the sign in front of the square root should be taken as minus instead. Converting this {\color{black} speed} back to its dimensional form agrees with that in \cite{C2006}. From $\eqref{eKdVorderO_shear}$, assuming undisturbed conditions in the far-field, we find
\begin{equation}
\Uzero = v \zzero.
\end{equation}

Using this result, the system at $O(\eps)$ is as follows:
\begin{align}
\label{eKdVorder1_shear}
\left\{
\begin{aligned}
\Uone_\xi - v \zone_\xi & = - \zzero_T + (U_2 - 2 v) \zzero \zzero_\xi, \\[2mm]
\dfrac{h_r U_2 + \rr U_1 - v(h_r + \rr)}{h_r} \Uone_\xi + \zone_\xi & = - \dfrac{v(h_r + \rr)}{h_r} \zzero_T - {\color{black} \hat c_0} \zzero \zzero_\xi - \dfrac{v^2(1 + h_r \rr)}{3} \zzero_{\xi\xi\xi},
\end{aligned}
\right.
\end{align}
with
\begin{equation}
\begin{aligned}
 \hat c_0 & =  {\color{black} \dfrac{-h_r \rr  U_1 U_2-\rr  U_1^2+v(2 h_r \rr  U_1+h_r \rr  U_2+3 \rr  U_1) + v^2 (h_r^2 - 3\rr - 2h_r \rr)}{h_r^2}.}
\end{aligned}
\end{equation}
Adding the first equation in system \eqref{eKdVorder1_shear} with the second one multiplied by $v$ and dividing by  $ -{v^2(h_r + \rr)}/{h_r} - 1$ gives the following  evolution equation for $\zzero$:
\begin{equation}
\label{mcc_kdv_shear}
\zzero_T + \tilde{\alpha} \zzero \zzero_\xi + \tilde{\beta} \zzero_{\xi\xi\xi} = 0,
\end{equation}
where
\begin{equation}
\label{mcc_kdv_coeffs_shear}
\begin{aligned}
\tilde{\alpha} & = 
{\color{black} \frac{-U_2 h_r^2 + v(2 h_r^2 -U_1 U_2 h_r \rho_r - U_1^2 \rho_r) + \rho_r v^2 (2 U_1 h_r + U_2 h_r + 3 U_1) + v^3(h_r^2 - 3\rr - 2 h_r \rr)}{h_r^2 + v^2 h_r (h_r + \rr)}},
\\ \tilde{\beta} & = \frac{v^3 h_r (1 + h_r \rr)}{3h_r + 3 v^2 (h_r + \rr)}.
\end{aligned}
\end{equation}
From the first equation in $\eqref{eKdVorder1_shear}$, after substituting in for $\zzero_T$ term using the KdV equation \eqref{mcc_kdv_shear} and assuming undisturbed conditions as $\xi \to \infty$, we obtain
\begin{equation}
\Uone = \hat c_1 \zone + \hat c_2 {\zzero}^2 + \hat c_3 \zzero_{\xi\xi},
\end{equation}
with coefficients $\hat{c}_{1,2,3}$ {\color{black} given by} 
\begin{align}
\hat c_1 & = \frac{h_r + v^2 (h_r + \rr)}{ 2v(h_r + \rr) - h_r U_2 - \rr U_1}, \\
\hat c_2 & = 
{\color{black} - \frac{v^2 \left(h_r^2+4 h_r \rr +3 \rr \right) - v \left(h_r^2 U_2+2 h_r \rr  U_1+2 h_r \rr  U_2+3 \rr  U_1\right)+ \rr U_1 ( h_r U_2 + U_1)}{ 2h_r[2 v (h_r + \rr) - h_r U_2 - \rr U_1] }},
\\
\hat c_3 & = \dfrac{v^2 h_r}{3} \cdot \frac{1 + h_r \rr}{2 v (h_r + \rr) - h_r U_2 - \rr U_1}.
\end{align}

The system at the next order, $O(\eps^2)$, gives two equations which are too cumbersome to write down, but by taking the same linear combination between these equations as at $O(\eps)$, we derive the eKdV model for the truncation $\zhat = \zzero + \eps \zone$ in the form
\begin{equation}
\label{mcc_ekdv_shear}
\zhat_{T} + {\color{black} \tilde{\alpha} } \zhat \zhat _{\xi}  + \tilde{\beta} \zhat_{\xi \xi \xi} + \eps \big[ \tilde{\alpha}_1 \zhat^2 \zhat_{\xi} + \tilde{\gamma}_1 \zhat \zhat_{\xi \xi \xi} + \tilde{\gamma}_2 \zhat_{\xi} \zhat_{\xi \xi} + \tilde{\beta}_1 \zhat_{\xi \xi \xi \xi \xi} \big] = 0,
\end{equation}
{\color{black} where we have omitted any $O(\eps^2)$ terms.} The coefficients in the $O(\eps)$ extension are listed in Appendix A. \\

We can take $\Omega_{1,2} \to 0$ in order to reduce to the eKdV equation without current which has been derived in \cite{CC1999} (albeit in dimensional form). Alternatively to the $\Omega$ limits, we can also take $\rr \to 0$ to obtain the eKdV for surface waves with an underlying current. Here, all $h_r,U_1$ terms disappear, 
leaving us with coefficients depending only on $U_2$ that determines the strength of the underlying linear shear current. By applying the limits $U_{1,2} \to 0$, $\rr \to 0$ simultaneously, we reduce to the eKdV equation as derived from the Green-Naghdi equations for surface waves (see \cite{STCK2024}). \\

The current-free reduction is obtained by taking $\Omega_{1,2} \to 0$ giving us the eKdV equation
\begin{align}
\label{mcc_ekdv}
\zhat_T + \alpha \zhat \zhat_\xi + \beta \zhat_{\xi\xi\xi} + \eps \bigg[ \alpha_1 \zhat^2 \zhat_\xi + \gamma_1 \zhat \zhat_{\xi\xi\xi} + \gamma_2 \zhat_\xi \zhat_{\xi\xi} + \beta_1 \zhat_{\xi\xi\xi\xi\xi} \bigg] = 0,
\end{align}
where the coefficients $\alpha,\beta$ are reduced from $\tilde{\alpha},\tilde{\beta}$ and are given by
\begin{equation}
\label{mcc_kdv_coeffs}
 \alpha = \dfrac{3v}{2} \cdot \dfrac{h_r^2 - \rr}{h_r(h_r + \rho_r)}, \qquad \beta = \dfrac{v}{6} \cdot \dfrac{h_r (1 + h_r \rr)}{h_r + \rr}, \quad \mbox{with} \quad v = \sqrt{ \dfrac{ h_r}{ h_r + \rr } }, 
\end{equation}
and the extension coefficients are
\begin{align}
\alpha_1 & = -\dfrac{3v}{8} \cdot \frac{h_r^4 + 8h_r \rr + 14 h_r^2 \rr + 8 h_r^3 \rr + \rr^2 }{h_r^2(h_r + \rho_r)^2}, \qquad \beta_1 = \dfrac{v}{24} \cdot \dfrac{h_r^2(1+h_r \rr)^2}{(h_r + \rr)^2}, \\[5mm]
\gamma_2 & = \dfrac{v}{24} \cdot \frac{23h_r^2 - 31 \rr - 8 h_r \rr + 8 h_r^2 \rr + 31 h_r^3 \rr - 23 h_r \rr^2}{(h_r + \rho_r)^2}, \\[5mm]
\gamma_1 & = \dfrac{v}{12} \cdot \dfrac{5h_r^2 - 7 \rr - 2 h_r \rr + 2h_r^2 \rr + 7h_r^3\rr - 5h_r\rr^2}{(h_r + \rho_r)^2}.
\end{align}

The quadratic nonlinearity coefficient $\alpha$ vanishes under the condition $\rr = h_r^2$. This is known  as the critical depth ratio (e.g.\ \cite{CC1999} and references therein). This condition is important as this is when the KdV equation loses the ability to accurately describe the nonlinear effects as the wave evolves. Under such a regime, we are motivated to look further to the next order and obtain the eKdV equation for internal waves incorporating the nonlinear effects in this case. However, the eKdV equation is also useful in the generic case as a model for the waves of moderate amplitude, {\color{black} which is shown in the subsequent sections.}

\section{Approximate solutions for solitary waves of eKdV equation}
\label{sec:Solitons}

In this section we consider the eKdV equation \eqref{mcc_ekdv} for the plane waves. In several physical contexts there have been developments in constructing {\color{black} solitary wave} solutions of the eKdV equation (see \cite{KST2018, HFMS2022} and references therein). In particular, approximate solutions of the eKdV equation have been constructed using the Kodama-Fokas-Liu near-identity transformations (NITs) \cite{K1985a, K1985b,FL1996}. Indeed,
the generic eKdV equation can be mapped to another equation of the same type by a nonlocal near-identity transformation
\begin{equation}
\zeta = B - \varepsilon \left ( a B^{2} + b B_{\xi\xi} + c B_{\xi} \int B \ {\rm d} {\xi} + d B_{T} \xi \right ),
\label{NIT0}
\end{equation}
where $a$, $b$, $c$, $d$ are arbitrary constants. Kodama used a transformation with $c=d=0$ to reduce the eKdV equation to the next member of the KdV hierarchy, while Fokas and Liu used the full transformation to reduce it to the KdV equation itself, with the accuracy up to $O(\varepsilon^2)$. \\

These and modified near-identity transformations for reducing the eKdV equation to the KdV equation have been successfully used to obtain particular approximate solutions to the eKdV equation from the known solutions of the KdV equation (e.g., solitary wave solutions  \cite{MS1996} and the undular bore solution \cite{MS2006}). Once the {\color{black} solitary wave} solution $B$ is obtained in the reduced model, applying the inverse near-identity transformation yields an approximate solution to the eKdV equation. \\

In \cite{GBK2020}, the near-identity transformations were used to map the eKdV equation to the Gardner equation instead of the KdV equation. We streamline this approach by introducing the near-identity transformation for the function $B(\xi,T)$ in the form
\begin{equation}
\label{NIT}
\zeta = B - \eps \bigg[ a B^2 + b B_{\xi\xi} + c B_\xi \bigg( \int_{{\color{black}\xi_0}}^\xi B(\xi',T) \ \D \xi' + f(T) \bigg) + d B_T \xi \bigg]
\end{equation}
where the coefficients $a,b,c,d$ must be chosen in a particular way depending on {\color{black} the coefficients of the eKdV equation ($\alpha,\beta,\alpha_1,\gamma_1,\gamma_2,\beta_1$)} in order to reduce to {\color{black} a Gardner-type equation}. In the transformation, $f(T)$ is an arbitrary function which will be {\color{black} determined later}. The required choice of coefficients in this transformation is
\begin{equation}
a=0, \quad b = \dfrac{5 \alpha \beta_1 + 3 \beta (\gamma_1 -  \gamma_2)}{6 \alpha \beta}, \quad  c = \dfrac{3 \beta \gamma_1 - 4\alpha \beta_1}{9 \beta^2}, \quad d = -\dfrac{\beta_1}{3 \beta^2}.
\end{equation}

Hence, we deduce the Gardner-type equation as
\begin{equation}
\label{mG}
B_T + \alpha B B_\xi + \beta B_{\xi\xi\xi} + \eps \alpha_2 B^2 B_\xi - \eps c f'(T) B_{\xi} = 0,
\end{equation}
where 
\begin{equation}
\alpha_2 = \alpha_1 - \frac{c \alpha}{2} + d \alpha^2 = \frac{18 \alpha_1 \beta^2 - 2 \alpha^2 \beta_1 - 3 \alpha \beta \gamma_1}{18 \beta^2},
\label{alpha2}
\end{equation}
{\color{black} provided 
\begin{equation}
\varepsilon (-4 \alpha \beta_1 + 3 \beta \gamma_1) [\alpha B(\xi_0, t)^2 + 2 \beta B_{xx}(\xi_0, t)] B_x = 0. 
\label{remainder}
\end{equation}
For solitary waves,  we choose $\xi_0 = -\infty$ to ensure that this remainder term vanishes.} 
Reduction from \eqref{mG} to the Gardner equation
\begin{equation}
\label{G}
B_T + \alpha B B_{\tilde \xi }+ \beta B_{{\tilde \xi} {\tilde \xi} {\tilde \xi}} + \eps \alpha_2 B^2 B_{\tilde \xi }= 0.
\end{equation} 
is then possible by virtue of the transformation
\begin{equation}
\tilde{\xi} = \xi + \eps c f(T).
\label{txidef}
\end{equation}

The Gardner equation \eqref{G} in variables $(\tilde{\xi},T)$ admits a one-parameter family of {\color{black} solitary wave} solutions in the form
\begin{equation}
\label{GSol}
B(\tilde{\theta}) = \dfrac{M}{1 + F \cosh( G \tilde{\theta} )}, \quad \tilde{\theta} = \tilde{\xi} - \tilde{v} T,
\end{equation}
where $M$ is a free parameter, and
\begin{align}
F  = \sqrt{1 + \eps \dfrac{M \alpha_2}{\alpha}}, \quad
G  = \sqrt{\dfrac{M \alpha}{6 \beta}}, \quad
\tilde{v}  = \dfrac{M \alpha}{6}.
\end{align}
This means that the corresponding solution to the Gardner-type equation \eqref{mG} has the form
\begin{equation}
\label{mGSol}
B(\theta) = \dfrac{M}{1 + F \cosh(G \theta)}, \quad \theta = \xi - \tilde{v} T + \eps c f(T).
\end{equation}
We note the phase correction compared to \cite{GBK2020} (also to \cite{B2023}, where the same formulae as in \cite{GBK2020} have been applied to internal waves in a two-layer fluid). Here, in the limit as $F \to 0$ (i.e. $M \to M^* = -\alpha/(\eps\alpha_2)$), the Gardner {\color{black} solitary wave} becomes wider with the top of the  {\color{black} solitary wave} being flatter, i.e. it becomes a ``table-top" soliton. Using this solution for $B$, we apply the near-identity transformation \eqref{NIT} and obtain an approximate solution to \eqref{mcc_ekdv}:
\begin{align}
\zeta = \dfrac{M}{1 + F \cosh (G \theta)} \bigg[ 1 & - \dfrac{\eps b F G^2 ( F \cosh(2 G \theta) - 2 \cosh(G \theta) - 3F )}{2 ( 1 + F \cosh(G \theta) )^2 } \nonumber \\ 
& + \dfrac{\eps F \sinh(G \theta)}{1 + F \cosh(G \theta)} \bigg( G c f(T) - G d \tilde{v} \xi + \dfrac{2 c M \omega(\theta)}{\sqrt{1-F^2}} \bigg) \bigg],
\end{align}
where
\begin{equation}
\omega(\theta) = \mathrm{arctanh} \bigg[ \sqrt{\dfrac{1-F}{1+F}} \tanh \bigg( \dfrac{G \theta}{2} \bigg) \bigg].
\end{equation}

At this stage, for $\zeta$ to be a travelling wave solution,  it should only be a function of $\theta$. This yields
\begin{equation}
G c f(T) - G d \tilde{v} \xi = C \theta + D
\end{equation}
for some constants $C,D$. Indeed, when we choose $C = - G d \tilde{v}$ and $D=0$, the above relation simplifies (after using the definition of $\theta$) as follows:
\begin{equation}
G c f(T) - G d \tilde{v} \xi = G \theta d \tilde{v}^2 T - \eps G \theta c d \tilde{v} f(T) - G d \tilde{v} \xi \quad \iff \quad f(T) = \dfrac{d \tilde{v}^2}{c(1 + \eps d \tilde{v})} T.
\end{equation}
The function $f(T)$ is hence defined as a linear function of $T$ and this choice ensures we obtain a travelling wave solution of the eKdV equation via the near-identity transformation. Therefore, the approximate solution $\zeta = \zeta(\theta)$ of the eKdV equation \eqref{mcc_ekdv} is given by
\begin{align}
\label{ekdv_soliton_sol_travelling}
\zeta = \dfrac{M}{1 + F \cosh (G \theta)} \bigg[ 1 & - \dfrac{\eps b F G^2 ( F \cosh(2 G \theta) - 2 \cosh(G \theta) - 3F )}{2 ( 1 + F \cosh(G \theta) )^2 } \nonumber \\ 
& + \dfrac{\eps F \sinh(G \theta)}{1 + F \cosh(G \theta)} \bigg( -G d \tilde{v} \theta  + \dfrac{2 c M \omega(\theta)}{\sqrt{1-F^2}} \bigg) \bigg],
\end{align}
{\color{black} with the accuracy up to $O(\eps^2)$.}

\section{Numerical modelling of solitary waves}
\label{sec:PlaneWaveNumerics}

In this section, we compare numerical simulations of the current-free MMCC model, the derived eKdV equation, and also two types of Gardner equations: truncated, and improved, extending the preliminary comparisons developed in  \cite{B2023}.  Here, the truncated Gardner equation is the eKdV equation \eqref{mcc_ekdv}, but with $\gamma_{1,2} = \beta_1 = 0$, whereas the improved Gardner equation is precisely \eqref{G} that we obtain by applying the near-identity transformation. \\

\begin{figure}
\centering
\includegraphics[scale=0.55]{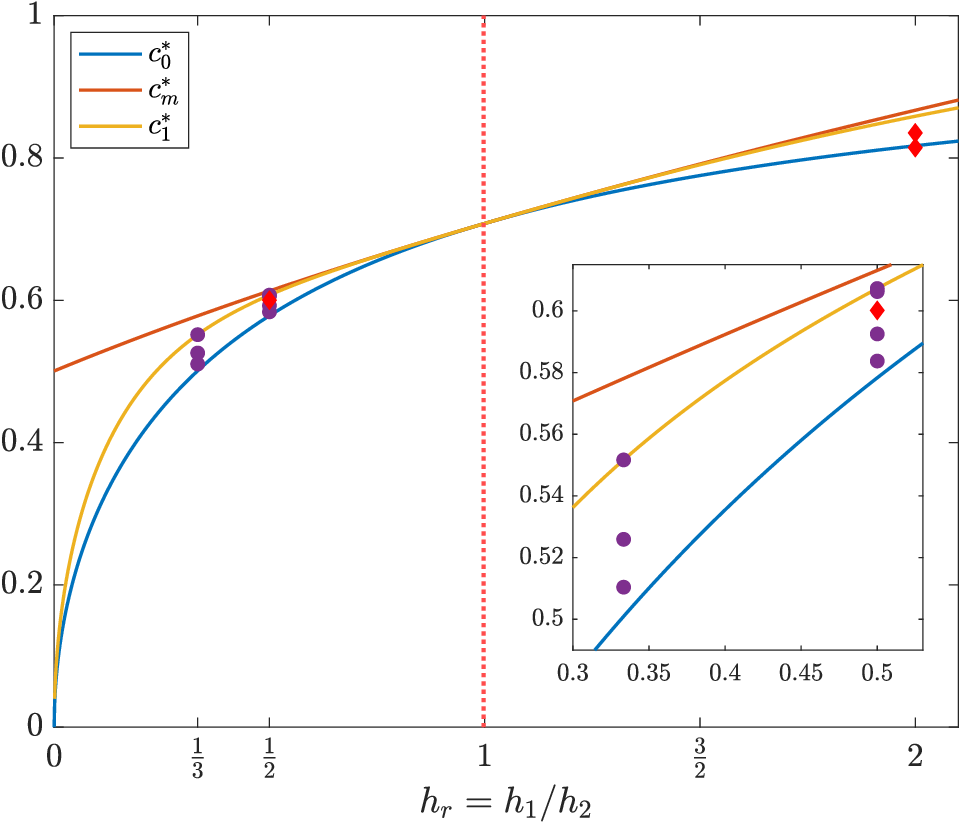}
\caption{The velocity $c^*_m$ of the maximum amplitude MMCC soliton, the linear long wave velocity $c_0^*$, and the maximum velocity $c_1^*$ of the eKdV soliton corresponding to $M=M^*$ over $h_r$ when $\rr = 1.005^{-1}$. {\color{black} Note that $c_1^*$ is plotted for $\eps = 0.15$.} The vertical dotted line represents the critical depth ratio $h_r = \sqrt{\rr}$. {\color{black} The purple dots correspond to the parameter values at which approximate eKdV solitary waves were selected as initial conditions for the time simulations in this study. The red dimonds correspond to the cnoidal waves considered in Section~\ref{sec:CnoidalWaveNumerics}, when $\eps = 0.1$ and $\rr = 1.005^{-1}$.}}
\label{MCC_speeds}
\end{figure}

To begin, Figure \ref{MCC_speeds} shows the comparison between velocities of $c_m^*, c_0^*,$ and $c_1^*$ for fixed $\eps = 0.15$  {\color{black} and} $\rr = 1.005^{-1}$,  {\color{black} but} varying $h_r$. Here, $c_m^*$ denotes the  {\color{black} speed} of the MMCC  {\color{black} solitary wave} having maximum amplitude. This is given by Choi \& Camassa \cite{CC1999} and we write its form in dimensionless variables of this study as:
\begin{equation}
c_m^* = \dfrac{ \sqrt{1 + h_r} }{1 + \sqrt{\rr}}.
\end{equation}
Also, $c_1^*$ denotes the limiting  {\color{black} speed}  of the eKdV  {\color{black} solitary wave} corresponding to $M=M^*$ (such that $F(M^*) = 0$), which can be written as 
\begin{equation}
c_1^* = \left [ c_0^* + \eps \tilde{v} - \dfrac{\eps^2 d \tilde{v}^2}{1 + \eps d \tilde{v}}  \right ]_{M = M^*}, 
\end{equation}
where  $c_0^*$ represents the linear long wave  {\color{black} speed}, which in our previous derivation we have denoted as $v$. 
The critical depth ratio $h_r = \sqrt{\rr}$, at which  the nonlinear coefficient of the KdV equation, $\alpha$, vanishes, is  indicated in this figure by the vertical dotted line. The {\color{black} purple} dots represent the parameter values where approximate eKdV  {\color{black} solitary waves} were chosen as initial conditions for the time simulations in this study. Two cases were considered: $h_r=1/2$, which is relatively close to criticality (note that $c^*_m$ and $c^*_1$ are very close at this values, which we use as a definition of 'closeness to criticality'), and $h_r=1/3$, which is further away from the criticality, but is still {\color{black} roughly} within the range of validity of the eKdV equation since $h_r \sim O(1)$.  {\color{black} The red dimonds correspond to the cnoidal waves considered in Section~\ref{sec:CnoidalWaveNumerics}.} \\

\begin{figure}
\centering
\includegraphics[scale=0.7]{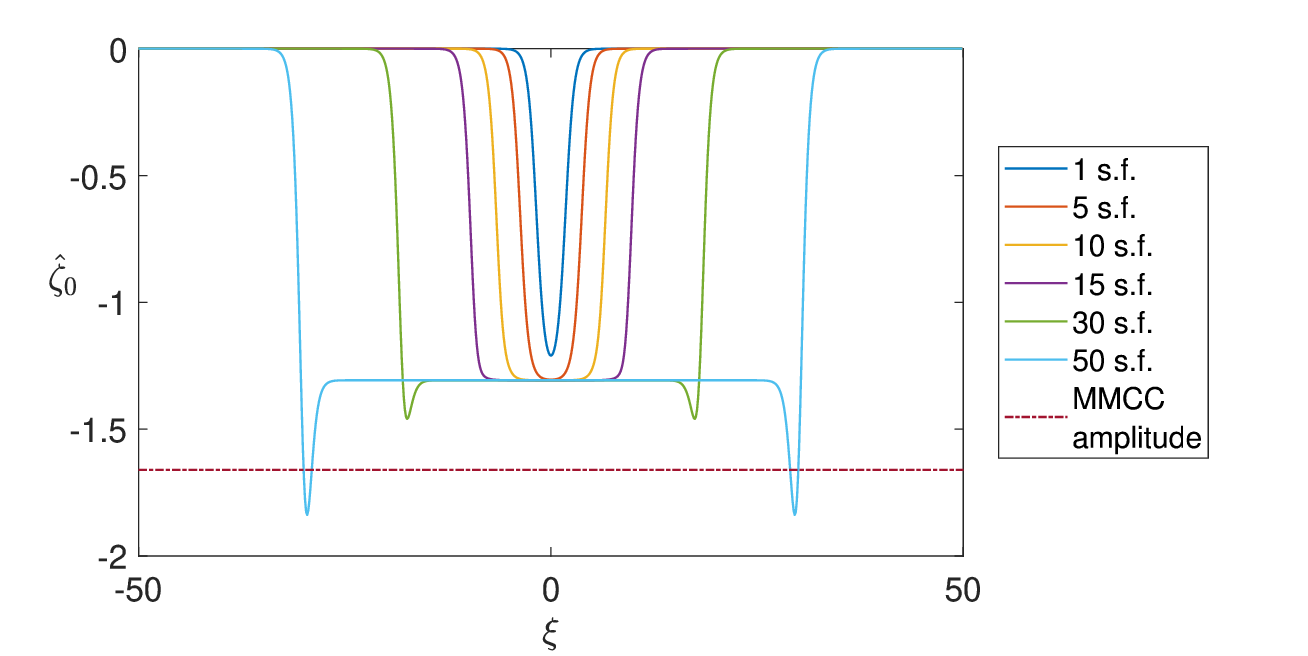}
\caption{Approximate solutions \eqref{ekdv_soliton_sol_travelling} of the eKdV equation for  $\eps = 0.15$,  $h_r = 0.5$, and $\rr = 1.005^{-1}$. Different curves represent different cases for how close $M$ is to $M^*$, with the thinnest profile representing $M^*$ truncated to $1$ significant figure (s.f.), and gradually and respectively increasing in width along with the number of significant figures chosen. The widest profile corresponds to $M$ equal to $M^*$ truncated to $50$ s.f. with horns at the boundaries of the table-top reaching well beyond the limiting MMCC amplitude.}
\label{M*approach}
\end{figure}

The Gardner  {\color{black} solitary wave} \eqref{GSol} has the property that in the limit $F \to 0$ (or, equivalently, $M\to M^*$) it becomes table-top. 
The approximate travelling-wave solutions \eqref{ekdv_soliton_sol_travelling} of the eKdV equation \eqref{mcc_ekdv}  are shown in Figure \ref{M*approach} as $M\to M^*$.
We observe that for small $M$ (e.g.\ taking a few significant figures of $M^*$), the profile resembles a KdV-type  {\color{black} solitary wave}, and as we take more digits of $M^*$ the  {\color{black} solitary wave} becomes table-top at around $15$ digits. Continuing this limit further reveals, as the table-top becomes wider, the emergence of horns at the boundaries of the table-top region. These horns continue to grow in amplitude as $F\to 0$, reaching well beyond the maximum amplitude of the MMCC  {\color{black} solitary wave} known from \cite{CC1999}. \\

\begin{figure}
\centering
\includegraphics[width=0.4\textwidth]{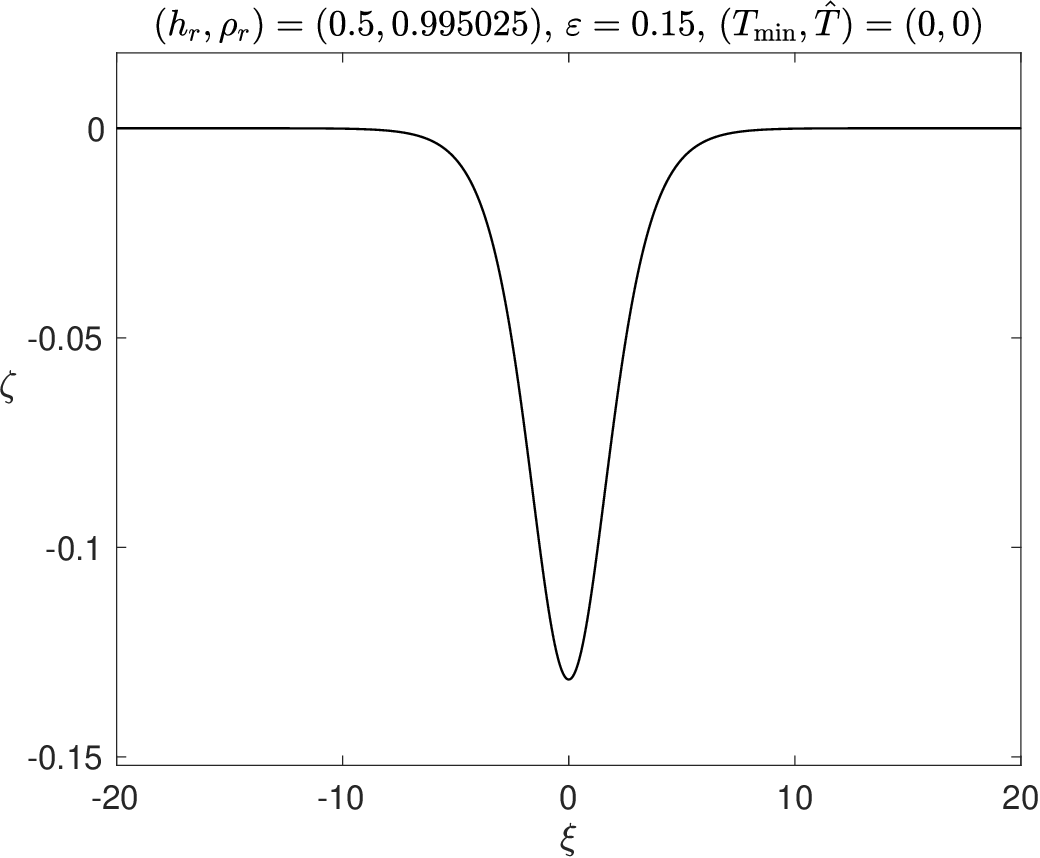}\quad
\includegraphics[width=0.4\textwidth]{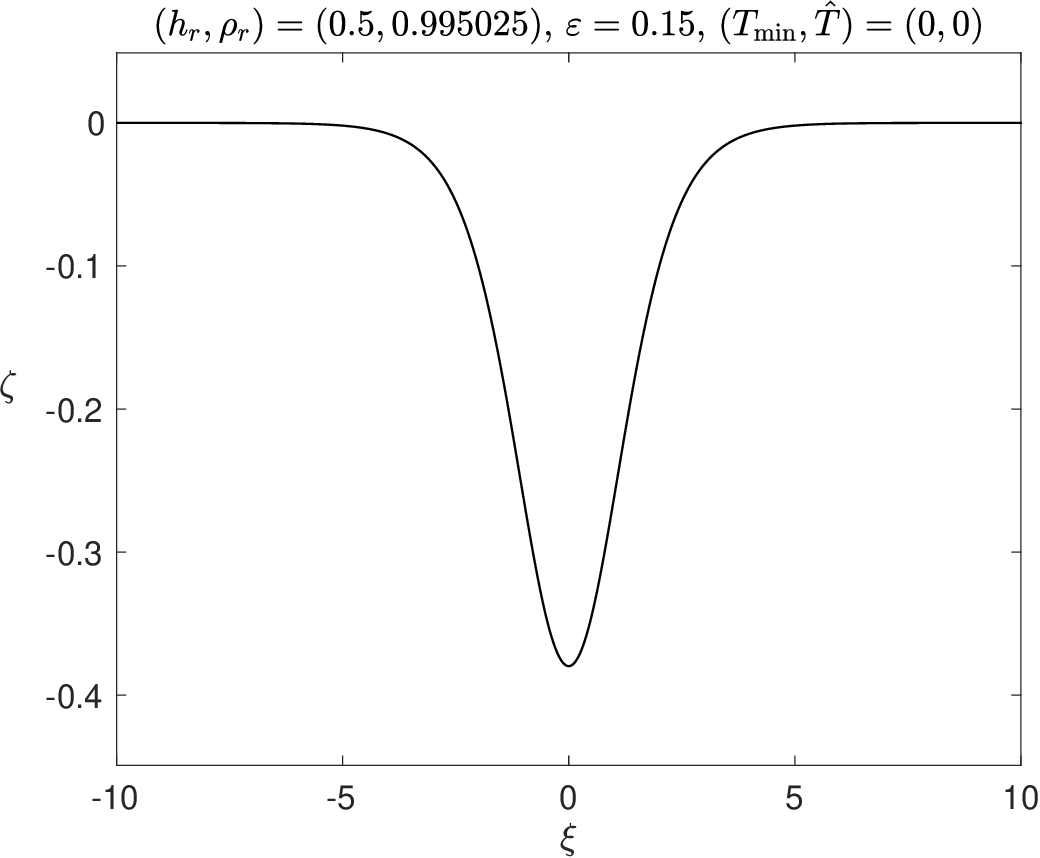}\\[0.2cm]
\includegraphics[width=0.4\textwidth]{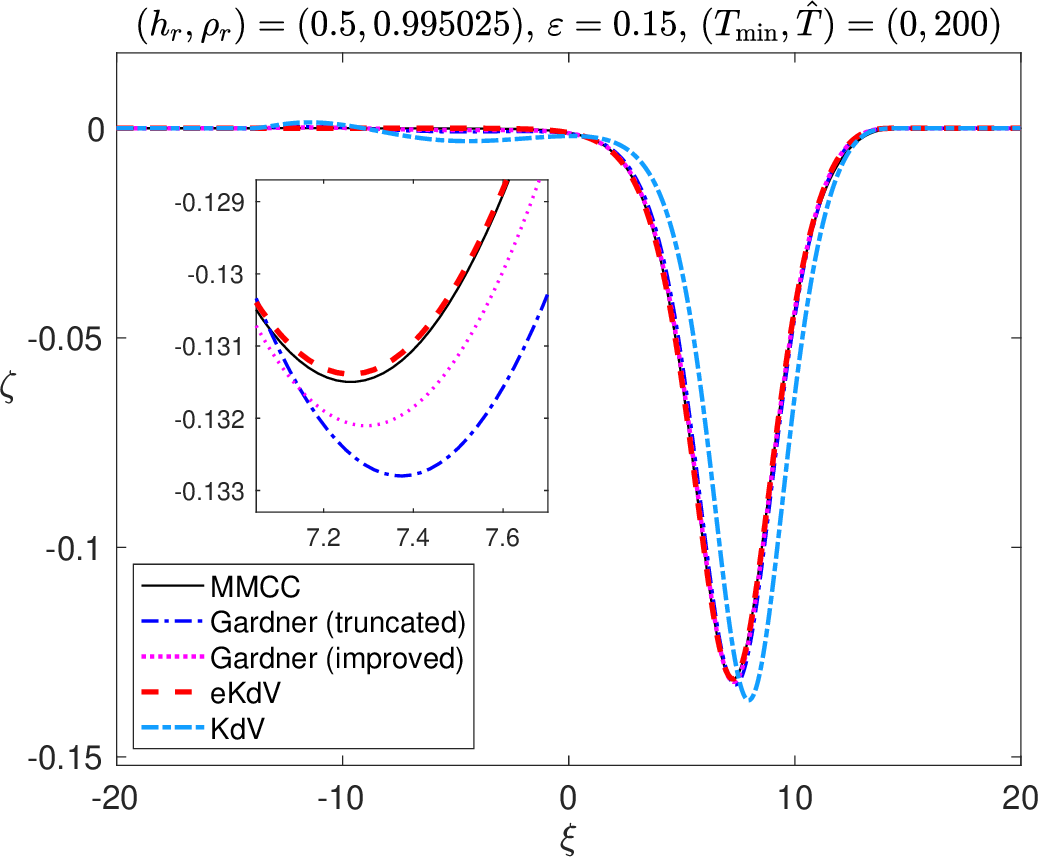}\quad
\includegraphics[width=0.4\textwidth]{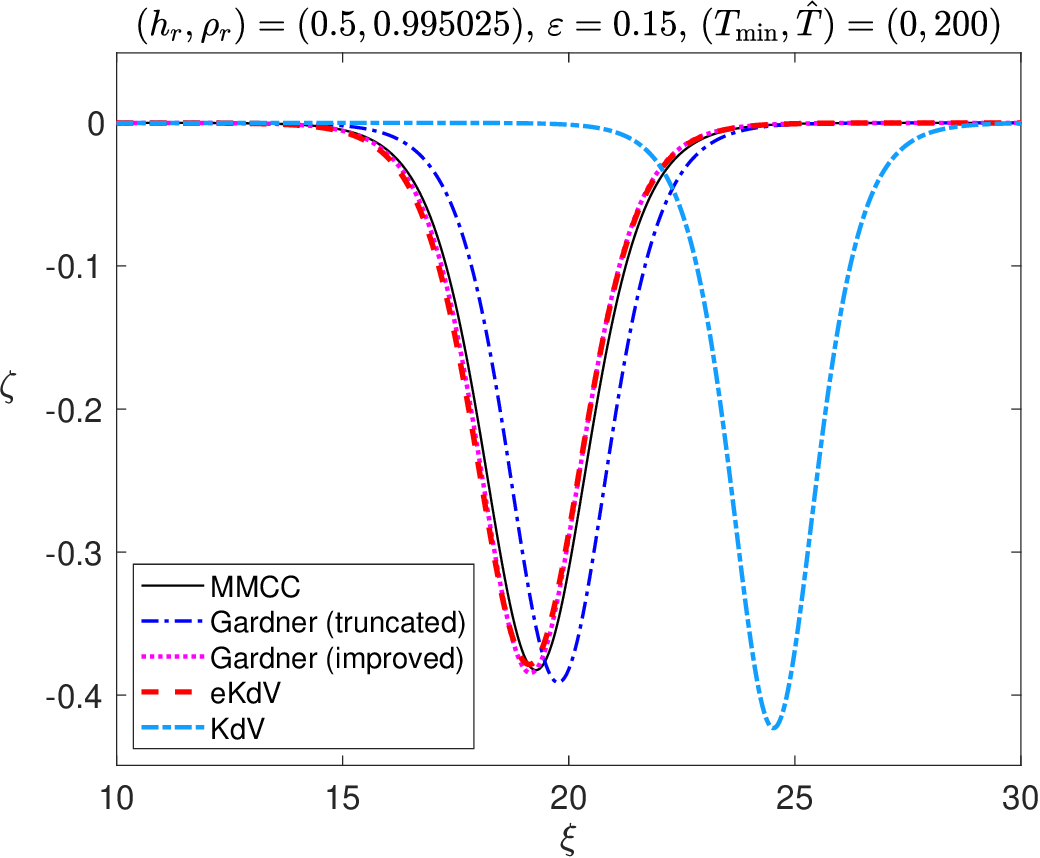}\\[0.2cm]
\includegraphics[width=0.4\textwidth]{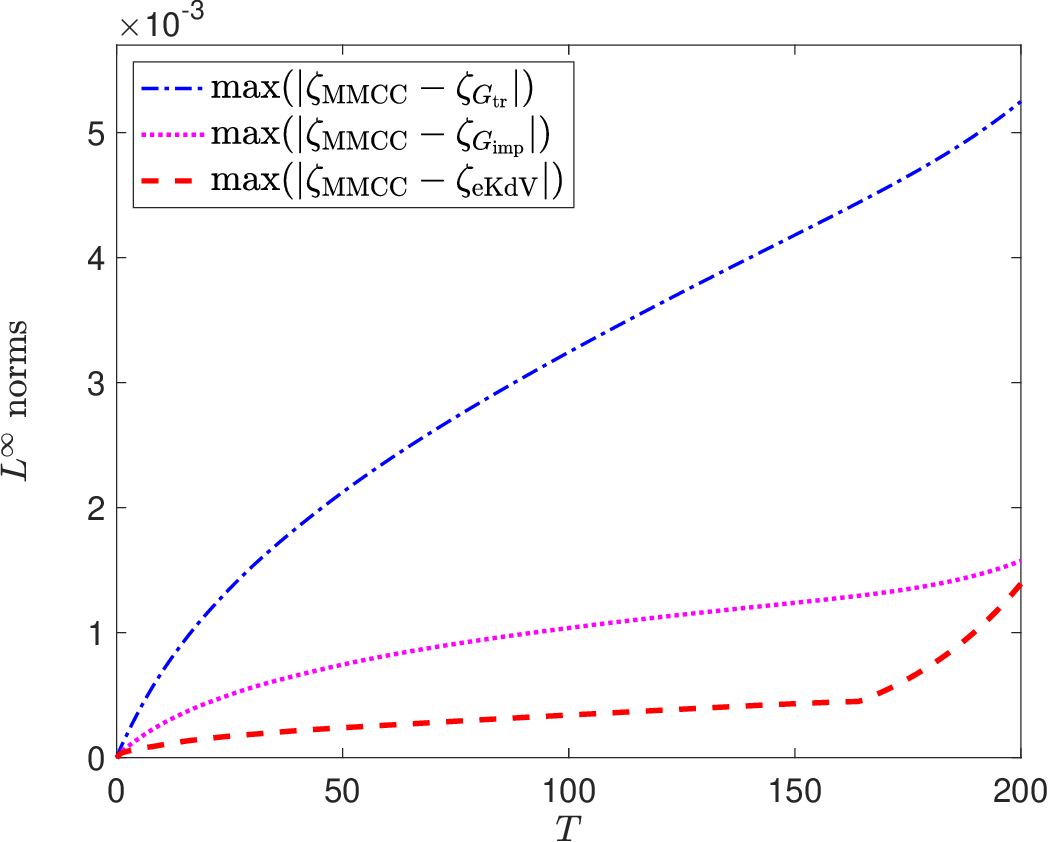}\quad
\includegraphics[width=0.4\textwidth]{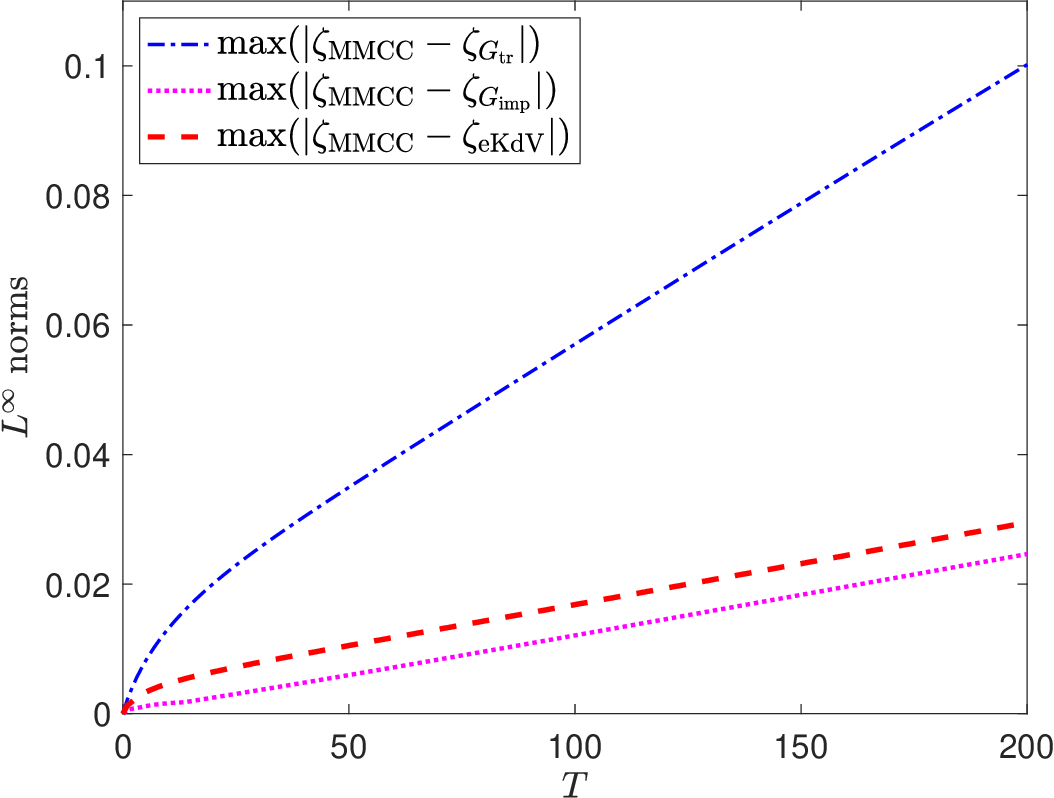}
\caption{Initial (top panels) and final  (middle panels) solution profiles  ($T=0$ and $T=200$, respectively) of the MMCC, truncated Gardner, eKdV and KdV models for $M=-0.25$ (left panels) and $M=-0.65$ (right panels) when $\eps = 0.15$,  $h_r = 0.5$, and $\rr = 1.005^{-1}$. {\color{black}  For reference, $M^* \approx -1.307$. The bottom panels show the evolutions of the $L^\infty$ norms of the differences between the solution profile obtained from the MMCC model and the solution profiles obtained from the eKdV and Gardner models, as indicated in the legends (note that $\zeta_\mathrm{MMCC}$, $\zeta_\mathrm{eKdV}$, $\zeta_{G_\mathrm{tr}}$ and $\zeta_{G_\mathrm{imp}}$ correspond to the solutions of the MMCC, eKdV, Gardner truncated and Gardner improved models).} 
}
\label{Msmall}
\end{figure}

\begin{figure}
\centering
\includegraphics[width=0.4\textwidth]{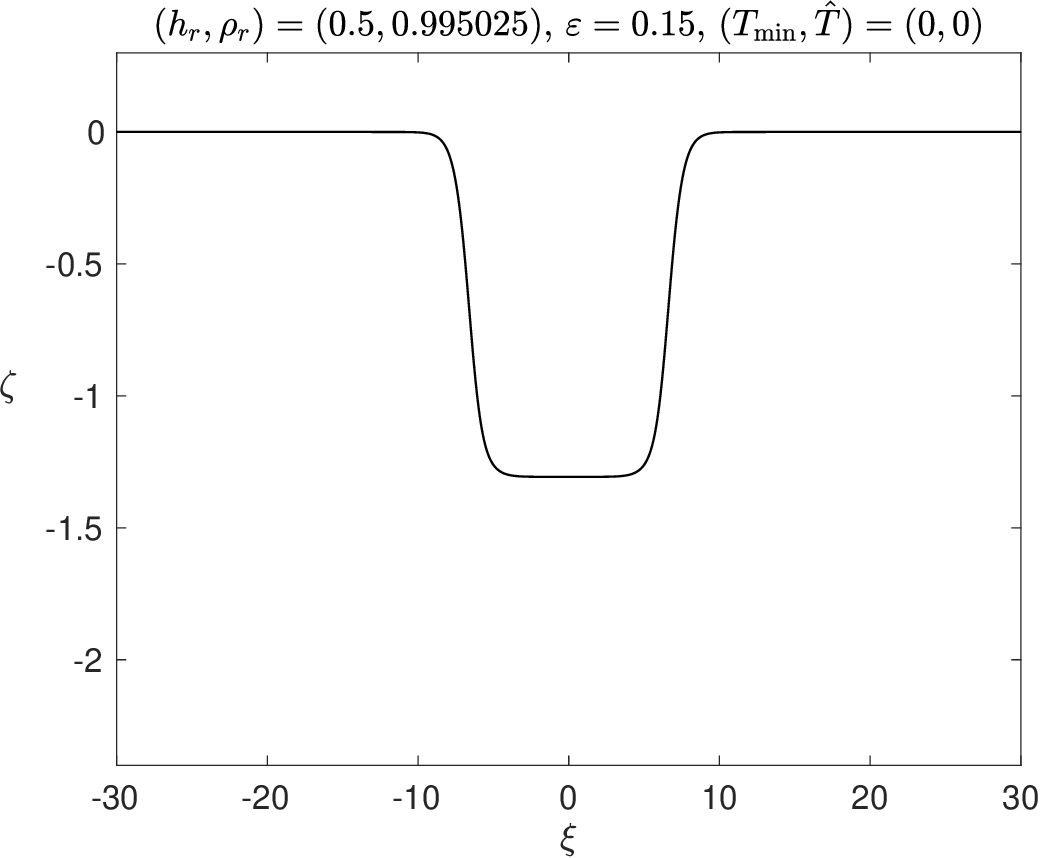}\quad
\includegraphics[width=0.4\textwidth]{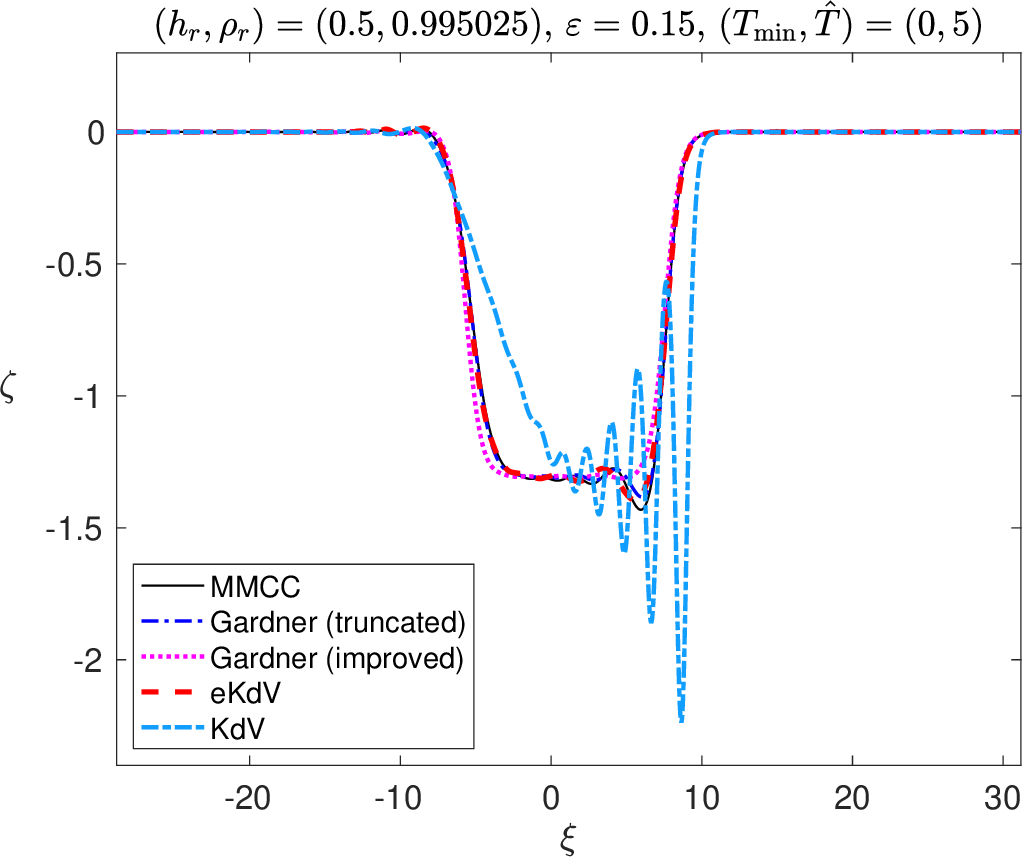}\\[0.2cm]
\includegraphics[width=0.4\textwidth]{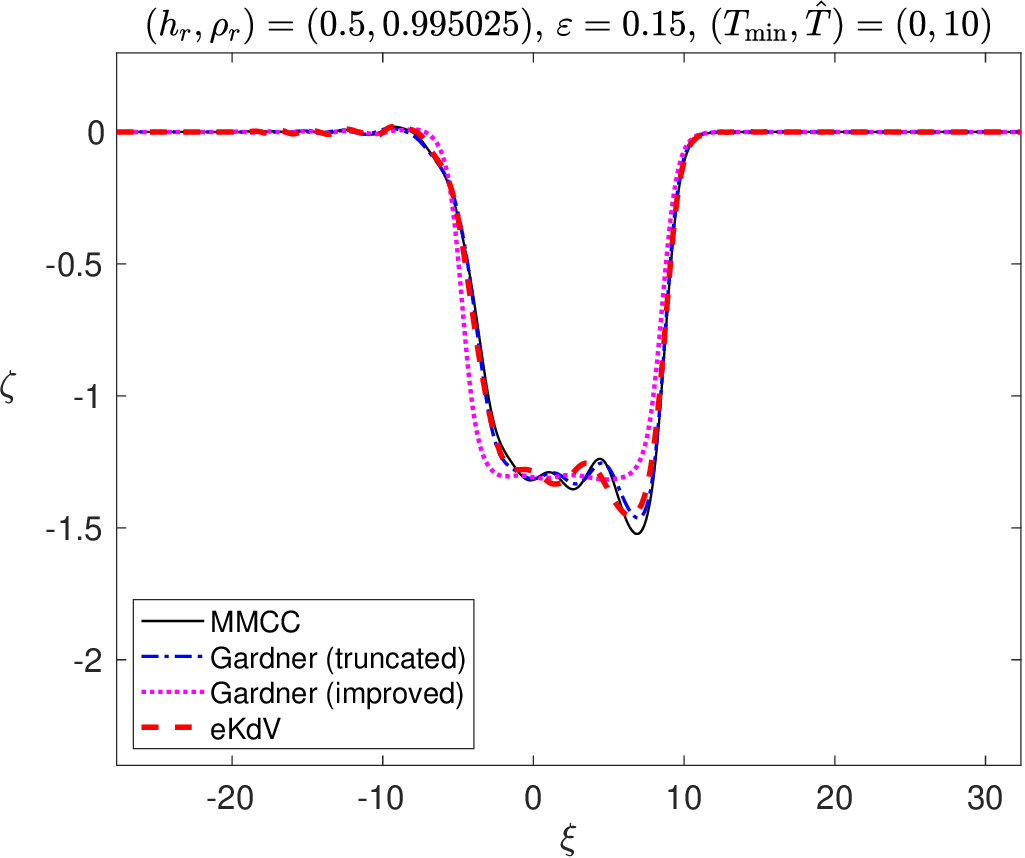}\quad
\includegraphics[width=0.4\textwidth]{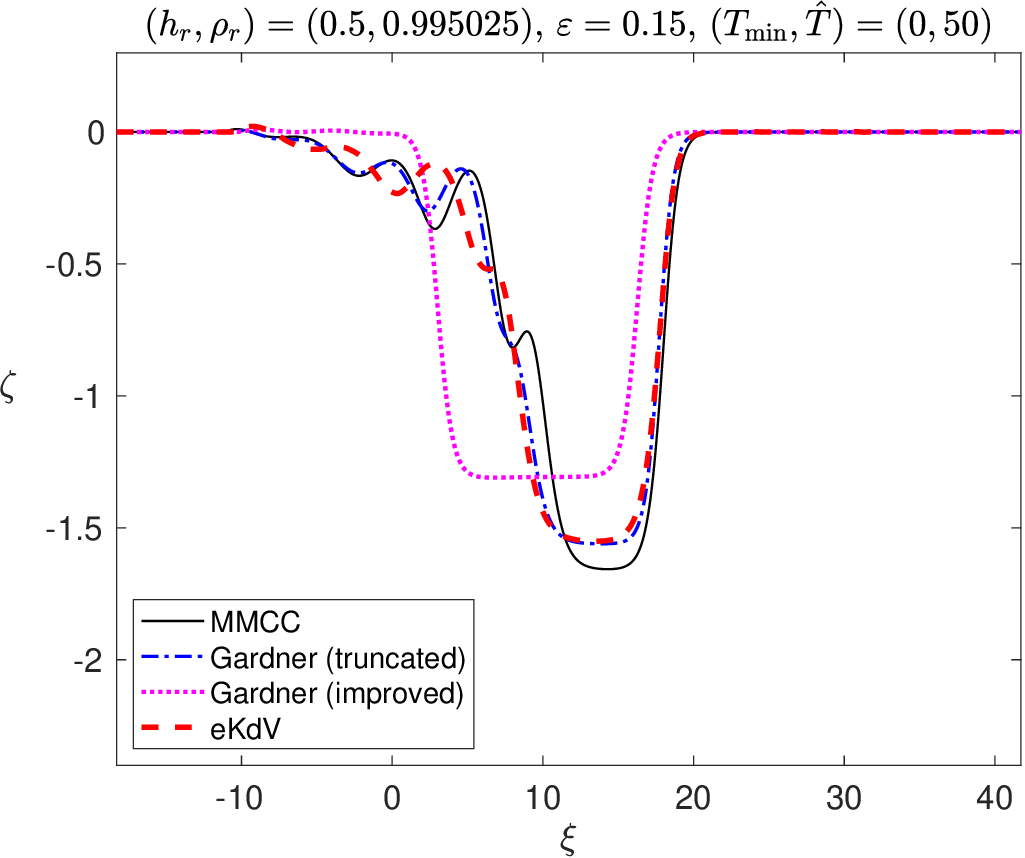}\\[0.2cm]
\includegraphics[width=0.4\textwidth]{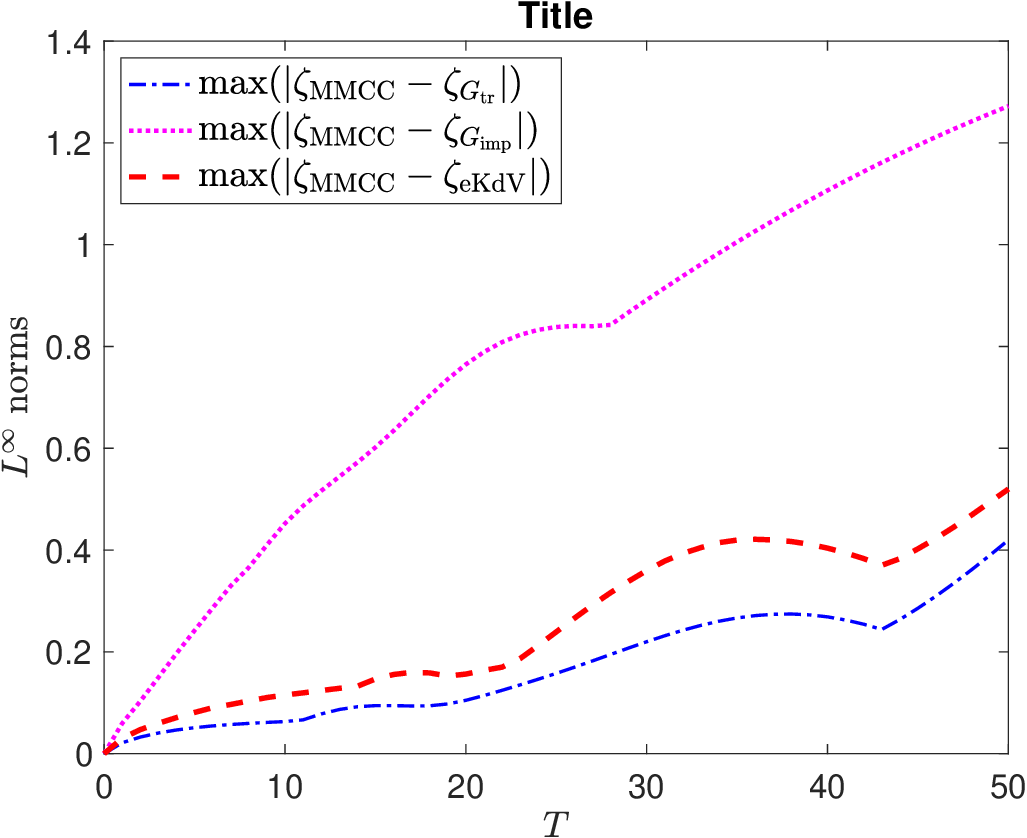}
\caption{Initial ($T=0$), intermediate ($T=5$ and $T=10$), and final ($T=50$) solution profiles of the MMCC, truncated Gardner, improved Gardner, eKdV and KdV models for the $M$ value corresponding to $10$ significant figures of $M^*$ {\color{black} (for reference, $M^* \approx -1.307$)}, 
when $\eps = 0.15$,  $h_r = 0.5$, and $\rr = 1.005^{-1}$.  
{\color{black} The bottom panel show the evolutions of the $L^\infty$ norms of the differences between the solution profile obtained from the MMCC model and the solution profiles obtained from the eKdV and Gardner models, as indicated in the legends (note that $\zeta_\mathrm{MMCC}$, $\zeta_\mathrm{eKdV}$, $\zeta_{G_\mathrm{tr}}$ and $\zeta_{G_\mathrm{imp}}$ correspond to the solutions of the MMCC, eKdV, Gardner truncated and Gardner improved models). }
}
\label{Mmoderate}
\end{figure}

We now proceed to initiate the MMCC model with approximate solution profiles \eqref{ekdv_soliton_sol_travelling} of the eKdV equation under different $M$ values. The details of the numerical scheme applied here can be found in Appendix B. Firstly, we test two initial conditions shown in Figure \ref{Msmall}. Here, the MMCC model is  {\color{black} initialised}  with the eKdV approximate solution profile \eqref{ekdv_soliton_sol_travelling} corresponding to either $M = -0.25$ or $M = -0.65$ (note, for reference, that $M^* \approx -1.307$ in this case). The figure demonstrates that the initial condition  {\color{black} sheds} some mass in its time evolution of the MMCC model, but has its shape largely unchanged, even over long time intervals. For comparison, the numerical solutions of the eKdV and  Gardner models, as well as the KdV model, are plotted, all  {\color{black} initialised} with the same initial profile. The results clearly show that the eKdV model approximates the MMCC model very well, with noticeable improvement over both the KdV and the truncated Gardner models as the amplitude of the initial profile increases. 
The results for the improved Gardner model are close to those of the eKdV model {\color{black} for small and moderate amplitude waves}, suggesting that we can use the improved Gardner model {\color{black} as a good approximation in this amplitude range}. The analytical approximation {\color{black} (\ref{ekdv_soliton_sol_travelling})} for the solitary wave (not shown in the figure) is close to the numerical solution of {\color{black} the parent system}. \\

As $M$ is increased {\color{black} to a value} close to $M^*$, a table-top forms, and Figure \ref{Mmoderate} {\color{black} shows the} time evolution of the MMCC, Gardner, KdV and eKdV models. For the regular KdV model, the initial condition leads to {\color{black} a highly oscillatory wave profile (a dispersive shock wave), showing, in comparison with  the MMCC model,}  the failure of the KdV approximation for the large-amplitude waves. On the contrary, in Figure \ref{Mmoderate}, we observe that for a considerable time {\color{black} other numerical solutions show} little evolution, and the approximate {\color{black} solitary wave solution of the eKdV equation is close to the numerical solutions of other models}. At $T = 50$, however, the initial condition {\color{black} evolves into a taller and narrower table-top solitary wave}. {\color{black} The truncated Gardner equation better approximates the MMCC model than the improved Gardner equation, while the computed solution of the improved Gardner equation remains almost unchanged and close to the analytical approximation of the solitary wave of the eKdV equation. 
We note that the truncated Gardner solitary wave is slightly slower than the MMCC one as it lags behind, which results in a slight phase shift over a long fast-time period (note that this figure displays slow time instead). It can be also observed that the truncated Gardner model underestimates the amplitude of the MMCC solitary wave. Interestingly, the eKdV model behaves similarly to the truncated Gardner equation, showing that the range of validity of the eKdV model is greater than that of the analytical approximation obtained using the near-identity transformation. The MMCC evolution also shows that the initial solitary wave becomes thinner over time by shedding mass towards the left side of the simulation domain. The amplitude of the MMCC solitary wave grows toward the maximum amplitude for this choice of parameters. Over long-time evolution, the amplitude approaches this maximum level, with all trailing waves exiting the simulation domain.} \\

\begin{figure}
\centering
\includegraphics[scale=0.4]{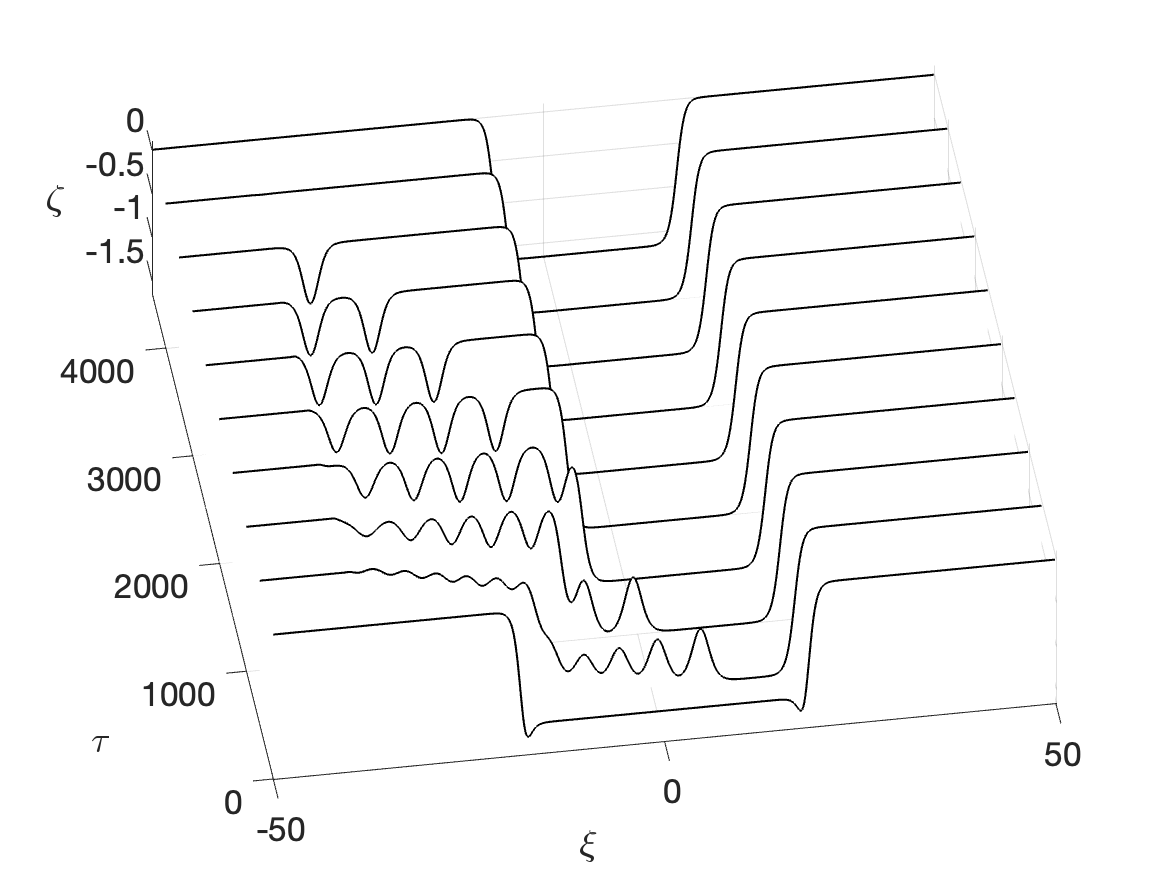}\quad
\includegraphics[scale=0.4]{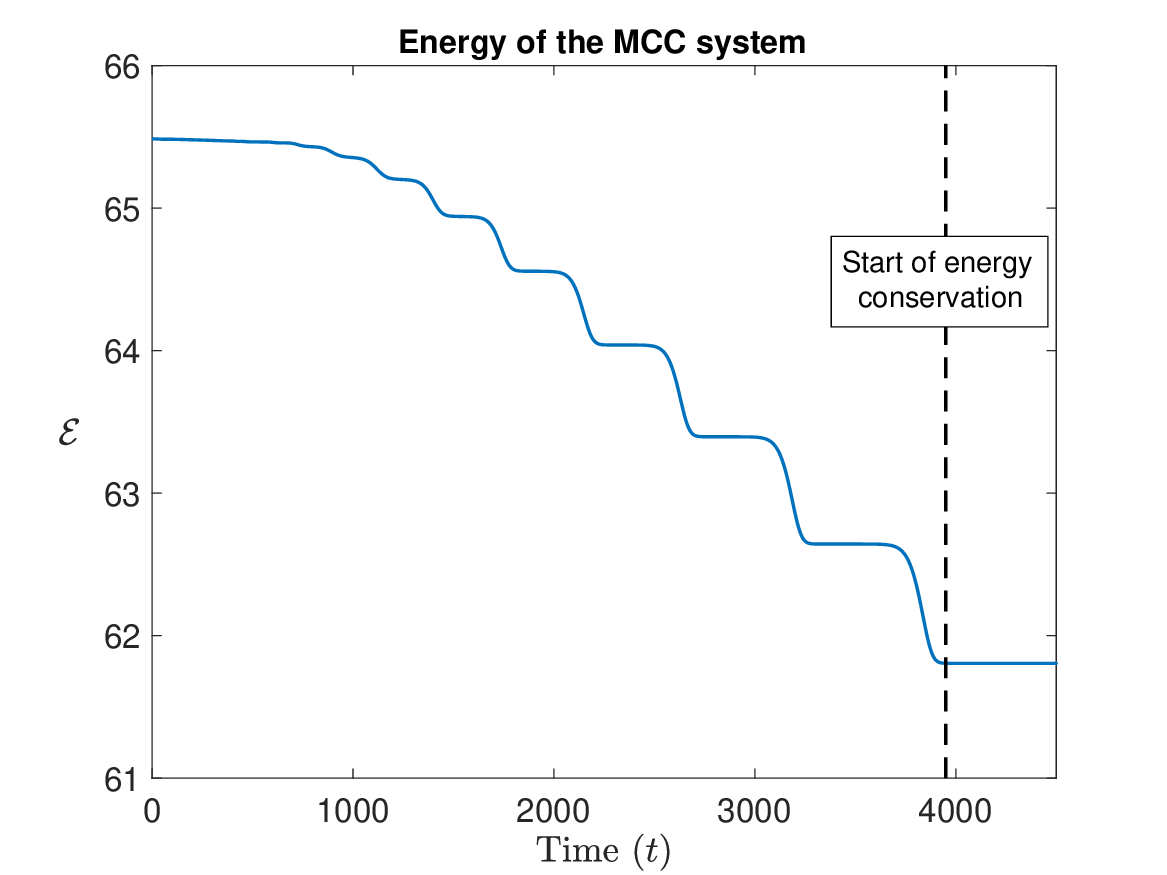}
\caption{Left panel demonstrates the time evolution of the current-free MMCC model (equations \eqref{dimless_shear_mcc1}, \eqref{dimless_shear_mcc2}) when $\eps = 0.15$,  $h_r = 0.5$, and $\rr = 1.005^{-1}$. Here the computational domain is $\xi \times \tau = [-50,50] \times [0, 4500]$. The initial condition is \eqref{ekdv_soliton_sol_travelling} with $M$ set as the truncation of $M^*$ at 30 significant figures {\color{black} (for reference, $M^* \approx -1.307$)}. Velocity of the reference frame is $v+\delta v$ where $\delta v = 0.03515$. The right panel tracks the time evolution of the energy $\mathcal{E}$ of this numerical simulation. }
\label{Mhorns_mcc}
\end{figure}

Next, we examine the time evolution when the initial condition {\color{black} is a solitary wave approximation with} horns. In particular, we consider the case where $M$ is the truncation of $M^*$ after $30$ significant figures, resulting in horns that remain below the maximum MMCC amplitude. The evolution under the MMCC model is shown in Figure \ref{Mhorns_mcc}. This figure reveals that, during the evolution, the right-hand horn grows in both amplitude and width. As this occurs, a significant amount of mass is then expelled towards the left-hand side of the simulation domain. Notably, the right-hand horn grows to precisely match the maximum amplitude of the MMCC solitary waves. In the long-time behaviour, once all {\color{black} excited trailing} waves exit the simulation, the system stabilises into a {\color{black} noticeably narrower} table-top {\color{black} solitary wave} solution of the MMCC model, which retains its form as it propagates further. To ensure the validity of the simulation, we check the energy of this simulation against the energy of the MMCC model, as provided in \cite{CC1999}.  For a flat-bottom topography and a rigid lid, the MMCC model conserves the dimensional energy $\bar{\mathcal{E}}$ over time, where

\begin{equation}
\bar{\mathcal{E}} = \dfrac{1}{2} \int (\rho_2 - \rho_1) g \zeta^2 + \sum_{i = 1}^2 \rho_i \bigg( \eta_i \bar{\bfu}_i^2 + \frac{1}{3} \eta_i^3 \bar{\bfu}^2_{ix} \bigg) \ \mathrm{d} x.
\end{equation}
After applying the scalings for the dimensionless variables \eqref{scalings}, we obtain
\begin{align}
\mathcal{E} & = \dfrac{\lambda}{2} \int \rho_2 (1 - \rr) g a^2 (\zeta^*)^2 + \sum_{i = 1}^2 \rho_i \bigg( h_2 \eta^*_i \eps^2 c^2 (\bar{\bfu}_i^*)^2 + \frac{1}{3} h_2^3 (\eta_i^*)^3 \frac{\eps^2 c^2}{\lambda^2} (\bar{\bfu}_{i x^*}^*)^2 \bigg) \ \mathrm{d} x^* \nonumber \\
& = \dfrac{\lambda \rho_2 (1 - \rr) g a^2}{2} \int (\zeta^*)^2 + \dfrac{h_2 \eps^2 c^2}{\rho_2 (1 - \rr) g a^2} \sum_{i = 1}^2 \rho_i \bigg( \eta^*_i (\bar{\bfu}_i^*)^2 + \frac{\delta^2}{3} (\eta_i^*)^3 (\bar{\bfu}_{i x^*}^*)^2 \bigg) \ \mathrm{d} x^* \nonumber \\
& = \dfrac{\lambda \rho_2 (1 - \rr) g a^2}{2} \int (\zeta^*)^2 + \dfrac{h_2 \eps^2 g h_2 (1-\rr)}{\rho_2 (1 - \rr) g \eps^2 h_2^2} \sum_{i = 1}^2 \rho_i \bigg( \eta^*_i (\bar{\bfu}_i^*)^2 + \frac{\delta^2}{3} (\eta_i^*)^3 (\bar{\bfu}_{i x^*}^*)^2 \bigg) \ \mathrm{d} x^* \nonumber \\
& = \dfrac{\lambda \rho_2 (1 - \rr) g a^2}{2} \int (\zeta^*)^2 + \dfrac{1}{\rho_2} \sum_{i = 1}^2 \rho_i \bigg( \eta^*_i (\bar{\bfu}_i^*)^2 + \frac{\delta^2}{3} (\eta_i^*)^3 (\bar{\bfu}_{i x^*}^*)^2 \bigg) \ \mathrm{d} x^*
\end{align}

This leads us to the following definition of the dimensionless energy $\mathcal{E}$:
\begin{align}
\mathcal{E} & = \dfrac{2}{ \lambda g a^2 (\rho_2 - \rho_1)} \bar{\mathcal{E}} \\ & = \dfrac{1}{2} \int (\zeta^*)^2 + \dfrac{1}{\rho_2} \sum_{i = 1}^2 \rho_i \bigg( \eta^*_i (\bar{\bfu}_i^*)^2 + \frac{\delta^2}{3} (\eta_i^*)^3 (\bar{\bfu}_{i x^*}^*)^2 \bigg) \ \mathrm{d} x^*, \nonumber \\
& = \dfrac{1}{2} \int (\zeta^*)^2 + \rr  \bigg( \eta^*_1 (\bar{\bfu}_1^*)^2 + \frac{\eps}{3} (\eta_1^*)^3 (\bar{\bfu}_{1 x^*}^*)^2 \bigg) + \bigg( \eta^*_2 (\bar{\bfu}_2^*)^2 + \frac{\eps}{3} (\eta_2^*)^3 (\bar{\bfu}_{2 x^*}^*)^2 \bigg)\ \mathrm{d} x^*, \nonumber
\end{align}

and this is precisely the quantity shown to be conserved in Figure \ref{Mhorns_mcc} towards the end of the simulation after all left-propagating waves {\color{black} leave the computational domain (to be precise, the energy still slightly decays due to a low-pass filter applied to regularise the MMCC model, but it is approximately conserved).}  Hence, in this case the approximate solution of the eKdV equation provided a {\color{black} good initial condition} for the generation of the table-top solitary wave of the MMCC model.\\

\begin{figure}
\centering
\includegraphics[width=0.29\textwidth]{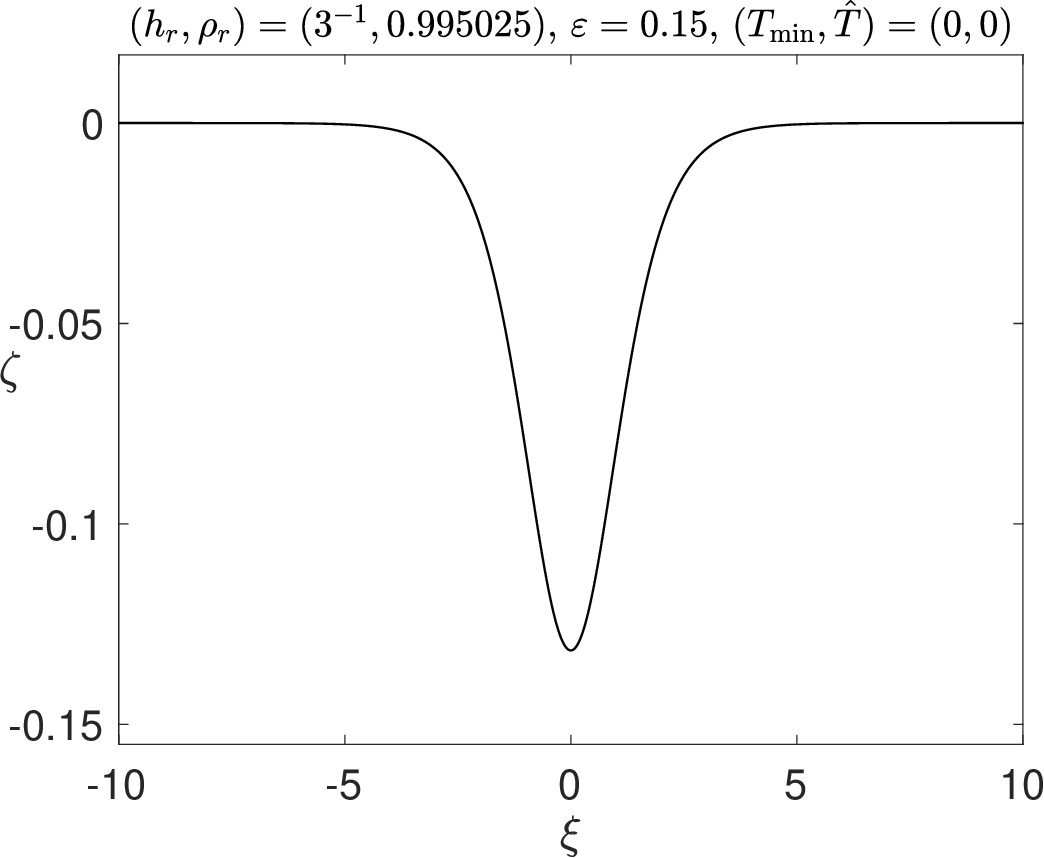}\,
\includegraphics[width=0.285\textwidth]{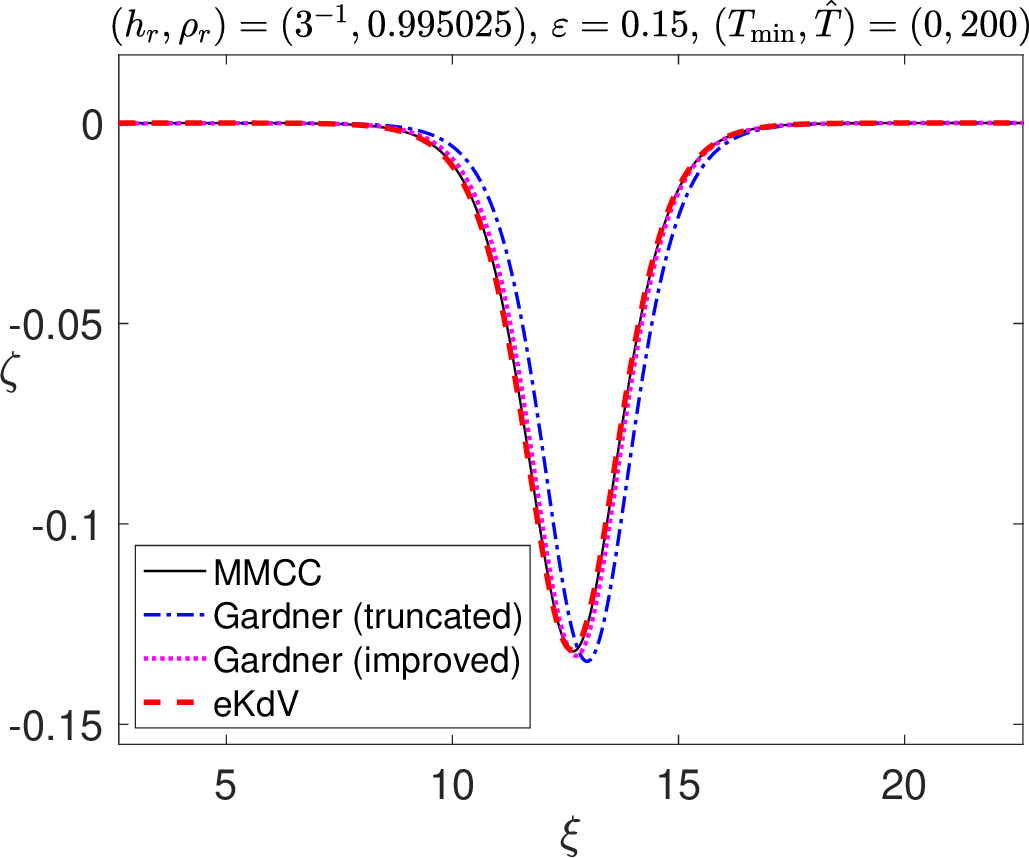}\,
\includegraphics[width=0.31\textwidth]{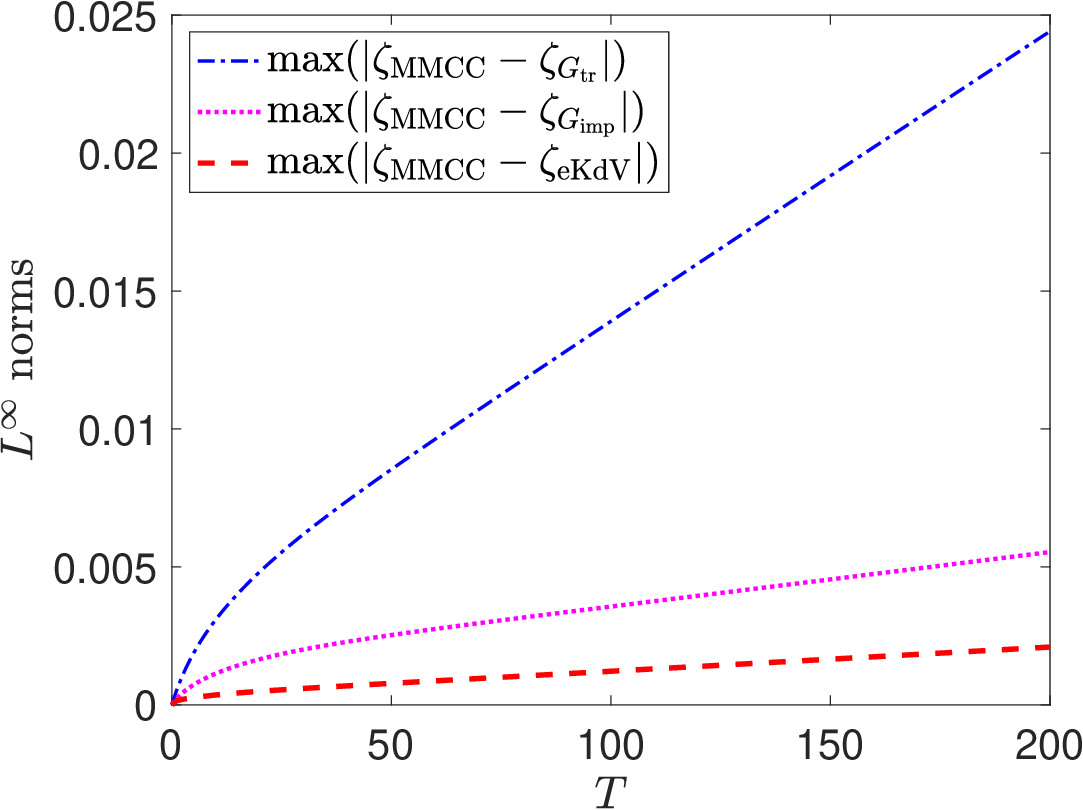}\\[0.2cm]
\includegraphics[width=0.29\textwidth]{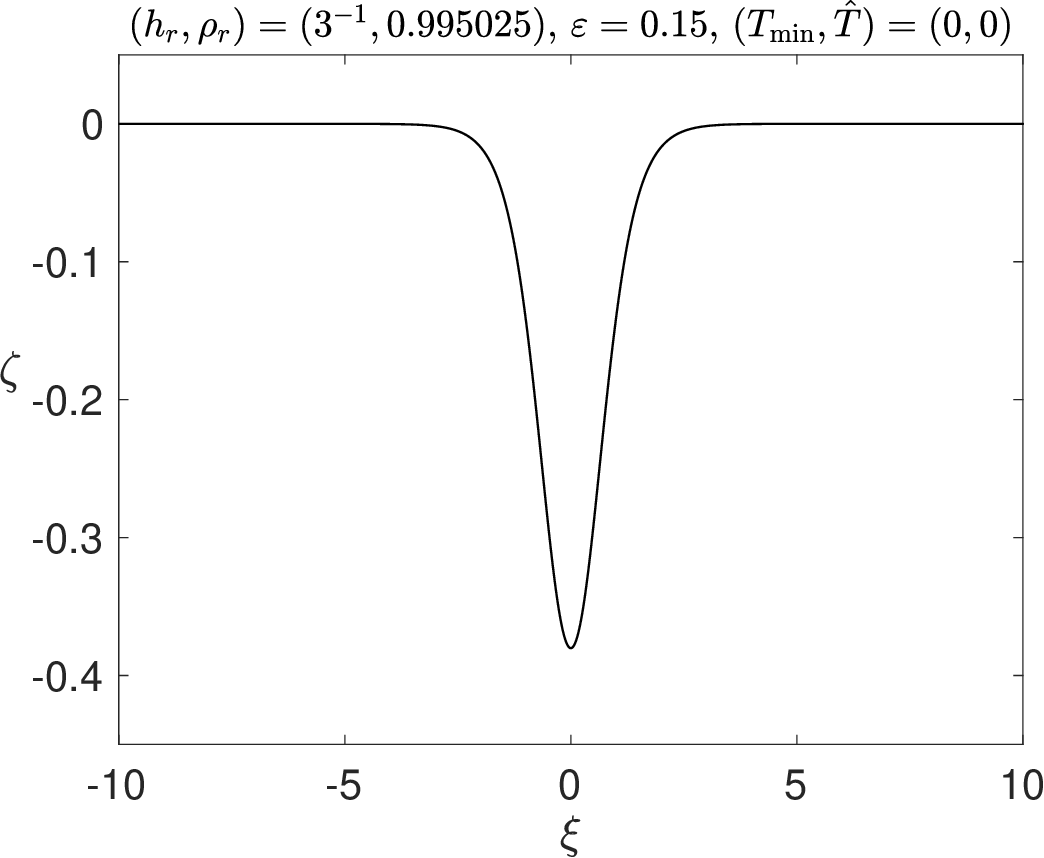}\,
\includegraphics[width=0.285\textwidth]{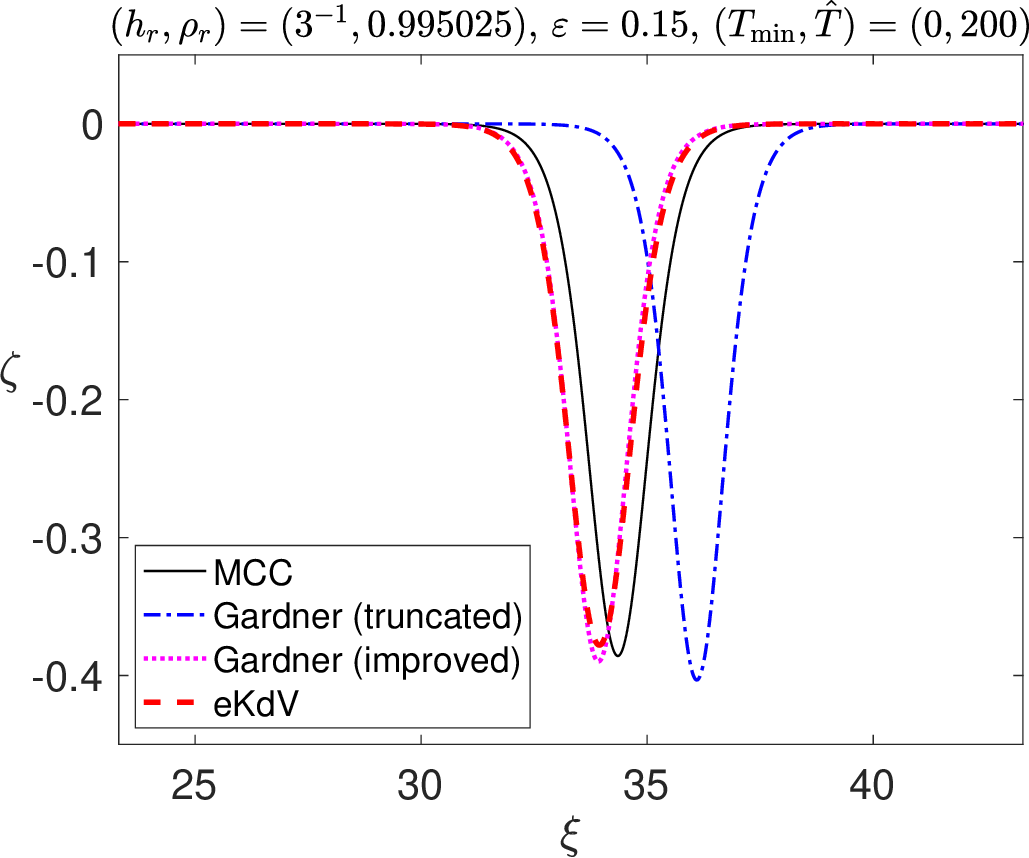}\,\,
\includegraphics[width=0.303\textwidth]{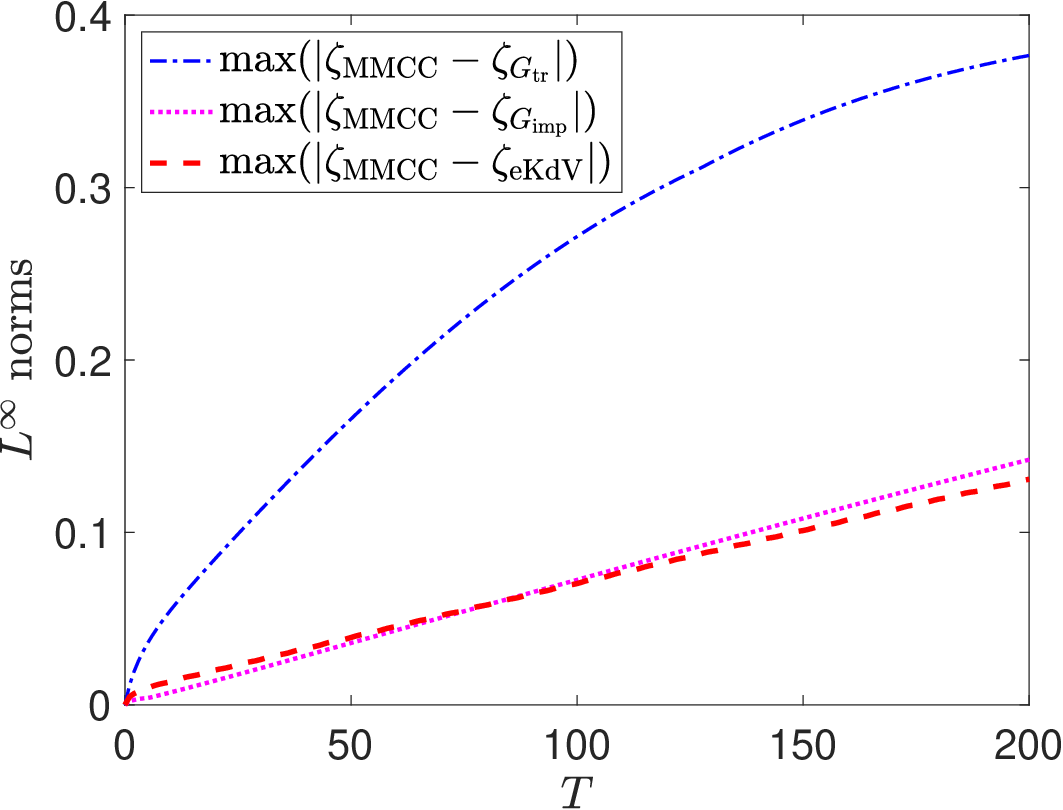}\\[0.2cm]\includegraphics[width=0.285\textwidth]{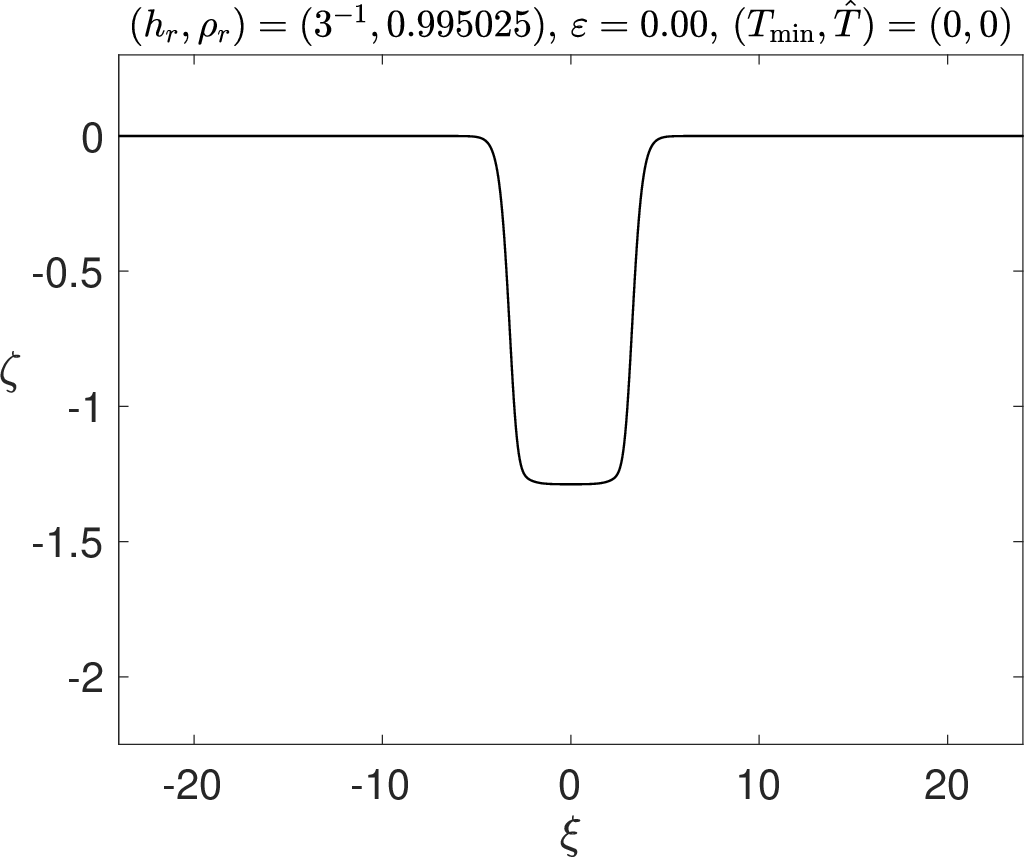}\,
\includegraphics[width=0.285\textwidth]{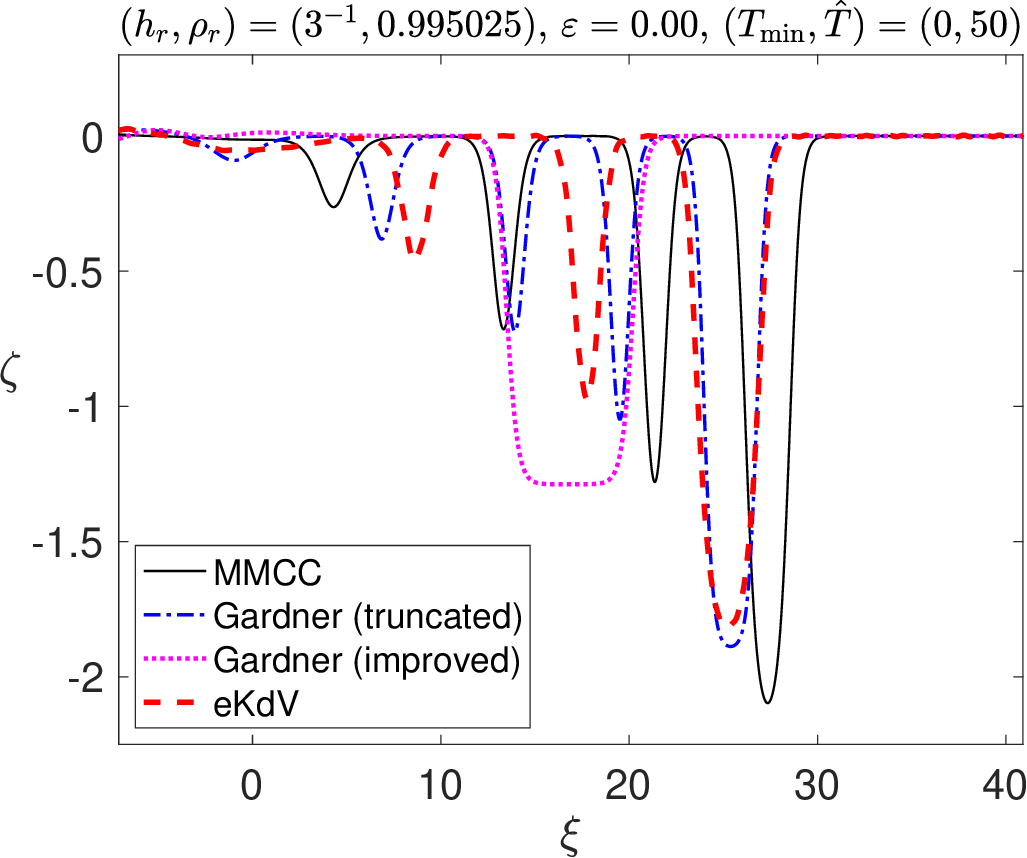}\,
\includegraphics[width=0.3\textwidth]{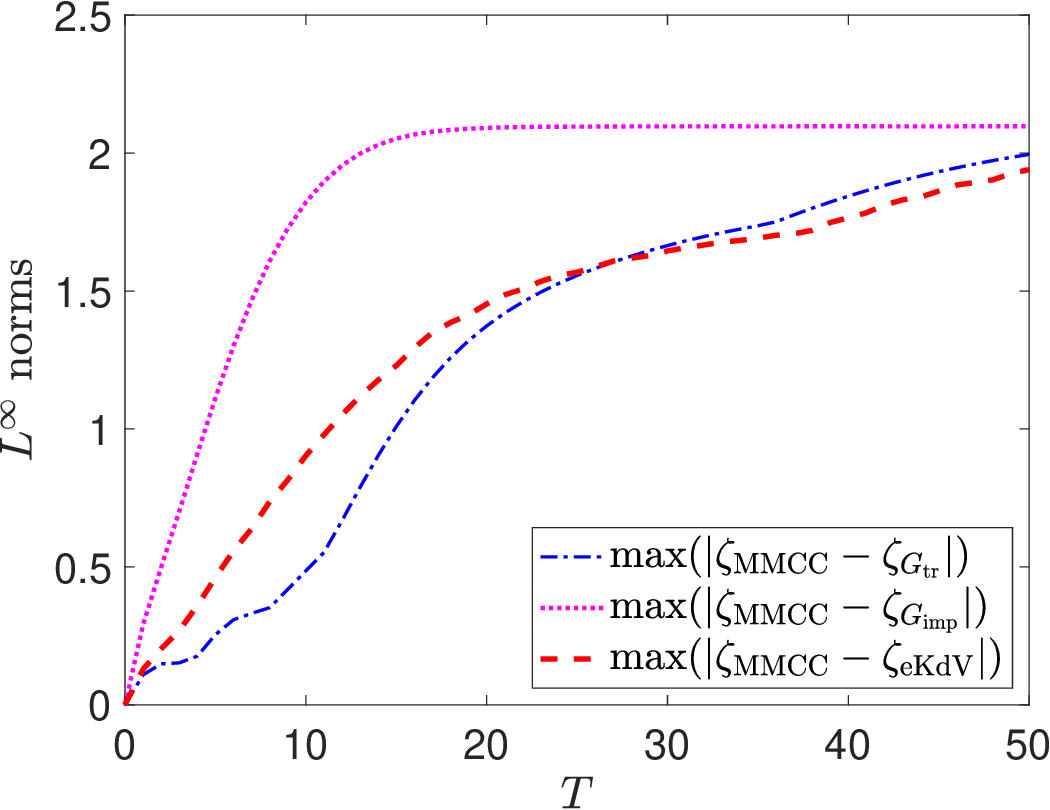}\\[0.2cm]
\includegraphics[width=0.295\textwidth]{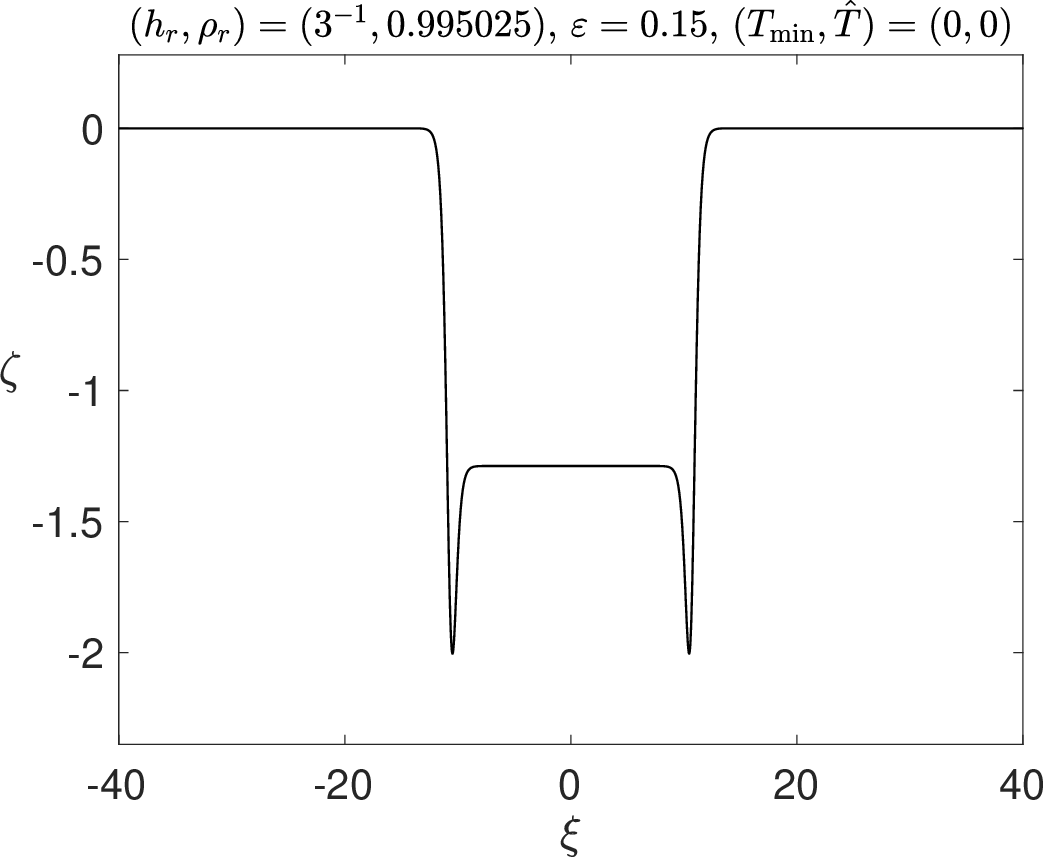}\,
\includegraphics[width=0.285\textwidth]{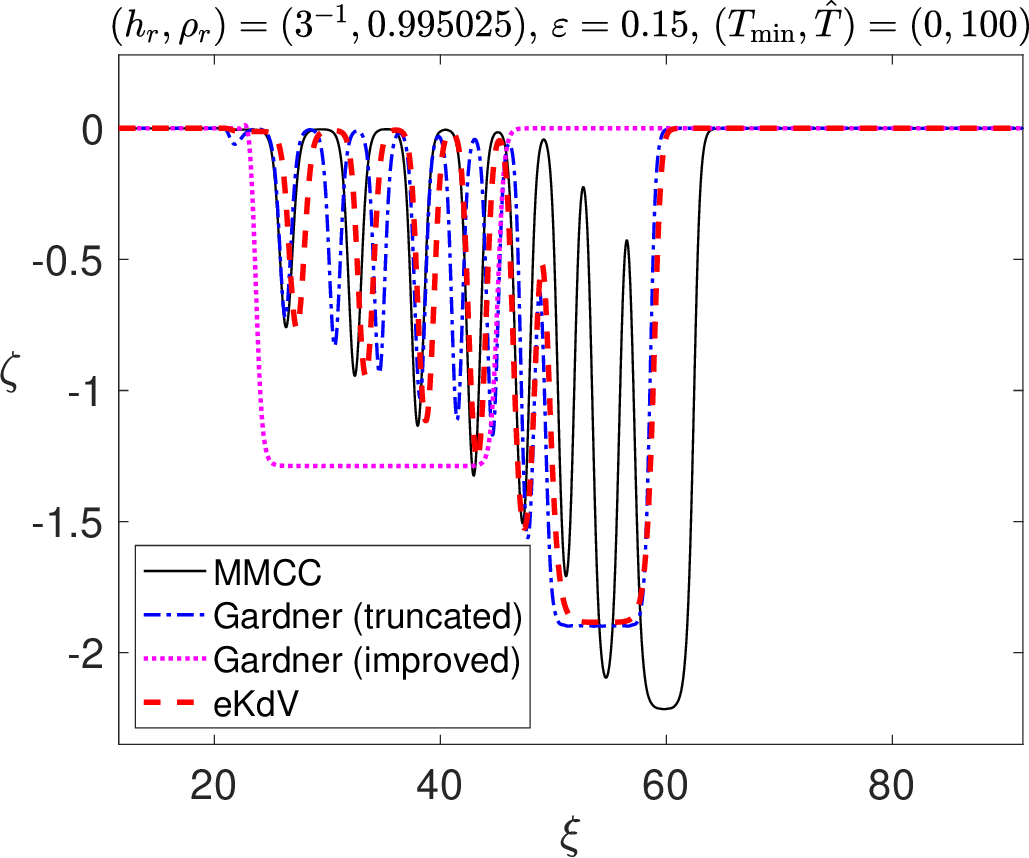}\,
\includegraphics[width=0.3\textwidth]{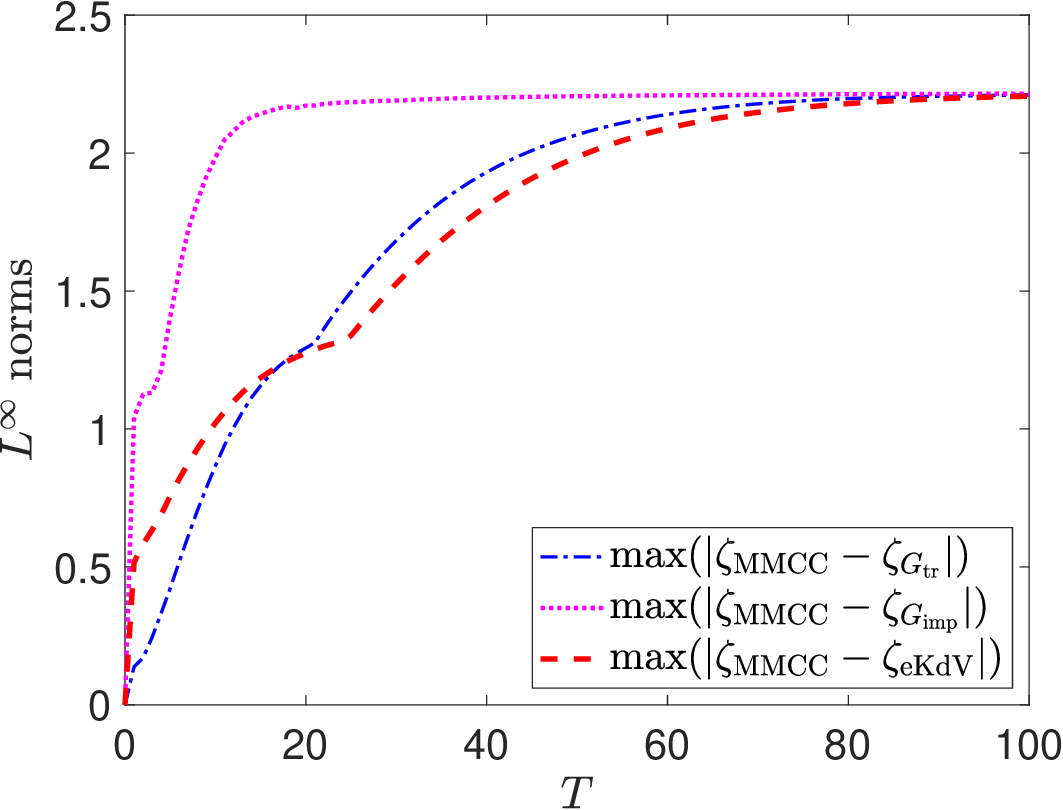}
\caption{Comparison of MMCC, Gardner, and eKdV models under the parameters $\eps = 0.15$,  $h_r = 1/3$, and $\rr = 1.005^{-1}$. Here the computational domain is $\xi \times \tau = [-40,40] \times [0, 4500]$. The initial condition for all models is \eqref{ekdv_soliton_sol_travelling} with $M$ set as either $-0.25$ (top row), $-0.65$ (second row), or the truncation of $M^*$ at $8$ decimal places (third row) or $30$ decimal places (bottom row). The initial profiles are shown in the left panels. The middle panels in the first and second rows correspond to $T=200$, whereas in the third and fourth rows the middle panel correspond to $T=50$ and $T=100$, respectively. 
 {\color{black} The right panels in each row show the evolutions of the $L^\infty$ norms of the differences between the solution profile obtained from the MMCC model and the solution profiles obtained from the eKdV and Gardner models, as indicated in the legends (note that $\zeta_\mathrm{MMCC}$, $\zeta_\mathrm{eKdV}$, $\zeta_{G_\mathrm{tr}}$ and $\zeta_{G_\mathrm{imp}}$ correspond to the solutions of the MMCC, eKdV, Gardner truncated and Gardner improved models).}
}
\label{Mhorns_mcc_comparison_2}
\end{figure}

Figure \ref{Mhorns_mcc_comparison_2} shows what happens if we {\color{black} use the same values for} $M$ as before, but with a depth ratio $h_r = 1/3 < 1/2$. Here, for the case {\color{black} when} $M$ is truncated at $30$ significant figures of $M^*$, the initial profile becomes much narrower, and the {\color{black} amplitudes of the horns are greater}. This increases the difficulty of the numerical simulation, as the wavenumber spectrum requires considerably more filtering in order to avoid the development of instabillity in the MMCC runs. This figure shows that the emergence of the MMCC solitary wave takes longer than in the previous case, {\color{black} and the table-top barely forms at $T=100$.}  Nevertheless, the characteristics remain the same: the MMCC simulation eventually converges to {\color{black} its  true solitary wave solution.} In comparison, the Gardner models perform worse at predicting the MMCC behaviour in this case. They tend to underestimate the {\color{black} solitary wave} amplitude and overestimate its width, and also exhibit a more significant phase shift. {\color{black} By comparing the results at smaller $M$ values ($M=-0.25$ and $M=-0.65$),  it becomes clear that the eKdV model now has comparable performance to the improved Gardner model for the waves of moderate amplitude and to the truncated Gardner model for the waves of large amplitude. The analytical approximations are close to the solitary waves of the parent system for the waves of small and moderate amplitude, while they are closer to the numerical solution of the improved Gardner equation for the waves of large amplitude.
} Numerical simulations with MMCC model at $h_r$ values below $1/3$ presented significant numerical challenges 
and have therefore been beyond the scope of the present study. \\

To summarise, {\color{black} the improved Gardner equation {\color{black} provides a good approximation to} the MMCC solution for the waves of small and moderate amplitude, while the truncated Gardner model {\color{black} can be considered a reasonable approximation for large amplitude waves, for both} $h_r = 1/2$ and $h_r = 1/3$. In all cases, the eKdV equation produces numerical results comparable to the best performing Gardner model.}

\section{Approximate solutions for cnoidal waves of eKdV equation}
\label{sec:Cnoidal}

Following \cite{K2012} for the discussion of the cnoidal wave solution of the Gardner equation, we consider the Gardner equation in canonical form, i.e.
\begin{equation}
\label{GardnerCanonical}
\hat{B}_{\hat{T}} + 6 \hat{B} \hat{B}_{\hat{\xi}} + \hat{B}_{\hat{\xi}\hat{\xi}\hat{\xi}} - 6 \eps \hat{B}^2 \hat{B}_{\hat{\xi}} = 0,
\end{equation}
which can be reduced,
upon the introduction of
 a characteristic variable $\hat{\theta} = \hat{\xi} - \hat{V} \hat{T}$, to
\begin{equation}
\hat{B}_{\hat{\theta}}^2 = \eps \hat{B}^4 - 2 \hat{B}^3 - \hat{V} \hat{B}^2 + \hat{\kappa}_1 \hat{B} + \hat{\kappa}_2 := \hat{Q}(\hat{B}),
\end{equation}
where $\hat{Q}$ is quartic ``potential'' and $\hat{\kappa}_{1,2}$ are arbitrary constants. We impose that $\hat{Q}$ has four real roots
\begin{equation}
\label{GardnerCanonicalRoots}
\hat{B}_1 \leq \hat{B}_2 \leq \hat{B}_3 \leq \hat{B}_4,
\end{equation}
whereby three of these are independent, and the last one can be found by noting that the sum of roots must be $2/\eps$. This then fixes the constants $\hat{V}, \hat{\kappa}_1, \hat{\kappa}_2$ as:
\begin{align}
\hat{V} & = \eps (\hat{B}_1 \hat{B}_2 + \hat{B}_1 \hat{B}_3 + \hat{B}_1 \hat{B}_4 + \hat{B}_2 \hat{B}_3 + \hat{B}_2 \hat{B}_4 + \hat{B}_3 \hat{B}_4), \\
\hat{\kappa}_1 & = -\eps ( \hat{B}_1 \hat{B}_2 \hat{B}_3 + \hat{B}_2 \hat{B}_3 \hat{B}_4 + \hat{B}_3 \hat{B}_4 \hat{B}_1 + \hat{B}_4 \hat{B}_1 \hat{B}_2 ), \\
\hat{\kappa}_2 & = \eps \hat{B}_1 \hat{B}_2 \hat{B}_3 \hat{B}_4,
\end{align}
in accordance with Vieta's formulae. The cnoidal wave solution of \eqref{GardnerCanonical} can then be written down as
\begin{equation}
\label{GardnerCnoidal}
\hat{B} = \hat{B}_2 + \dfrac{(\hat{B}_3 - \hat{B}_2) \mathrm{cn}^2(\hat{\Gamma} \hat{\theta},\hat{m})}{1 - \frac{\hat{B}_3 - \hat{B}_2}{\hat{B}_4 - \hat{B}_2} \mathrm{sn}^2(\hat{\Gamma} \hat{\theta},\hat{m})},
\end{equation}
where $\mathrm{cn},\mathrm{sn}$ are Jacobi elliptic functions, and 
\begin{align}
\hat{\Gamma} = \dfrac{\sqrt{\eps(\hat{B}_3 - \hat{B}_1)(\hat{B}_4 - \hat{B}_2)}}{2}, \qquad \hat{m} = \frac{(\hat{B}_3 - \hat{B}_2)(\hat{B}_4 - \hat{B}_1)}{(\hat{B}_4 - \hat{B}_2)(\hat{B}_3 - \hat{B}_1)}.
\end{align}

Using scalings, it is possible to convert \eqref{GardnerCanonical} to \eqref{G} therefore obtaining the cnoidal wave solutions of \eqref{G}. To do so, we use the following change of variables:
\begin{equation}
\label{transforms}
\hat{B} = \dfrac{-\alpha_2}{\alpha} B, \qquad \hat{\xi} = \dfrac{\alpha}{-\alpha_2} \sqrt{\dfrac{-\alpha_2}{6 \beta}} \xi , \qquad \hat{T} = \beta \bigg( \dfrac{\alpha}{-\alpha_2} \sqrt{\dfrac{-\alpha_2}{6 \beta}} \bigg)^3 T.
\end{equation}
By using the substitution $\theta = \xi - V T$, we can reduce \eqref{G} to the nonlinear equation of the form
\begin{equation}
B_\theta^2 = -\dfrac{\alpha_2}{6 \beta} Q(B), \quad Q(B) := \eps B^4 + \dfrac{2\alpha}{\alpha_2} B^3 - \dfrac{6V}{\alpha_2} B^2 + \kappa_1 B + \kappa_2.
\end{equation}
Then, from \eqref{GardnerCnoidal} combined with the transformation given by \eqref{txidef}, 
the cnoidal wave solution of \eqref{mG} can be written as
\begin{equation}
\label{cnoidalsol}
B(\theta) = B_2 + \dfrac{(B_3 - B_2) \mathrm{cn}^2(\Gamma \theta,m)}{1 - \frac{B_3 - B_2}{B_4 - B_2} \mathrm{sn}^2(\Gamma \theta,m)},
\end{equation}
where 
\begin{align}
\theta & = \xi + \dfrac{\alpha_2 V}{6} T + \eps c f(T), \qquad \Gamma = \sqrt{ \dfrac{- \eps \alpha_2 (B_3-B_1)(B_4-B_2)}{24\beta} }, \\
m & = \frac{(B_3 - B_2)(B_4 - B_1)}{(B_4 - B_2)(B_3 - B_1)} = \hat{m},
\end{align}
and
\begin{equation}
B_1 \leq B_2 \leq B_3 \leq B_4 \quad {\color{black} \mbox{for} \quad  {-\alpha_2}/{\alpha} >0}
\end{equation}
{\color{black} (with the opposite ordering otherwise)} are the zeroes of $Q(B)$ with
\begin{align}
V & = \eps (B_1 B_2 + B_1 B_3 + B_1 B_4 + B_2 B_3 + B_2 B_4 + B_3 B_4), \\
\kappa_1 & = -\eps ( B_1 B_2 B_3 + B_2 B_3 B_4 + B_3 B_4 B_1 + B_4 B_1 B_2 ), \\
\kappa_2 & = \eps B_1 B_2 B_3 B_4.
\end{align}

Solution \eqref{cnoidalsol} corresponds to periodic motions in the region $B_2 \leq B \leq B_3$. Thus, we can vary $B_{1,2,3}$ and write
\begin{equation}
B_4 = -\dfrac{2 \alpha}{\eps \alpha_2} - B_1 - B_2 - B_3
\end{equation}
to examine periodic solutions numerically. Furthermore, the wavelength $\hat{L}$ of \eqref{GardnerCnoidal} is given as
\begin{equation}
\hat{L} = \dfrac{4 K(\hat{m})}{\sqrt{\eps (\hat{B}_3 - \hat{B}_1)(\hat{B}_4 - \hat{B}_2)}},
\end{equation}
where $K(\hat{m})$ is the complete elliptic integral of the first kind. After performing the appropriate scalings, we find the wavelength $L$ of \eqref{cnoidalsol} as
\begin{equation}
\label{wavelength}
L = \dfrac{4 K(m)}{\sqrt{\eps (B_3 - B_1)(B_4 - B_2)}} \cdot \sqrt{\dfrac{6\beta}{-\alpha_2}} = \dfrac{2K(m)}{\Gamma}.
\end{equation}

Having the cnoidal wave solution \eqref{cnoidalsol} of the equation \eqref{mG}, {\color{black} we choose $B_2 = 0$, and consider the periodic motions with the amplitude equal to $B_3$ (positive for the waves of elevation and negative for the waves of depression).  In what follows, we apply the near-identity transformation \eqref{NIT}, integrating from $\xi_0 = 0$. The remainder term (\ref{remainder}) does not vanish exactly, but it is small, provided the waves are sufficiently long. We obtain an} approximate cnoidal wave solution of the eKdV equation \eqref{mcc_ekdv}:
\begin{align}
\label{ekdv_cnoidal_sol}
A = \dfrac{B_2 B_3 - B_2 B_4 S^2 - B_3 B_4 C^2}{B_2 C^2 + B_3 S^2 - B_4} + \eps \dfrac{(B_3 - B_2)(B_4 - B_2)(B_4 - B_3)}{3 (B_2 C^2 + B_3 S^2 - B_4)^3} A^*,
\end{align}
where $A^*$ is given by
\begin{align*}
A^* & = C D S ( B_2 C^2 + B_3 S^2 - B_4 ) \bigg[ \Gamma \{ 6 c f(T) + \xi ( 6 c B_2 + d V \alpha_2 ) \} \\
& + 6 c (B_3 - B_2) \dfrac{D}{|C|} \dfrac{\Pi(\pi_1, \pi_2, \pi_3)}{H(\theta)} \bigg] - 6 b \Gamma^2 C^2 \bigg[ \{ 3 ( B_3 - B_2 ) S^2 + B_4 - B_2 \} D^2 \\
& - m S^2 ( B_2 C^2 + B_3 S^2 - B_4 ) \bigg] -6 b \Gamma^2 D^2 S^2 ( B_2 C^2 + B_3 S^2 - B_4),
\end{align*}
and we have introduced the following shorthand notations for the Jacobi elliptic functions
\begin{align}
S = \mathrm{sn}(\Gamma \theta,m), \quad C = \mathrm{cn}(\Gamma \theta,m), \quad D = \mathrm{dn}(\Gamma \theta,m).
\end{align}
Here, $\Pi$ is the incomplete elliptic integral of the third kind with the parameters
\begin{align*}
\pi_1 = \dfrac{ B_3 - B_4 }{(m-1)(B_4 - B_2)}, \quad \pi_2 = \arcsin\bigg( \sqrt{m-1} \mathrm{sc}(\Gamma \theta,m) \bigg), \quad \pi_3 = \dfrac{1}{1-m},
\end{align*}
and $H$ is a function of an additional Jacobi function $\mathrm{sc}$ defined as follows:
\begin{align}
H = H(\theta) = \sqrt{m-1} \sqrt{ 1 + (1-m) \mathrm{sc}^2 (\Gamma \theta,m) }.
\end{align}
To ensure this is a travelling wave, we may impose that
\begin{equation}
6c f(T) + ( 6 c B_2 + \alpha_2 d V ) \xi = ( 6 c B_2 + \alpha_2 d V ) \theta \quad \iff \quad f(T) = \dfrac{\alpha_2 V (6 c B_2 + \alpha_2 d V)}{6c(6 - \eps \alpha_2 d V - 6 \eps c B_2)} T,
\end{equation}
which leads to the travelling cnoidal wave solution in the form
\begin{align}
\label{ekdv_cnoidal_sol_travelling}
A(\theta) = \dfrac{B_2 B_3 - B_2 B_4 S^2 - B_3 B_4 C^2}{B_2 C^2 + B_3 S^2 - B_4} + \eps \dfrac{(B_3 - B_2)(B_4 - B_2)(B_4 - B_3)}{3 (B_2 C^2 + B_3 S^2 - B_4)^3} \bar{A}^*(\theta),
\end{align}
where 
\begin{align}
\bar{A}^*(\theta) & = C D S ( B_2 C^2 + B_3 S^2 - B_4 ) \bigg[ ( 6 c B_2 + d V \alpha_2 ) \Gamma \theta + 6 c (B_3 - B_2) \dfrac{D}{|C|} \dfrac{\Pi(\pi_1, \pi_2, \pi_3)}{H(\theta)} \bigg] \nonumber \\
& - 6 b \Gamma^2 C^2 \bigg[ \{ 3 ( B_3 - B_2 ) S^2 + B_4 - B_2 \} D^2 - m S^2 ( B_2 C^2 + B_3 S^2 - B_4 ) \bigg] \nonumber \\
& -6 b \Gamma^2 D^2 S^2 ( B_2 C^2 + B_3 S^2 - B_4).
\end{align}

This cnoidal wave solution reduces to the previously obtained {\color{black} solitary wave} solution in the limit $m \to 1$.
Let us therefore consider the limit {\color{black} $B_1 \to B_2$ with  $B_2 = 0$.} This allows us to obtain
\begin{align*}
B & = \dfrac{B_3 B_4 \sech^2(\Gamma \theta)}{B_4 - B_3 \tanh^2(\Gamma \theta)}  = \dfrac{B_3 B_4}{B_4 \cosh^2(\Gamma \theta) - B_3 \sinh^2(\Gamma \theta)} \\[2mm]
& = \dfrac{2 B_3 B_4}{(B_4 - B_3) \cosh(2\Gamma \theta) + B_4 + B_3} 
 = \dfrac{\frac{2 B_3 B_4}{B_4 + B_3}}{1 + \frac{B_4 - B_3}{B_4 + B_3} \cosh(2\Gamma \theta)},
\end{align*}
where we have used the identities
\begin{equation}
\cosh^2(\Gamma \theta) = \dfrac{\cosh(2\Gamma \theta) + 1}{2}, \qquad \sinh^2(\Gamma \theta) = \dfrac{\cosh(2\Gamma \theta) - 1}{2}.
\end{equation}

As $B_{3,4}$ are restricted by a sum of roots condition, we can write
\begin{equation}
B_4 = - \dfrac{2 \alpha}{\eps \alpha_2} - B_3,
\end{equation}
and the solution takes the form
\begin{equation}
B = \dfrac{ M }{1 + F \cosh( G \theta)}, \qquad \theta = \xi - \tilde{v} T + \eps c f(T),
\end{equation}
where the choice
\begin{equation}
M = \dfrac{2 B_3 B_4}{B_4 + B_3 }, \quad F = \dfrac{B_4 - B_3}{B_4 + B_3}, \quad G = 2 \Gamma, \quad \tilde{v} = - \dfrac{\alpha_2 V}{6}
\end{equation}
identifies our cnoidal wave reduction with the previously obtained solution \eqref{mGSol}.
This cnoidal wave reduction depends on a single parameter $B_3$, and in order to identify this parameter with the old parameter $M$ we can invert the above relation for $M$ to obtain
\begin{equation}
\label{B3_MF_relation}
B_3 = -  \dfrac{\alpha}{\eps \alpha_2} \underbrace{ \bigg( 1 - \sqrt{1 + \dfrac{\eps \alpha_2}{\alpha} M} \bigg) }_{= 1 - F} = \dfrac{ M }{ 1 + F },
\end{equation}
where the choice of the sign for the solution of a quadratic for $B_3$ is such that $B_3 \to 0$ as $M \to 0$. 
\\

\section{Numerical modelling of cnoidal waves}
\label{sec:CnoidalWaveNumerics}

\begin{figure}
\centering
\includegraphics[width=0.6\textwidth]{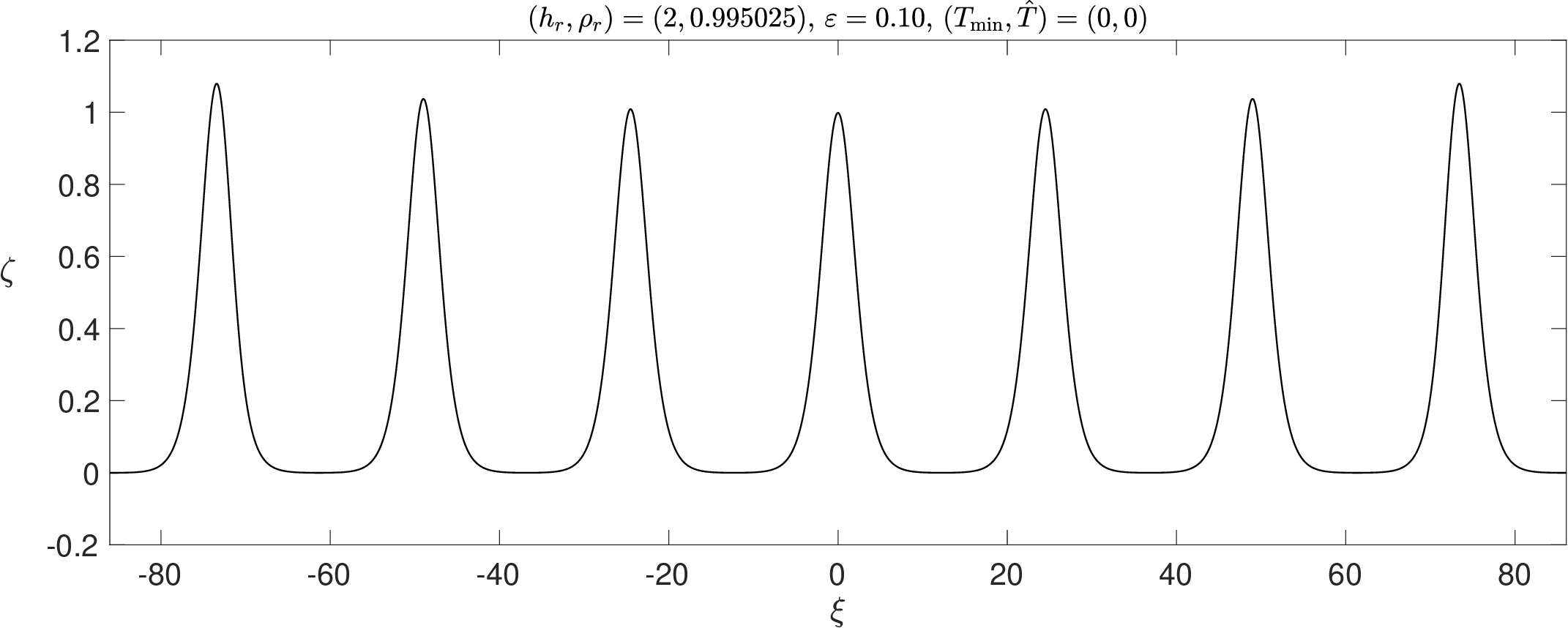}\\[0.2cm]
\includegraphics[width=0.6\textwidth]{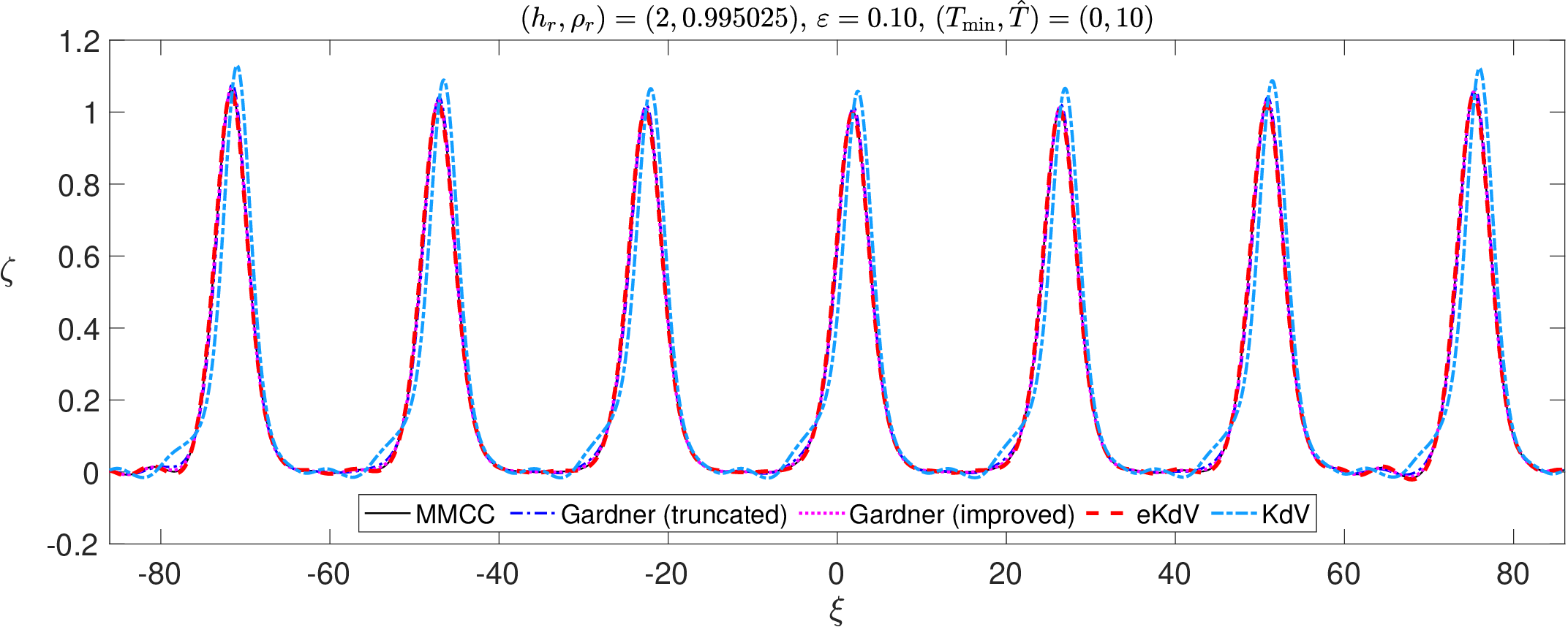}\\[0.2cm]
\includegraphics[width=0.6\textwidth]{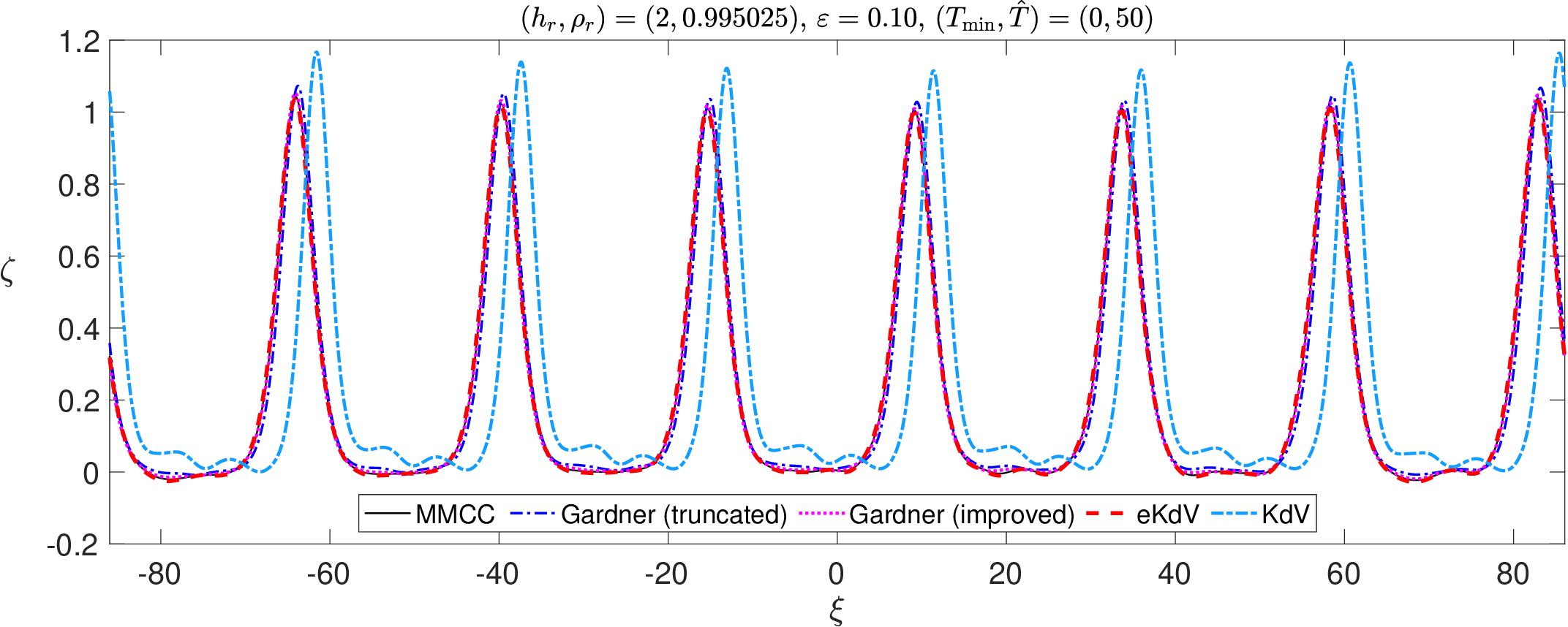}\\[0.2cm]
\includegraphics[width=0.6\textwidth]{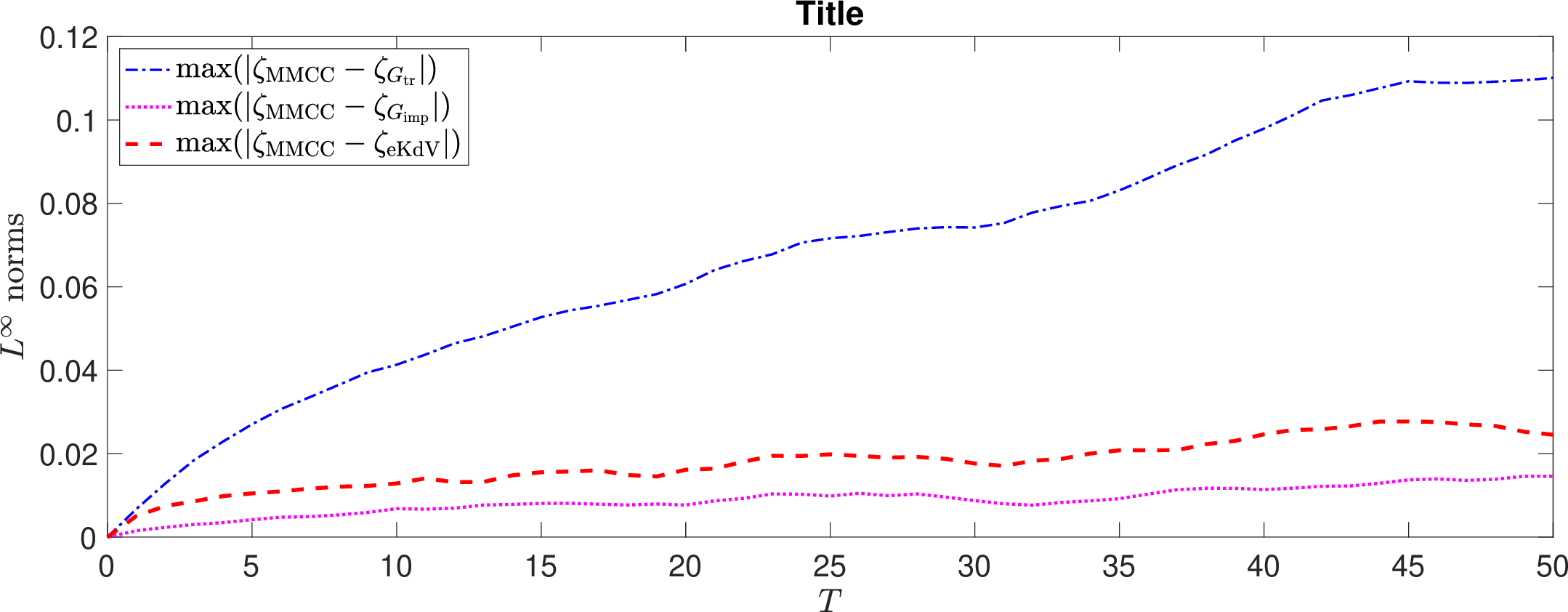}
\caption{Comparison of numerical results from the MMCC, eKdV, and Gardner models. All have been initiated with the cnoidal wave \eqref{ekdv_cnoidal_sol} truncated with $7$ peaks. The three times at which the comparisons take place are $T = 0, 10, 50$. Here, the parameters are set as {\color{black} $\eps= 0.10$}, $\rho_r = 1.005^{-1}$, $h_r = 2,$ and $(B_1, B_2, B_3) = (-0.001, 0, 1)$. {\color{black} The bottom panel shows the evolution of the $L^\infty$ norms of the differences between the solution profile obtained from the MMCC model and the solution profiles obtained from the eKdV and Gardner models, as indicated in the legends (note that $\zeta_\mathrm{MMCC}$, $\zeta_\mathrm{eKdV}$, $\zeta_{G_\mathrm{tr}}$ and $\zeta_{G_\mathrm{imp}}$ correspond to the solutions of the MMCC, eKdV, Gardner truncated and Gardner improved models).} 
}
\label{Cnoidal_numerics_7peaks}
\end{figure}

\begin{figure}
\centering
\includegraphics[width=0.45\textwidth]{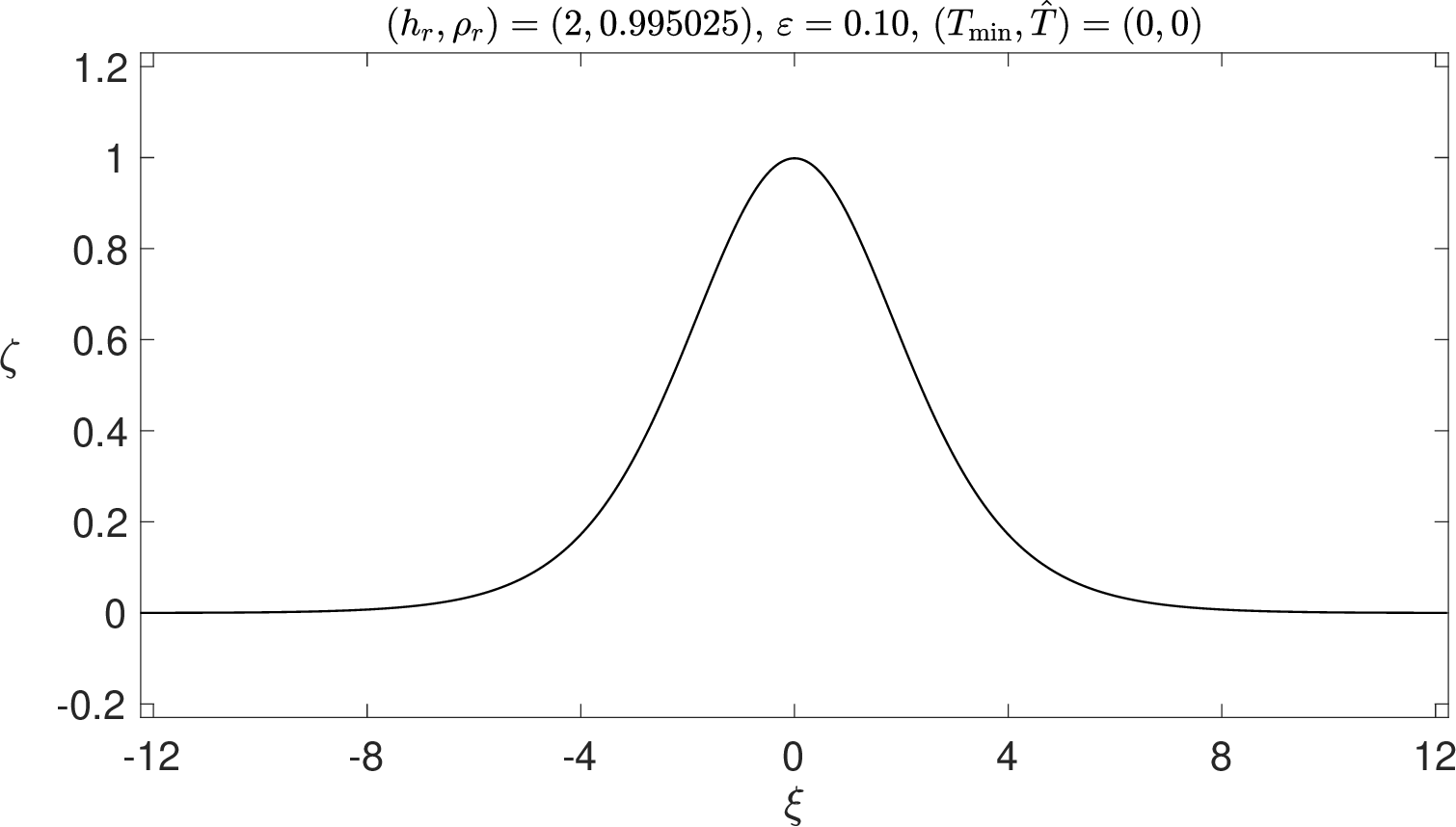}\quad
\includegraphics[width=0.45\textwidth]{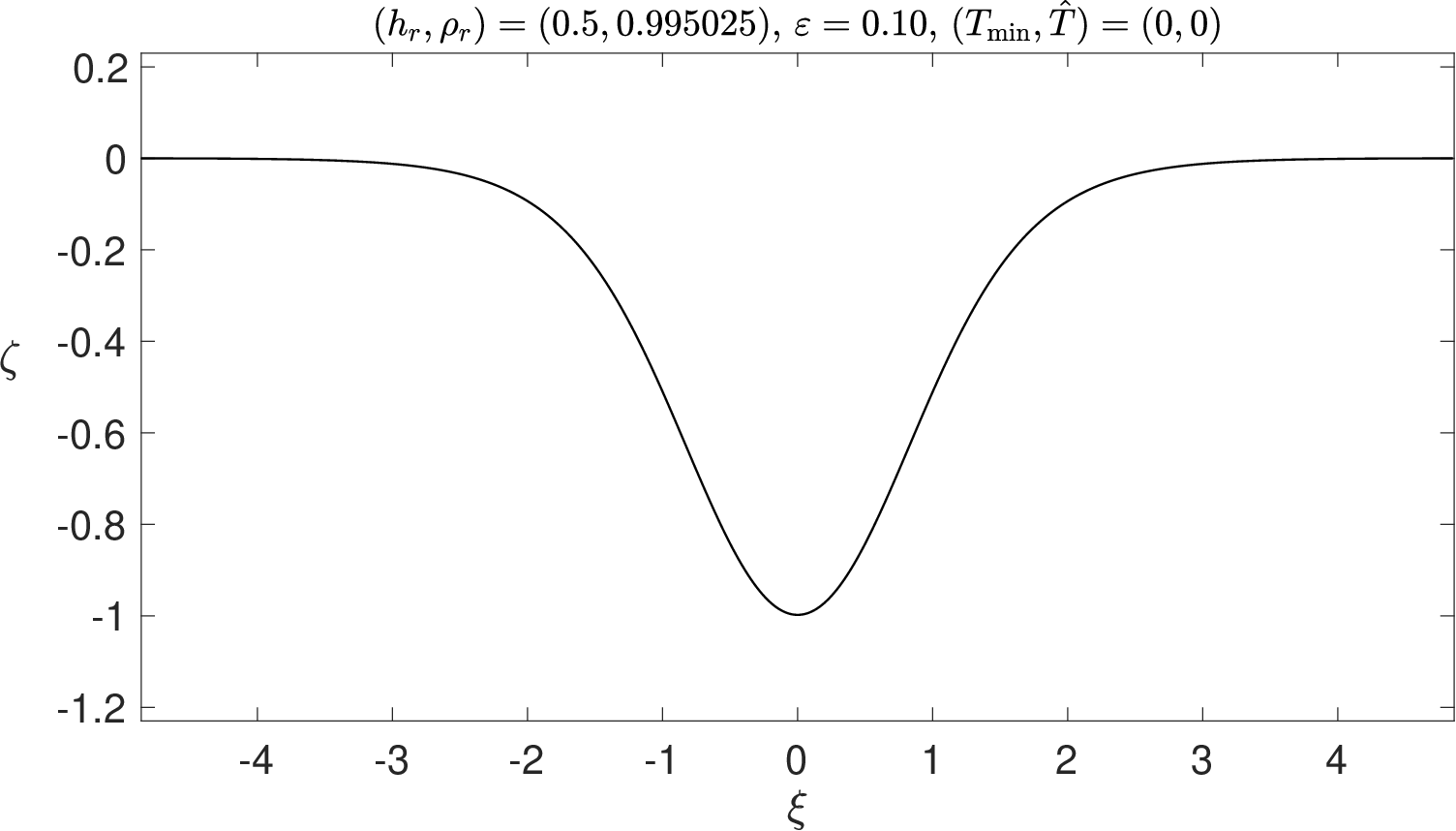}\\[0.2cm]
\includegraphics[width=0.45\textwidth]{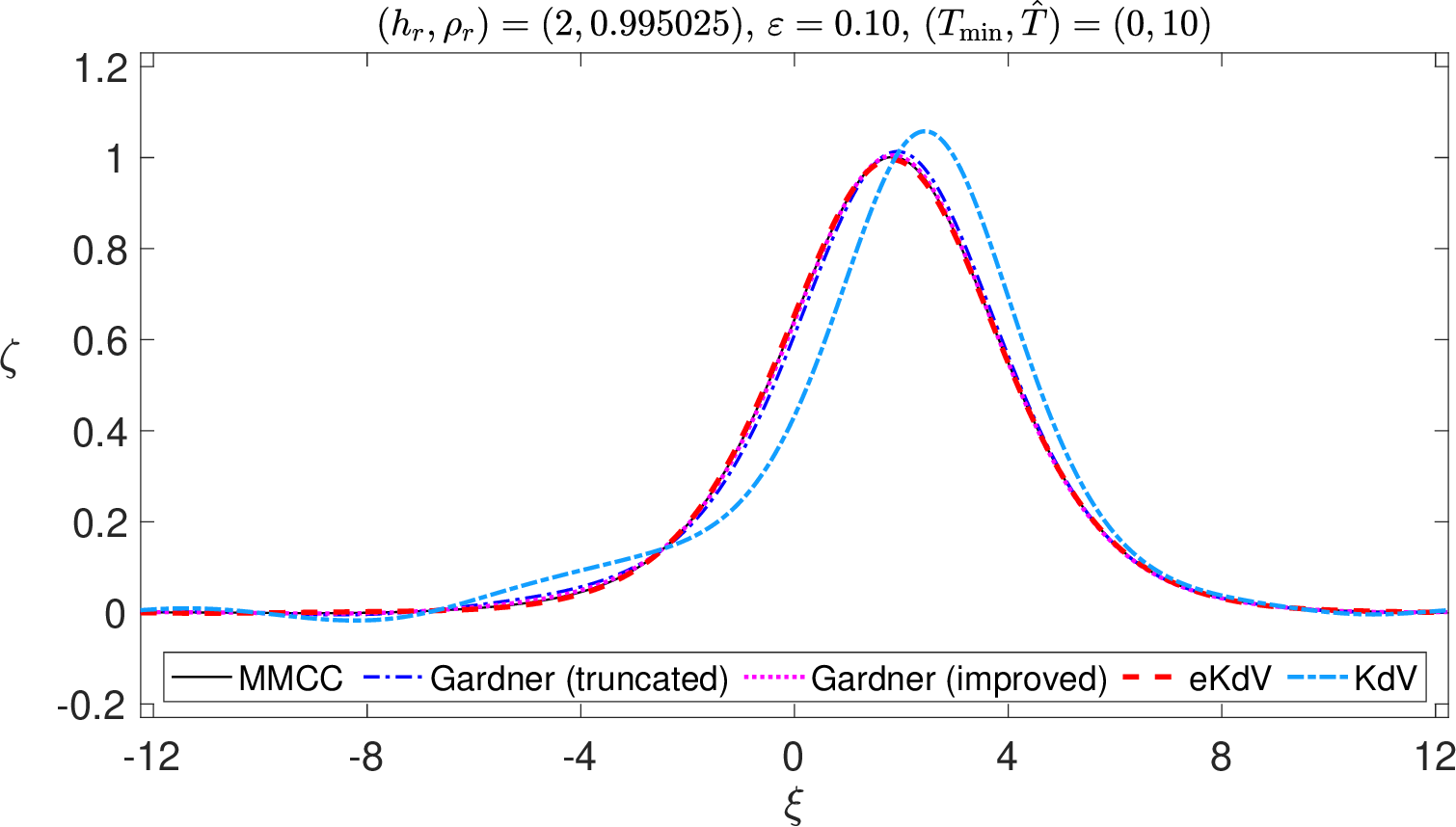}\quad
\includegraphics[width=0.45\textwidth]{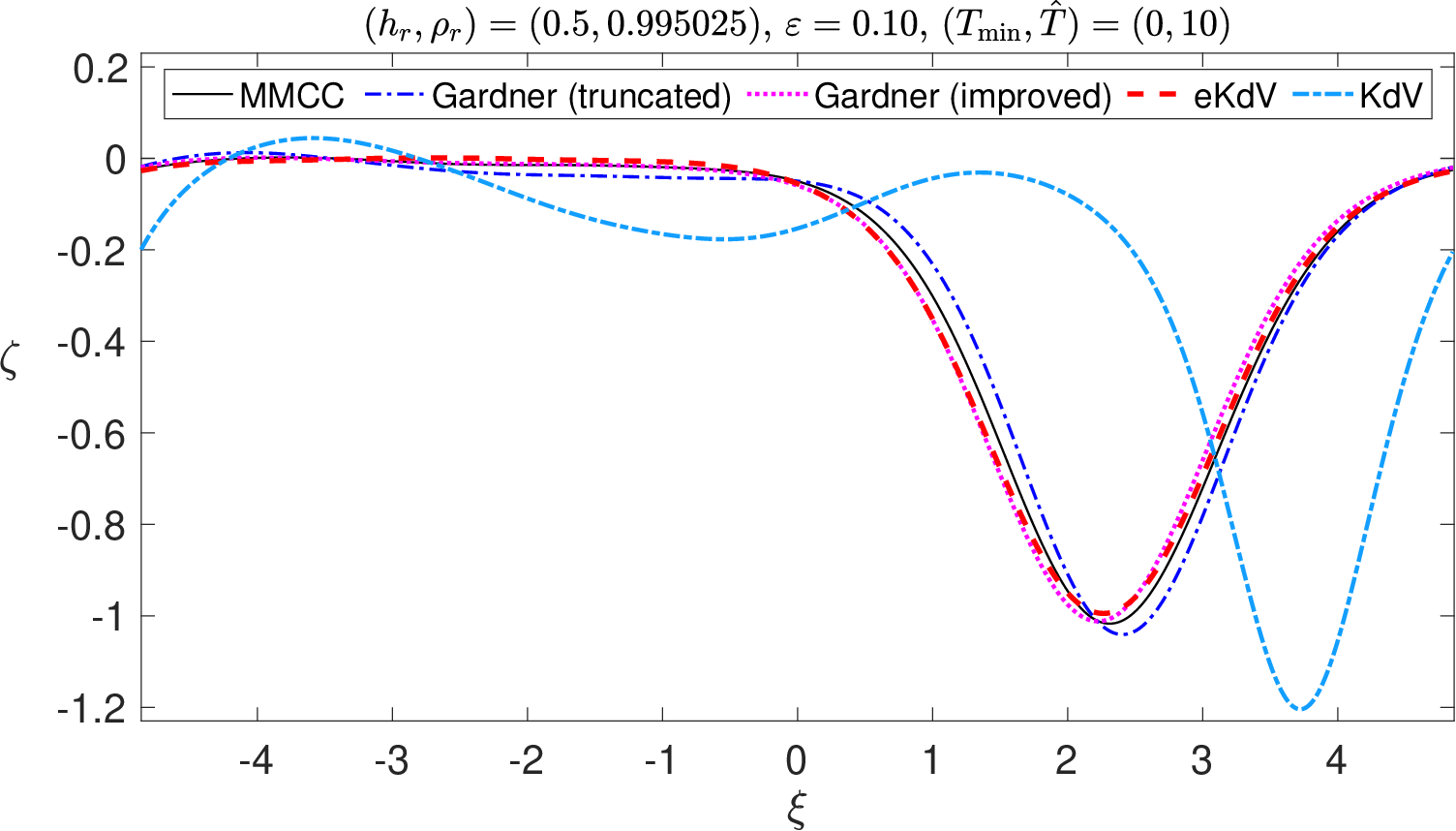}\\[0.2cm]
\includegraphics[width=0.45\textwidth]{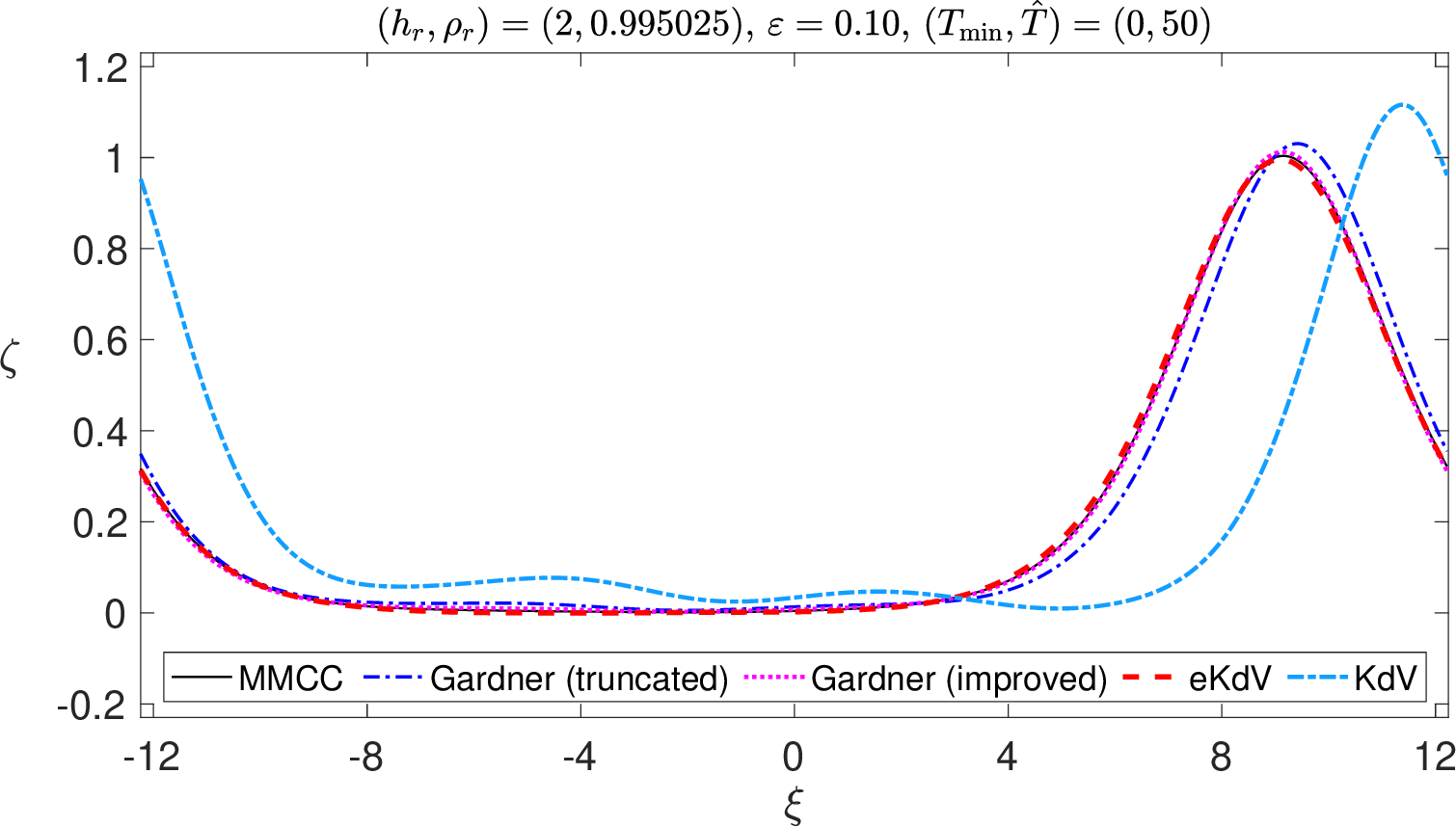}\quad
\includegraphics[width=0.45\textwidth]{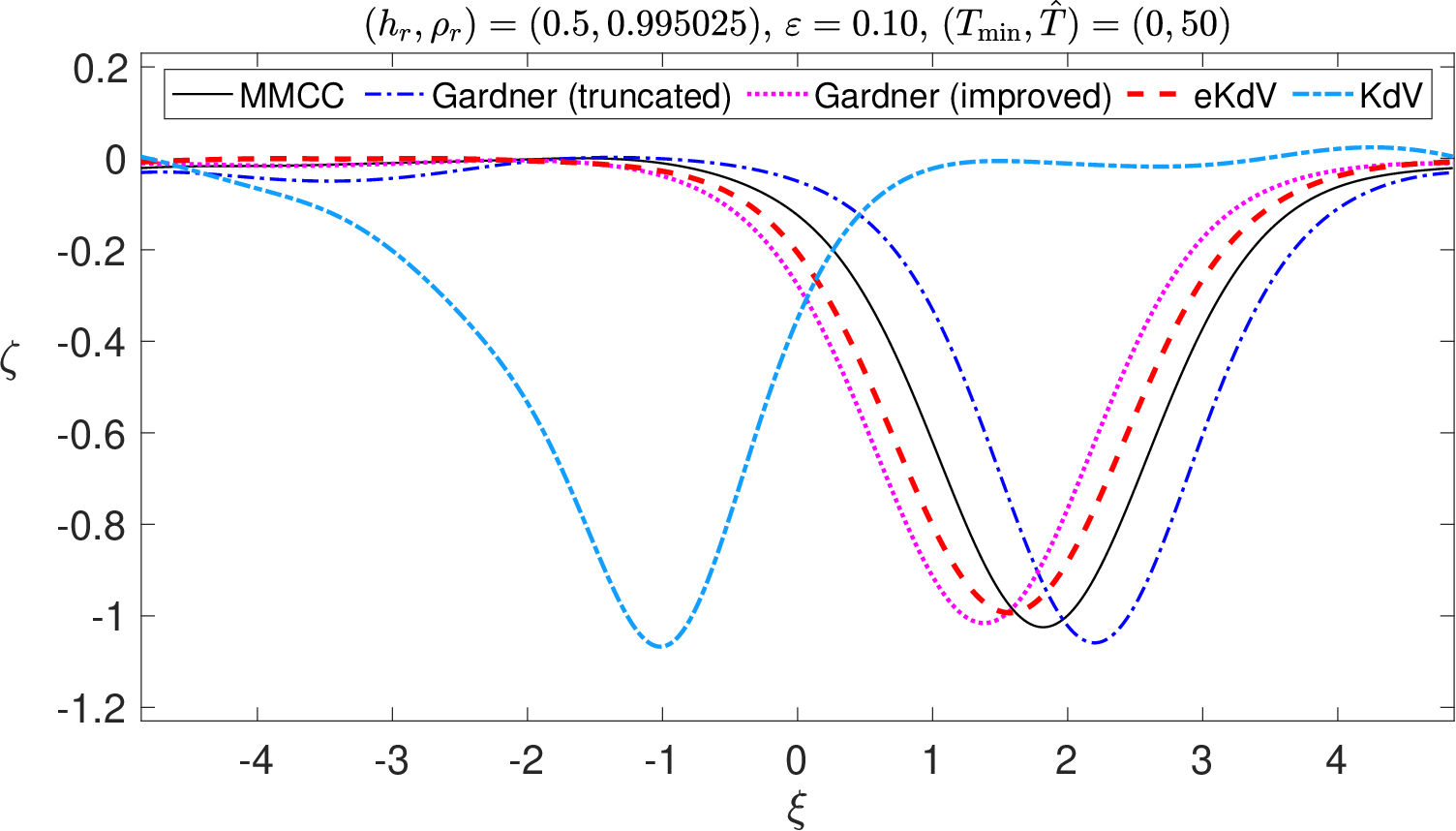}\\[0.2cm]
\includegraphics[width=0.45\textwidth]{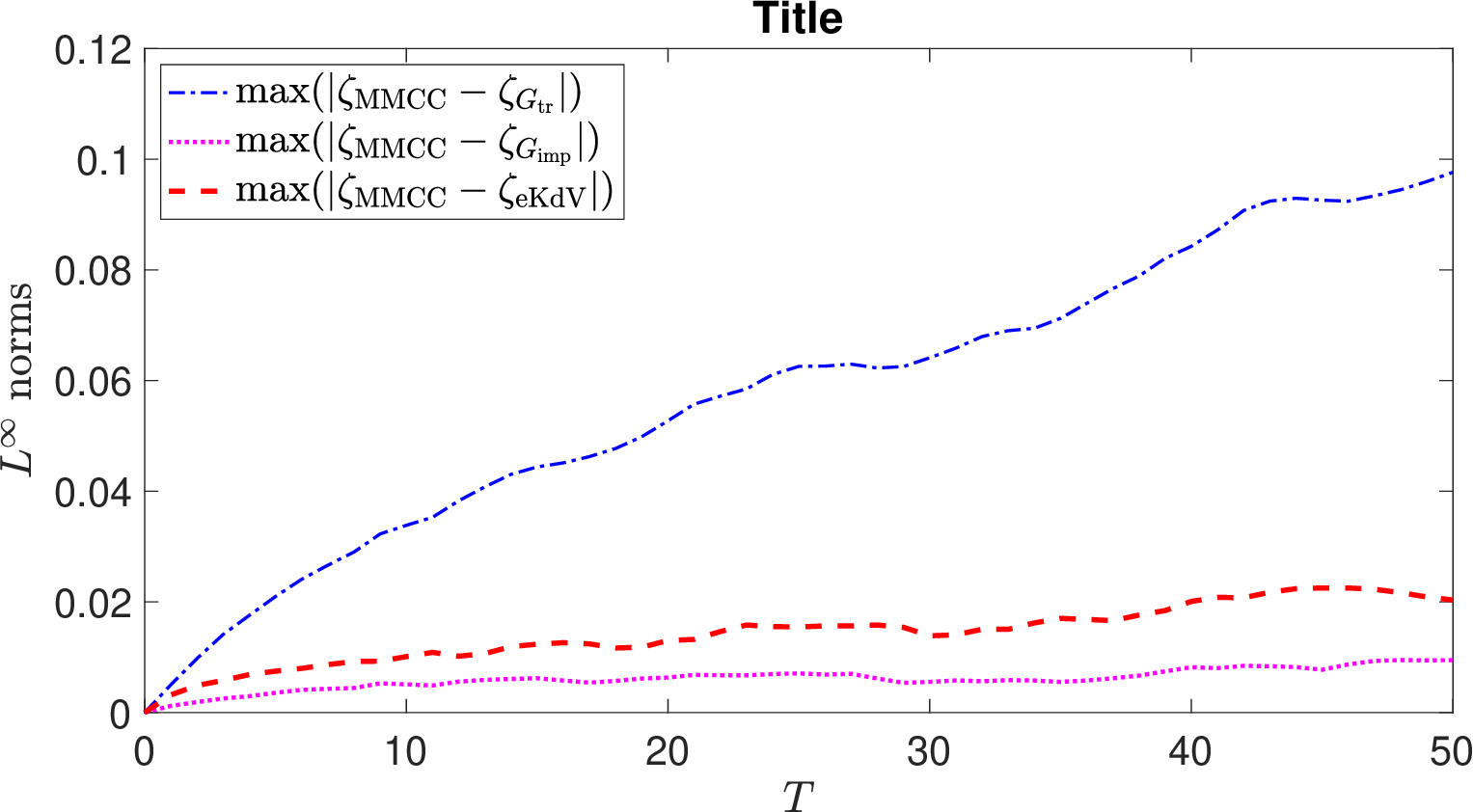}\quad
\includegraphics[width=0.45\textwidth]{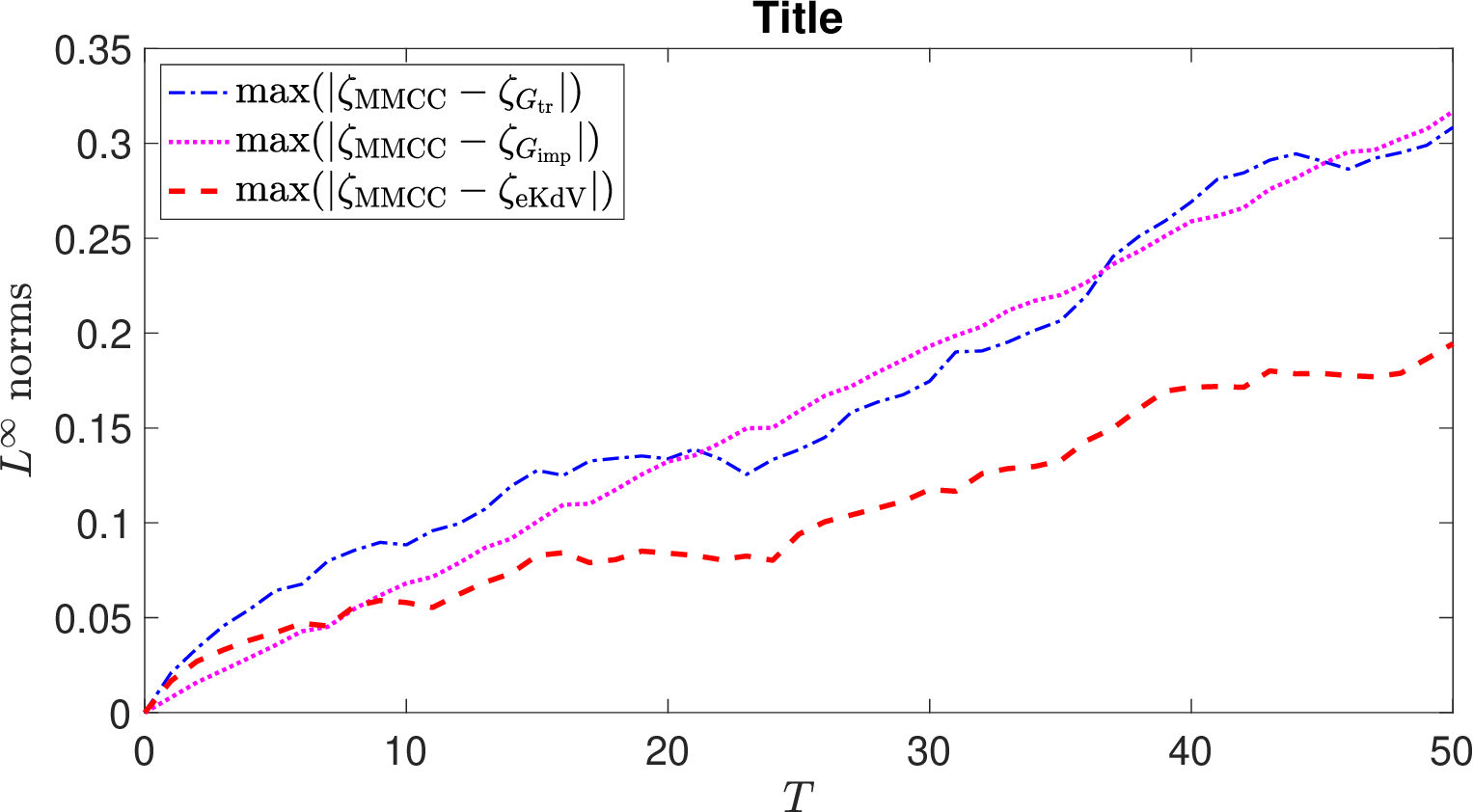}\\[0.2cm]
\caption{Comparison of numerical results from the MMCC, eKdV, and Gardner models. All have been initiated with the cnoidal wave \eqref{ekdv_cnoidal_sol} of elevation truncated with $1$ peak. The parameters are set as $\eps= 0.10$, $\rho_r = 1.005^{-1}$, $h_r = 2,$ and $(B_1, B_2, B_3) = (-0.001, 0, 1)$ for the left panels {\color{black} (waves of elevation)}, or {\color{black} $\eps= 0.10$, $\rho_r = 1.005^{-1}$, $h_r = 1/2,$} and $(B_1, B_2, B_3) = (0.001, 0, -1)$ for the right panels {\color{black} (waves of depression)}. The top row shows the initial conditions. The second that third rows correspond to $T = 10$ and $50$, respectively.  {\color{black} The bottom panels show the evolution of the $L^\infty$ norms of the differences between the solution profile obtained from the MMCC model and the solution profiles obtained from the eKdV and Gardner models, as indicated in the legends (note that $\zeta_\mathrm{MMCC}$, $\zeta_\mathrm{eKdV}$, $\zeta_{G_\mathrm{tr}}$ and $\zeta_{G_\mathrm{imp}}$ correspond to the solutions of the MMCC, eKdV, Gardner truncated and Gardner improved models).} }
\label{Cnoidal_numerics_1peak}
\end{figure}

The cnoidal wave solution \eqref{ekdv_cnoidal_sol} can be modelled numerically by imposing periodic boundary conditions. However, it is important to note that the constructed solution is only approximately periodic:  the amplitude grows indefinitely as one moves further away from the central peak. This behaviour, illustrated in Figure \ref{Cnoidal_numerics_7peaks}, introduces some ambiguity regarding the {\color{black} optimal} truncation of the constructed analytical approximation, when we use it as {\color{black} an initial-condition} for our numerical runs. {\color{black} Here, we perform experiments by comparing the symmetric truncation at the minimum value around the seven central peaks in Figure \ref{Cnoidal_numerics_7peaks} and one central peak in {\color{black} the left panels of} Figure \ref{Cnoidal_numerics_1peak}.   For these comparisons, we chose $h_r = 2$ and $B_1 = -0.001, B_2 = 0, B_3 =1\ (\mbox{with} \ B_4 = 6.896; m = 0.9991)$, which means we are close to the {\color{black} solitary wave} regime, and internal waves are waves of elevation. In these numerical simulations, we observe that the eKdV and improved Gardner models demonstrate very good agreement with the MMCC model, noticeably outperforming the truncated Gardner model, which exhibits phase shifts and amplitude growth. The truncation around a single central peak produces better results than truncation around seven peaks, with significantly fewer oscillations around the troughs of the cnoidal wave. Remarkably, both the eKdV and improved Gardner models {\color{black} are shown} to maintain excellent agreement with the MMCC model for the long simulation times. \\

In {\color{black} the right panels of} Figure \ref{Cnoidal_numerics_1peak} we revisit the case $h_r = 1/2$ considered in section 4, and {\color{black} perform} similar numerical runs using the approximate cnoidal waves solution truncated around a single central peak, choosing $B_1 = 0.001, B_2 = 0, B_3 =-1\ (\mbox{with} \ B_4 = -2.922; m = 0.9993$; waves of depression). In this case, the improvement offered by the eKdV and improved Gardner equations is marginal, when compared to the truncated Gardner model, which indicates that this is a borderline case for the validity of both models. Indeed, in our problem formulation the variables have been nondimensionalised using the depth of the bottom layer, and $h_r = 1/2$  means that the bottom layer is twice deeper than the top layer, while $h_r = 2$ means the opposite. Hence, although the value of $\eps$ is the same in all panels of Figure \ref{Cnoidal_numerics_1peak},  the ratio of the wave amplitude to the depth of the shallower layer and the ratio of the depth of the bottom later to the wavelength are smaller in the left panels. We also note that all three reduced models still produce very good approximations around the slow time value $T = 10$ in the right panels. }\\

Next, for {\color{black} $h_r = 2$} we test the approximation in the case which is further away from the {\color{black} solitary wave} limit. For the previous simulation, the elliptic modulus value was $m \approx 0.9991$. We now {\color{black} choose $B_1 = -1, B_2 = 0, B_3 = 1\ (\mbox{with} \ B_4 = 7.895; m = 0.5633$).} The results for this case are shown in Figure \ref{Cnoidal_numerics_3peaks_small_m} {\color{black} using a symmetric truncation at minimum values around the three central peaks.} Here, we observe that a smaller elliptic modulus produces a wave which does not have `flat' troughs compared to the previous near-{\color{black} solitary wave} case of $m\approx 1$. Nevertheless, {\color{black} the improved Gardner model continues to perform remarkably well in approximating the numerical results for the MMCC model, with the eKdV equation also producing a good approximation and significantly outperforming the truncated Gardner equation.  Interestingly, we see a noticeable improvement in the evolution of the wave over the seven-peak case in Figure \ref{Cnoidal_numerics_7peaks}. Indeed, while the seven-peak truncation used in Figure  \ref{Cnoidal_numerics_7peaks} developes small oscillations near the troughs of the wave at $T=50$, the three-peak truncation used in  Figure  \ref{Cnoidal_numerics_3peaks_small_m} does not have this feature at least up to the final simulation time of $T=50$, leading to a very good approximation to the cnoidal wave of the MMCC model. The simulation initiated using a truncation around a single peak produces very similar results and is not shown here. As before, the analytical approximations (not shown in the figures) are close to the numerical solution of the parent system for the waves of small and moderate amplitude, and they are closer to the numerical solution of the improved Gardner equation for the waves of greater amplitude and smaller wavelength  in the right panels of Figure \ref{Cnoidal_numerics_1peak}.
\\

\begin{figure}[h!]
\centering
\includegraphics[width=0.6\textwidth]{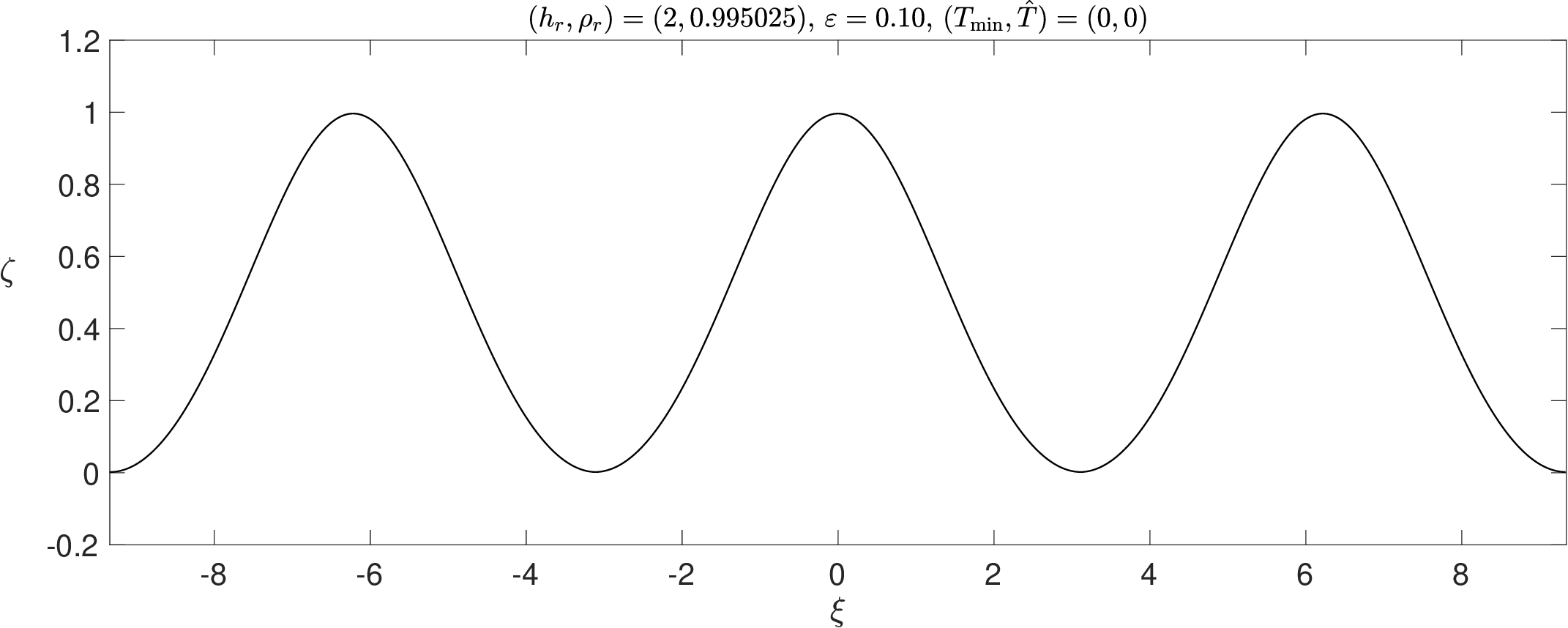}\\[0.2cm]
\includegraphics[width=0.6\textwidth]{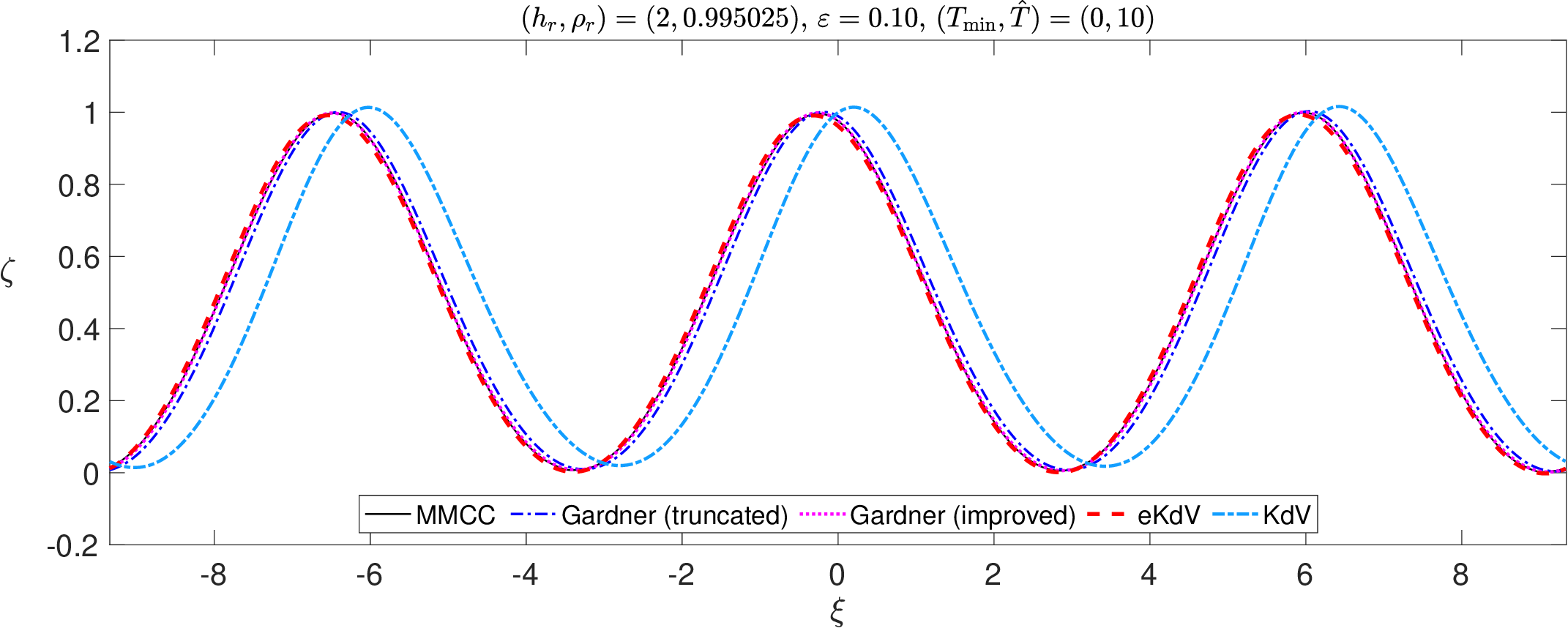}\\[0.2cm]
\includegraphics[width=0.6\textwidth]{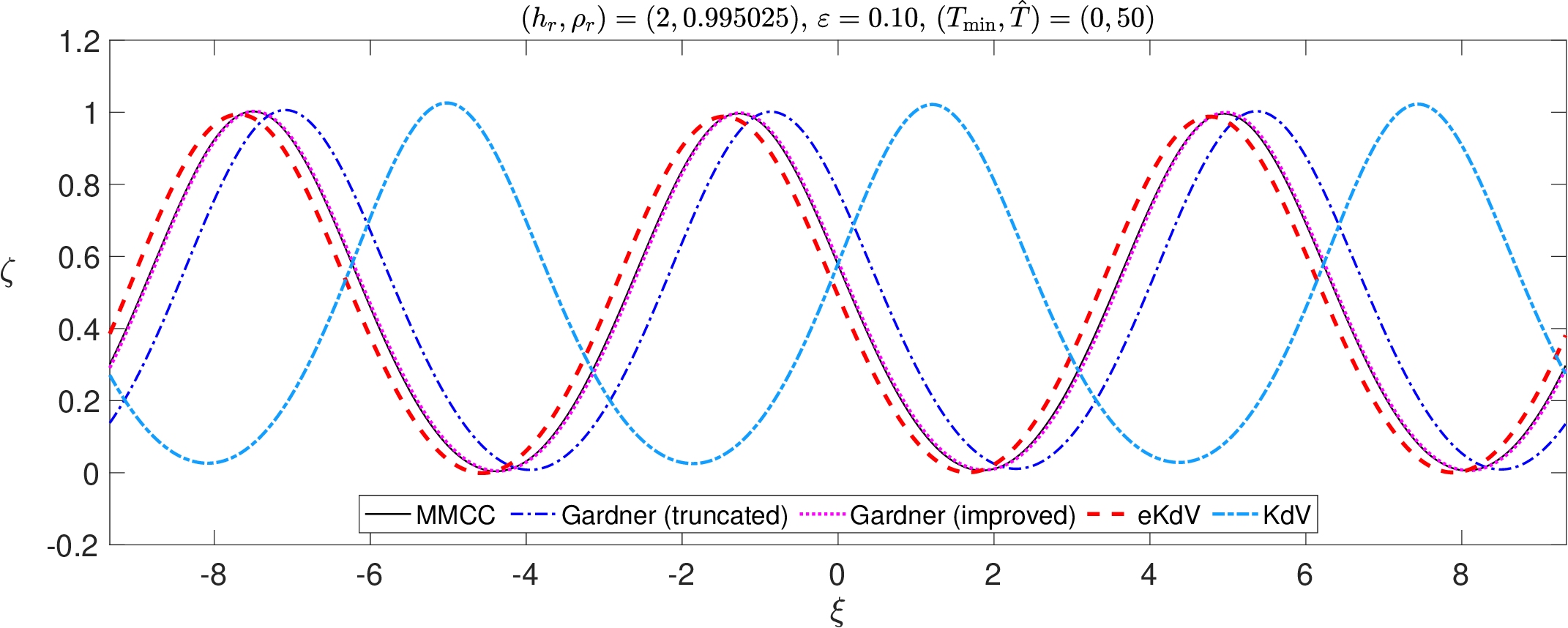}\\[0.2cm]
\includegraphics[width=0.6\textwidth]{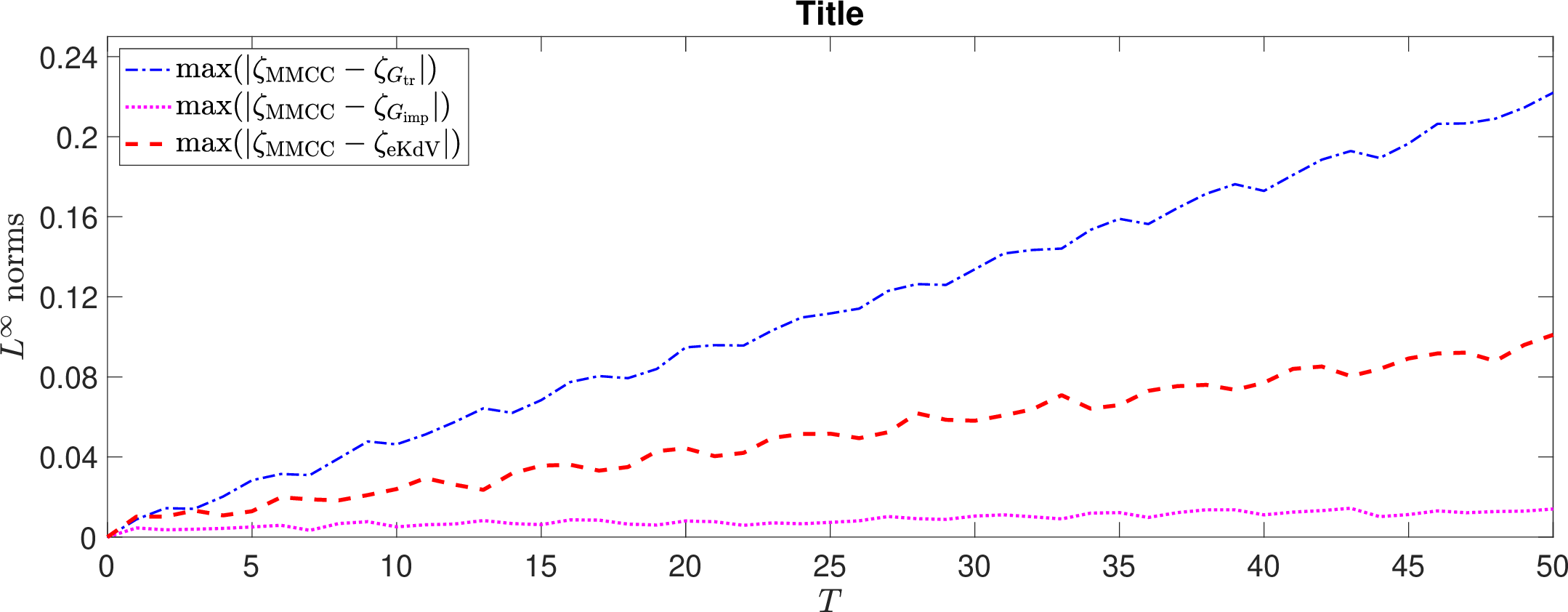}
\caption{Comparison of numerical results from the MMCC, Gardner, and eKdV models. All have been initiated with the cnoidal wave \eqref{ekdv_cnoidal_sol} truncated with $3$ peaks. Top left panel shows the initial condition, followed by top right panel showing the comparison at $T=10$, and the bottom panel showing it at $T=50$. Here, the parameters are set as {\color{black} $\eps = 0.10$}, $\rho_r = 1.005^{-1}$, $h_r = 2,$ and $(B_1, B_2, B_3) = (-1, 0, 1)$. {\color{black} The bottom panel shows the evolution of the $L^\infty$ norms of the differences between the solution profile obtained from the MMCC model and the solution profiles obtained from the eKdV and Gardner models, as indicated in the legends (note that $\zeta_\mathrm{MMCC}$, $\zeta_\mathrm{eKdV}$, $\zeta_{G_\mathrm{tr}}$ and $\zeta_{G_\mathrm{imp}}$ correspond to the solutions of the MMCC, eKdV, Gardner truncated and Gardner improved models).} }
\label{Cnoidal_numerics_3peaks_small_m}
\end{figure}

To summarise, the analytical approximation for cnoidal waves constructed in the previous section was used as {\color{black} an initial condition} for all reduced models and the parent system. The subsequent evolution was very close to the evolution of cnoidal waves of the MMCC system, provided the initial condition was truncated around one central peak (as well as several peaks, as long as there was no noticeable amplitude growth within this region).}

\section{Conclusions}
\label{sec:Conclusions}

In this paper we {\color{black} have} considered internal travelling waves in a two-layer fluid within the scope of a strongly-nonlinear model for a two-layer fluid with the background linear shear currents  \cite{C2006}. Asymptotic multiple-scale expansions {\color{black} were used} to derive the extended Korteweg - de Vries (eKdV) equation. In the absence of shear flows, the derived equation coincides with the models derived in  \cite{CC1999} for the two-layer case and in e.g.\ \cite{HFS2022} for the single-layer case.
Simple analytical approximations to its solitary and cnoidal wave solutions {\color{black} were constructed} by  applying a Kodama-Fokas-Liu near-identity transformation \cite{K1985a, K1985b, FL1996} to the derived {\color{black} eKdV} equation and mapping it to a Gardner equation, extending the results in \cite{GBK2020}. This improved Gardner equation has a different cubic nonlinearity coefficient and an additional transport term compared to the truncated Gardner equation. \\

In a case study, we {\color{black} tested performance of the reduced models by comparing their numerical solutions  with the results of direct numerical simulations for the parent (MMCC) system, which is known to be ill-posed due to} the intrinsic short-wave Kelvin-Helmholtz (KH)  instability 
\cite{JC2002,JC2008}. Nevertheless, with a careful choice of {\color{black} a low-pass filter to suppress unstable short waves}, we {\color{black} successfully conducted} a limited number of numerical experiments and {\color{black} compared} the results. In particular, {\color{black} when}  the constructed analytical approximations {\color{black} were used} as initial conditions (pre-conditioners),  {\color{black} they were} found to evolve into the MMCC solitary wave and its cnoidal wave.   \\

Within the scope of the cases considered in our studies, it was found that {\color{black} the improved Gardner equation gives a better approximation for the waves of small and moderate amplitude, while the truncated Gardner equation performs better for the waves of large amplitude, especially near the criticality. The performance of the eKdV equation is consistently close to the best performing Gardner model, which suggests the use of the eKdV equation as a reasonable universal model. In all cases, it can be concluded that the eKdV equation provides a good approximation to  the strongly-nonlinear MMCC model with a fraction of the computational cost, for the modelling of moderate amplitude waves. In this range, the eKdV equation has a very clear advantage over the simpler KdV equation, particularly in the long-time evolution.  The analytical approximations constructed in the paper were also successfully used as  initial conditions (pre-conditioners) for the generation of the large amplitude solitary waves of the MMCC model.}
  \\

It is  important to note that for $h_r < 1/3$ it was difficult to find a suitable filtering of the wavenumber spectra, and the simulations within the parent system very quickly broke down, not allowing us to test the models in this range.  It would be interesting to investigate the range of validity of the constructed models and approximations further, perhaps using solvers developed for continuous stratification in order to avoid the KH instability. {\color{black} In our simulations the range of validity of the eKdV equation turned out to be wider than that of the analytical approximations obtained using the near-identity transformations, presumably, } since the latter impose additional smallness requirements on the amplitude parameter. More numerical tests are required to fully clarify the range of validity of the eKdV equation and analytical approximations for the description of internal waves. Also, the eKdV equation {\color{black} considered in this paper} is developed under the assumption that $h_r \sim O(1)$. It could be useful to develop  models for other regimes, for example, when $h_r \sim O(\varepsilon)$. Finally, the analytical approximations developed in our paper can be used to analyse  the dependence of solutions on parameters of the problem, for example, the strength and direction of currents, which goes beyond the scope of our present study.
\bigskip

{\bf Acknowledgements}
\medskip

We thank Ricardo Barros and Zakhar Makridin for useful discussions. Nerijus Sidorovas, Dmitri Tseluiko and Karima Khusnutdinova
acknowledge the UKRI funding: the PhD project of Nerijus Sidorovas is funded by the Engineering and Physical Sciences Research
Council (EPSRC, project reference 2458723). {\color{black} Wooyoung Choi was supported by the US National Science Foundation through Grant
No. DMS-2108524.}  Karima Khusnutdinova and Wooyoung Choi also thank the London Mathematical
Society and Loughborough Institute of Advanced Studies for the partial support of their collaborative research.
\bigskip
\bigskip
\pagebreak

\appendix{\bf Appendix A: Coefficients of the eKdV model with linear shear currents}
\medskip

The coefficients of the extended KdV model \eqref{mcc_ekdv_shear} are defined as:

{\color{black} 
\begin{align}
\tilde{\alpha}_1 & = \frac{1}{2 h_r^2 \hat c_{10}} \cdot \frac{\hat c_4 + \hat c_5 v + \hat c_6 v^2 + \hat c_7 v^3 + \hat c_8 v^4 + \hat c_9 v^5}{2 v (h_r + \rr) - (h_r U_2+ \rr U_1)} \\[5mm]
& \hat c_4 = 3 h_r^2 \rr  U_1 (h_r U_2+U_1) (h_r U_2+\rr  U_1) \\[5mm]
& \hat  c_5 = -3 h_r^5 U_2^2+2 h_r^4 \rr U_1 U_2^3-15 h_r^4 \rr U_1 U_2-2 h_r^4 \rr U_2^4-6 h_r^4 \rr U_2^2-6 h_r^3 \rr^2 U_1^2-6 h_r^3 \rr U_1^2 \nonumber \\[3mm] & \qquad +3 h_r^3 \rr^2 U_1^2 U_2^2+2 h_r^3 \rr U_1^2 U_2^2-5 h_r^3 \rr U_1 U_2^3-12 h_r^3 \rr^2 U_1 U_2-9 h_r^3 \rr U_1 U_2 \nonumber \\[3mm] & \qquad  +3 h_r^2 \rr^3 U_1^3 U_2+h_r^2 \rr^2 U_1^3 U_2-15 h_r^2 \rr^2 U_1^2-h_r^2 \rr^2 U_1^2 U_2^2-3 h_r^2 \rr U_1^2 U_2^2 \nonumber \\[3mm] & \qquad +3 h_r \rr^3 U_1^4-2 h_r \rr^2 U_1^4-h_r \rr^2 U_1^3 U_2 \\[5mm]
& \hat c_6 = -2 h_r^5 U_2^3+9 h_r^5 U_2+15 h_r^4 \rr U_1-10 h_r^4 \rr U_1 U_2^2+8 h_r^4 \rr U_2^3+30 h_r^4 \rr U_2-12 h_r^3 \rr^2 U_1^2 U_2 \nonumber \\[3mm] & \qquad  -h_r^3 \rr U_1^2 U_2+24 h_r^3 \rr^2 U_1+18 h_r^3 \rr U_1-9 h_r^3 \rr^2 U_1 U_2^2+19 h_r^3 \rr U_1 U_2^2+10 h_r^3 \rr U_2^3 \nonumber \\[3mm] & \qquad  +12 h_r^3 \rr^2 U_2+9 h_r^3 \rr U_2 -6 h_r^2 \rr^3 U_1^3+h_r^2 \rr^2 U_1^3-15 h_r^2 \rr^3 U_1^2 U_2+h_r^2 \rr^2 U_1^2 U_2 \nonumber \\[3mm] & \qquad +12 h_r^2 \rr U_1^2 U_2 +27 h_r^2 \rr^2 U_1+8 h_r^2 \rr^2 U_1 U_2^2 +12 h_r^2 \rr U_1 U_2^2-18 h_r \rr^3 U_1^3 \nonumber \\[3mm] & \qquad  +14 h_r \rr^2 U_1^3+11 h_r \rr^2 U_1^2 U_2+\rr^3 U_1^3 \\[5mm]
& \hat c_7 = 5 h_r^5 U_2^2-6 h_r^5-30 h_r^4 \rr+12 h_r^4 \rr U_1 U_2-13 h_r^4 \rr U_2^2-24 h_r^3 \rr^2-18 h_r^3 \rr+9 h_r^3 \rr^2 U_1^2 \nonumber \\[3mm] & \qquad -2 h_r^3 \rr U_1^2 +33 h_r^3 \rr^2 U_1 U_2-42 h_r^3 \rr U_1 U_2+6 h_r^3 \rr^2 U_2^2-44 h_r^3 \rr U_2^2 -18 h_r^2 \rr^2 -7 \rr^3 U_1^2 \nonumber \\[3mm] & \qquad +30 h_r^2 \rr^3 U_1^2-14 h_r^2 \rr^2 U_1^2 -12 h_r^2 \rr U_1^2+24 h_r^2 \rr^3 U_1 U_2-24 h_r^2 \rr^2 U_1 U_2-48 h_r^2 \rr U_1 U_2 \nonumber \\[3mm] & \qquad -8 h_r^2 \rr^2 U_2^2-12 h_r^2 \rr U_2^2+42 h_r \rr^3 U_1^2 -52 h_r \rr^2 U_1^2-27 h_r \rr^2 U_1 U_2 \\[5mm]
& \hat c_8 = -3 h_r^5 U_2-3 h_r^4 \rr U_1+21 h_r^4 \rr U_2-21 h_r^3 \rr^2 U_1+36 h_r^3 \rr U_1-18 h_r^3 \rr^2 U_2+87 h_r^3 \rr U_2 \nonumber \\[3mm] & \qquad -48 h_r^2 \rr^3 U_1+51 h_r^2 \rr^2 U_1+48 h_r^2 \rr U_1-12 h_r^2 \rr^3 U_2+27 h_r^2 \rr^2 U_2+48 h_r^2 \rr U_2 \nonumber \\[3mm] & \qquad -45 h_r \rr^3 U_1+90 h_r \rr^2 U_1+18 h_r \rr^2 U_2+12 \rr^3 U_1 \\[5mm]
& \hat c_9 = -6 (h_r+1) \rr (h_r+\rr) (h_r (3 h_r-4 \rr+8)+\rr) \\[2mm]
& \hat c_{10} = v^3 \left(h_r+\rr\right)^2+3 v h_r \left(h_r+\rr\right)-h_r (h_r U_2+ \rr U_1)
\end{align}
\begin{align}
\tilde{\gamma}_1 & = \frac{v^2}{3\hat c_{10}} \cdot \frac{\hat c_{11} + \hat c_{12} v + \hat c_{13} v^2 + \hat c_{14} v^3}{2 v (h_r + \rr ) - (h_r U_2+\rr U_1)} \\[5mm]
& \hat c_{11} = -2 h_r^4 \rr U_2^3-h_r^4 \rr U_2-h_r^3 \rr^2 U_1-2 h_r^3 \rr^2 U_1 U_2^2-2 h_r^3 \rr U_1 U_2^2-h_r^3 U_2-2 h_r^2 \rr^2 U_1^2 U_2 \nonumber \\[3mm] & \qquad -h_r^2 \rr U_1+2 h_r^2 \rr U_1 U_2^2+2 h_r \rr^2 U_1^2 U_2+2 h_r \rr U_1^2 U_2+2 \rr^2 U_1^3 \\[5mm]
& \hat c_{12} = 2 h_r^4 \rr+9 h_r^4 \rr U_2^2+2 h_r^3 \rr^2+6 h_r^3 \rr^2 U_1 U_2+8 h_r^3 \rr U_1 U_2+2 h_r^3 \rr^2 U_2^2+h_r^3 \rr U_2^2 -10 \rr^2 U_1^2 \nonumber \\[3mm] & \qquad +2 h_r^3+2 h_r^2 \rr-h_r^2 \rr^3 U_1^2+6 h_r^2 \rr^2 U_1^2+2 h_r^2 \rr^2 U_1 U_2-4 h_r^2 \rr U_1 U_2-6 h_r^2 \rr U_2^2 \nonumber \\[3mm] & \qquad -h_r \rr^3 U_1^2-2 h_r \rr^2 U_1^2-2 h_r \rr U_1^2-8 h_r \rr^2 U_1 U_2-8 h_r \rr U_1 U_2 \\[5mm]
& \hat c_{13} = -17 h_r^4 \rr U_2-7 h_r^3 \rr^2 U_1-8 h_r^3 \rr U_1-7 h_r^3 \rr^2 U_2-4 h_r^3 \rr U_2-5 h_r^3 U_2+3 h_r^2 \rr^3 U_1 \nonumber \\[3mm] & \qquad -10 h_r^2 \rr^2 U_1-3 h_r^2 \rr U_1+6 h_r^2 \rr^2 U_2+13 h_r^2 \rr U_2+8 h_r \rr^3 U_1+7 h_r \rr^2 U_1 \nonumber \\[3mm] & \qquad +10 h_r \rr U_1+8 h_r \rr^2 U_2+10 h_r \rr U_2+20 \rr^2 U_1 \\[5mm]
& \hat c_{14} = 12 h_r^4 \rr+10 h_r^3 \rr^2+4 h_r^3 \rr+8 h_r^3-2 h_r^2 \rr^3-6 h_r^2 \rr^2 +2 h_r^2 \rr-10 h_r \rr^3 \nonumber \\[3mm] & \qquad  -6 h_r \rr^2-14 h_r \rr-14 \rr^2
\end{align}
\begin{align}
\tilde{\gamma}_2 & = \frac{v^2}{3 \hat c_{10}} \cdot \frac{\hat c_{15} + \hat c_{16} v + \hat c_{17} v^2 + \hat c_{18} v^3}{2 v (h_r + \rr ) -h_r U_2-\rr U_1} \\[5mm]
& \hat c_{15} = -6 h_r^4 \rr U_2^3-h_r^4 \rr U_2-h_r^3 \rr^2 U_1-6 h_r^3 \rr^2 U_1 U_2^2-4 h_r^3 \rr U_1 U_2^2-2 h_r^3 U_2^3-h_r^3 U_2 \nonumber \\[3mm] & \qquad  -2 h_r^2 \rr^2 U_1^2 U_2-h_r^2 \rr U_1+2 h_r^2 \rr U_1 U_2^2 \nonumber \\[3mm] & \qquad  +2 h_r \rr^3 U_1^3+4 h_r \rr^2 U_1^2 U_2+6 h_r \rr U_1^2 U_2+6 \rr^2 U_1^3 \\[5mm]
& \hat c_{16} = 2 h_r^4 \rr+27 h_r^4 \rr U_2^2+2 h_r^3 \rr^2+17 h_r^3 \rr^2 U_1 U_2+16 h_r^3 \rr U_1 U_2+9 h_r^3 \rr^2 U_2^2+2 h_r^3 \rr U_2^2 \nonumber \\[3mm] & \qquad  +9 h_r^3 U_2^2+2 h_r^3+2 h_r^2 \rr-h_r^2 \rr^3 U_1^2+7 h_r^2 \rr^2 U_1^2+h_r^2 \rr^2 U_1 U_2-3 h_r^2 \rr U_1 U_2 \nonumber \\[3mm] & \qquad  -7 h_r^2 \rr U_2^2-10 h_r \rr^3 U_1^2-3 h_r \rr^2 U_1^2-9 h_r \rr U_1^2-16 h_r \rr^2 U_1 U_2 \nonumber \\[3mm] & \qquad -19 h_r \rr U_1 U_2-28 \rr^2 U_1^2 \\[5mm]
& \hat c_{17} = -46 h_r^4 \rr U_2-17 h_r^3 \rr^2 U_1-16 h_r^3 \rr U_1-26 h_r^3 \rr^2 U_2-8 h_r^3 \rr U_2-22 h_r^3 U_2+3 h_r^2 \rr^3 U_1 \nonumber \\[3mm] & \qquad  -11 h_r^2 \rr^2 U_1-9 h_r^2 \rr U_1+12 h_r^2 \rr^2 U_2+14 h_r^2 \rr U_2+25 h_r \rr^3 U_1+11 h_r \rr^2 U_1\nonumber \\[3mm] & \qquad +29 h_r \rr U_1+16 h_r \rr^2 U_2+20 h_r \rr U_2+49 \rr^2 U_1 \\[5mm]
& \hat c_{18} = 29 h_r^4 \rr+27 h_r^3 \rr^2+8 h_r^3 \rr+21 h_r^3-2 h_r^2 \rr^3-15 h_r^2 \rr^2+11 h_r^2 \rr-23 h_r \rr^3-10 h_r \rr^2 \nonumber \\[3mm] & \qquad  -31 h_r \rr-31 \rr^2 \\[5mm]
\tilde{\beta}_1 & = \frac{v^4 h_r^2 (1 + h_r \rr)^2}{9 \hat c_{10}} \cdot \frac{ 3 v (h_r + \rr) - 2 (h_r U_2 + \rr U_1) }{2 v (h_r + \rr ) -(h_r U_2+\rr U_1)}
\end{align}
}
\pagebreak 

\appendix{\bf Appendix B:  {\color{black} Numerical methods}}
\medskip

We numerically solve the MMCC model \eqref{dimless_shear_mcc1}, \eqref{dimless_shear_mcc2} and the eKdV model \eqref{mcc_ekdv_shear}. Both models are solved using techniques as described in \cite{STCK2024} in dimensionless variables. As the MMCC model is in the fixed reference frame, we can ensure evolution in the moving reference frame by making the following change:
\begin{equation}
\zeta_t \mapsto \zeta_t + (v + \delta v) \zeta_x, \qquad \bar{u}_{2t} \mapsto \bar{u}_{2t} + (v + \delta v) \bar{u}_{2x},
\end{equation}
where $v$ denotes the linear long wave speed (as defined in \eqref{long_wave_speed}), and $\delta v$ represents the extra boost in velocity so that the frame travels with an overvall velocity of $v+\delta v$. Likewise, the KdV-type models are defined in terms of $\xi$ and so these already travel in the moving reference frame of velocity $v$. Hence, we make the change 
\begin{equation}
\zeta_T \mapsto \zeta_T - \dfrac{\delta v}{\eps} \zeta_\xi,
\end{equation}
so that the overall frame velocity  is $v+\delta v$ too. This means variables $x,\xi$ after this adjustment can be discretised in the same way. We discretise the spatial domain $x,\xi \in [\ximin,\ximax]$ using $N$ points so that
\begin{equation}
\xi_j = \ximin + (j-1) \delta \xi, \qquad \delta \xi = \dfrac{\ximax-\ximin}{N}, \qquad 0\leq j \leq N, 
\end{equation}
denotes the $j^{\text{th}}$ node. Here, $\xi_N \neq \ximax$ because we impose periodic boundary conditions. The time domains for fast time $t \in [\tmin,\tmax]$ and slow time $T = \eps t$ are also discretised with time-steps $\delta t = 5 \cdot 10^{-2}$ and $\delta T = 10^{-3}$, respectively. Throughout the numerics, we always initialise at $t= \tmin = 0$. Both MMCC and eKdV models {\color{black} are initialised} with the same initial condition and in the subsequent time evolution, they shed some radiation towards the boundaries. We use a type of sponge function to absorb the radiation at the boundaries at both the MMCC and eKdV level, and precise details of this sponge layer can be found in \cite{STCK2024}. \\

Both models contain spatial derivatives which are {\color{black} handled} using spectral methods, and time-stepping occurs via the Runge-Kutta $4^{\text{th}}$ order (RK4) scheme. Under a certain choice of parameters, it is necessary to filter high wavenumbers where energy grows in time evolution to prevent {\color{black} the KH} instability. This is implemented at the level of the RK4 algorithm, as every newly produced solution can have wavenumbers $|k| > k^*$ omitted from its Fourier spectrum.

\pagebreak

\end{document}